\makeatletter
   \def\input@path{{./elsarticle}}
\makeatother

\documentclass[twocolumn, 5p]{elsarticle}

\PassOptionsToPackage{hyphens}{url}

\usepackage{afterpage}
\usepackage{amsmath}
\usepackage{amssymb}
\usepackage{array}
\usepackage{booktabs}
\usepackage[inline]{enumitem}
\usepackage[T1]{fontenc}
\usepackage{graphicx}
\usepackage[breaklinks,hidelinks]{hyperref}
\usepackage{longtable}
\usepackage{makecell}
\usepackage{multirow}
\usepackage{pdflscape}
\usepackage{pifont}
\usepackage{tablefootnote}
\usepackage{stix}
\usepackage{wrapfig}
\usepackage{xcolor}

\graphicspath{{figures}}

\newcolumntype{P}[1]{>{\centering\arraybackslash}p{#1}}

\hypersetup{
    colorlinks,
    linkcolor={red!50!black},
    citecolor={blue!50!black},
    urlcolor={blue!80!black}
}

\journal{Computers \& Security}

\begin{document}

\begin{frontmatter}

\title{Network Intrusion Datasets: A Survey, Limitations, and Recommendations}

\author[inst1,inst2]{Patrik Goldschmidt\corref{cor1}}
\ead{patrik.goldschmidt@kinit.sk}
\cortext[cor1]{Corresponding author}

\author[inst2,inst3]{Daniela Chud\'{a}}
\ead{daniela.chuda@kinit.sk}

\affiliation[inst1]{1
    organization={Faculty of Information Technology, Brno University of Technology},
    addressline={Bo\v{z}et\v{e}chova 1/2}, 
    city={Brno},
    postcode={612 00},
    country={Czech Republic}
}

\affiliation[inst2]{
    organization={Kempelen Institute of Intelligent Technologies},
    addressline={Sky Park Offices, Bottova 7939/2A}, 
    city={Bratislava},
    postcode={811 09},
    country={Slovakia}
}

\affiliation[inst3]{
    organization={Faculty of Electrical Engineering and Information Technology, Slovak University of
Technology in Bratislava},
    addressline={Ilkovi\v{c}ova 3}, 
    city={Bratislava},
    postcode={841 04},
    country={Slovakia}
}

\begin{abstract}
Data-driven cyberthreat detection has become a crucial defense technique in modern cybersecurity. Network defense, supported by Network Intrusion Detection Systems (NIDSs), has also increasingly adopted data-driven approaches, leading to greater reliance on data. 
Despite the importance of data, its scarcity has long been recognized as a major obstacle in NIDS research. In response, the community has published many new datasets recently. However, many of them remain largely unknown and unanalyzed, leaving researchers uncertain about their suitability for specific use cases.

In this paper, we aim to address this knowledge gap by performing a systematic literature review (SLR) of 89 public datasets for NIDS research. Each dataset is comparatively analyzed across 13 key properties, and its potential applications are outlined. Beyond the review, we also discuss domain-specific challenges and common data limitations to facilitate a critical view on data quality. To aid in data selection, we conduct a dataset popularity analysis in contemporary state-of-the-art NIDS research. Furthermore, the paper presents best practices for dataset selection, generation, and usage. By providing a comprehensive overview of the domain and its data, this work aims to guide future research toward improving data quality and the robustness of NIDS solutions.
\end{abstract}


\end{frontmatter}

\date{April 2025}


\section{Introduction}
\label{sec:intro}

``\emph{The only truly secure system is one that is powered off, cast in a block of concrete and sealed in a lead-lined room with armed guards -- and even then I have my doubts.}'' \citep{dewdney1989_comp_recreations}. The legendary quote by Eugene H. Spafford from 1989 is not any less relevant nowadays. Each year, cyberattacks like malware or denial of service (DoS) cause losses of trillions of USD \citep{esentire_cybercrime_2024_report}. With recent attacks on hospitals and medical devices, even people's lives are in
danger \citep{papaioannou2020_survey_threats_iomt}. Therefore, robust security is required to minimize an attack surface and mitigate potential harm.

In general, information technology security spans three tasks: prevention, detection, and reaction. As perfect prevention is unattainable and reaction implicitly assumes that the attack has already taken place and has been detected, a considerable focus must also be put on detecting cyber threats \citep{apruzzese2023_sok_pragmatic}. This detection is done by intrusion detection systems.

An Intrusion Detection System (IDS) is a device or software application monitoring the activity of a target system or a computer network to identify their unauthorized use, misuse, or abuse \citep{mukherjee1994_nids}. Depending on the attack detection paradigm, we distinguish between signature-based (react on known attack patterns) and anomaly-based (react on unknown deviation from the norm) systems. Such systems can operate on the host level -- Host-based IDS (HIDS) or the network -- Network-based IDS (NIDS). Additional IDSs characteristics include the ability to operate in real-time and whether they merely report the attack or actively participate in its mitigation \citep{thakkar2022_ids_survey, buczak2016_dm_ml_ids}.

An efficient IDS maximizes attack detection (ideally 100\%) while minimizing the false alarm rate, i.e., legitimate samples flagged as malicious (ideally 0\%). The performance of detection systems thus needs to be evaluated, analyzed, and compared using datasets containing malicious and legitimate samples. For this reason, public intrusion datasets play an important role in IDS research and development \citep{flood2024_bad_design_smells_nids_datasets}.
The importance of data has been further emphasized by the rise of data-driven approaches for intrusion detection, most notably machine learning (ML) \citep{boutaba2018_ml_networking_survey}.

Despite the dire need for data, a general consensus within the network intrusion detection (NID) domain over the past decade was that there had been a lack of quality datasets for evaluating NIDS proposals \citep{abt2014_are_we_missing_labels, malowidzki2015_network, silva2022_netsec_datasets_bias}. This factor has significantly contributed to the poor adoption of ML-based methods into real-world systems despite huge scientific efforts \citep{sommer2010_sok_outside}.

As a response, the community has finally realized the importance of the data and put more effort into addressing data-related issues. However, the latest survey focused on listing, analyzing, and comparing network intrusion datasets was published in 2019 by \citet{ring2019_nids_datasets_survey}. Since then, dozens of new datasets have been released, but no comprehensive study on NID datasets has been published. Many new datasets thus remain undiscovered and unanalyzed by the community.

As we show in this work, data scarcity may no longer be the primary issue. Instead, we argue that \emph{datasets exist, but the community is largely unaware of their existence and lacks knowledge of best practices for their creation and usage.} This can be largely attributed to the lack of standardized NID data repositories and the neglect of domain-specific properties, leading to biased evaluations and over-optimistic results presented in many recent studies \citep{arp2022_dos_donts_ml_security}.

\textbf{Contributions}: Aiming to address the issues with discovering and handling data for network intrusion detection, we consider the primary contributions of the paper as follows:

\begin{enumerate}[itemsep=5pt, parsep=0pt]
    \item We conduct a systematic, up-to-date comparative survey of public datasets for NIDS research and development published until 2023 (inclusive). This will introduce readers to the latest trends and possibilities for conducting experiments and method benchmarks. To the best of our knowledge, this survey covers the largest number of publicly available NID datasets (89).

    \item In order to aid in data selection, we extract 13~properties for each dataset by manually downloading and analyzing the data. We also study the popularity of datasets in state-of-the-art NIDS research. To enhance accessibility, we share our analysis notebooks on GitHub\footnote{\url{https://github.com/xGoldy/nid-datasets}}, allowing others to review dataset structures without downloading the full data.

    \item We synthesize findings on flaws of existing datasets and describe their relation to domain-specific properties. In this manner, we elaborate on differences between NID and other ML application domains to promote a deeper understanding of data and, thus, its more effective utilization.

    \item Finally, we advocate the correct creation and usage of the data by discussing data handling recommendations.
\end{enumerate}

The rest of the paper is composed as follows: We define the survey's scope and methodology in Section~\ref{sec:methodology}. Section~\ref{sec:domain_specs} elaborates on the NID domain-specific properties and their impact on data. Section~\ref{sec:data_props} describes 13~collected properties used to compare the datasets. The main part of this paper -- the data survey, is presented in Section~\ref{sec:nid_data_survey}, listing datasets in a tabular format, categorizing them, and briefly elaborating on their properties and potential use cases. This section also analyzes dataset popularity and trending developments in NID data research. We discuss recommendations for data selection, creation, and usage in Section~\ref{sec:recommendations}. Finally, sections~\ref{sec:future_directions}--\ref{sec:conclusions} outline future research directions, discuss related work, and conclude the paper.


\section{Survey Methodology}
\label{sec:methodology}

In order to ensure comprehensive, unbiased, and reproducible results, we followed the Systematic Literature Review (SLR) methodology for computer science \citep{kitchenhamm2007_guidelines_slr}. This section outlines the survey's scope and research questions, with further details provided in~\ref{asec:survey_methodology}.

\subsection{Scope of the Survey}
\label{ssec:meth_scope}

Our paper aims to provide a holistic view of the network intrusion datasets and their properties. Therefore, we do not narrow down on any specific environment or intrusion type but include datasets containing network traffic from various environments like the Internet of Things (IoT), Cyber-Physical Systems (CPSs), Software Defined Networks (SDNs), or Critical infrastructure Supervisory Control and Data Acquisition (SCADA).

Although not restricted to a particular environment, we strictly include only datasets containing network traces. Therefore, datasets without activity observable on a computer network (i.e., packets or flows) were excluded. As a result, the survey did not consider datasets that include only environment-specific features like sensor values typical for CPS or SCADA.

Given this scope, we emphasize that this survey is not supposed to be an exhaustive list of all available IDS datasets and, thus, should not serve as the sole data reference for specific use cases. Instead, we encourage consulting other surveys focused on specific cybersecurity subdomains as needed.

\subsection{Research Questions}
\label{ssec:meth_research_questions}

Concerning the paper's goal and scope, we define our research questions as follows:

\begin{enumerate}[leftmargin=1.2cm, topsep=3pt, itemsep=1pt]
    \item[\emph{RQ1:}] \emph{What datasets are available for the network intrusion detection research and development?}
    \item[\emph{RQ2:}] \emph{What are the key qualitative and quantitative properties of the identified datasets?}
    \item[\emph{RQ3:}] \emph{Regarding the data properties from RQ2, are there any prevalent patterns or trends in dataset creation?}
    \item[\emph{RQ4:}] \emph{What were the most popular datasets within the domain in the past years?}
\end{enumerate}

The given questions were addressed via a systematic review of scientific literature published until the year 2023 (inclusive) and results synthesis. After filtering, we downloaded, manually analyzed, and compared 89 relevant datasets for NIDS research.  More information about this process, i.e., the search and selection of relevant studies, is described in~\ref{asec:survey_methodology}.


\section{Domain-Specific Properties and Their Impact on Data}
\label{sec:domain_specs}

The research on intrusion detection (ID) initially revolved around static signatures or simple statistical methods. Nevertheless, they require significant manual interventions, such as maintaining signature databases and threshold values. Motivated to minimize the need for manual labor and scale to big data more efficiently, the research community has gradually adopted data-driven approaches for ID. Nowadays, AI, and especially ML, are crucial parts of ID research \citep{apruzzese2023_role_of_ml_cybersec}.

Despite its rapid adoption in research, ML penetrated into practical cybersecurity solutions much slower than in other domains, such as computer vision or natural language processing, which have been actively using ML for over a decade \citep{apruzzese2023_role_of_ml_cybersec}. In fact, only a fraction of real-world network monitoring and threat detection teams relied on machine learning in early 2020s \citep{alahmadi2022_99fp_study_soc_alarms}.

The discrepancy between academic research and the actual deployments of ML-based NIDSs was discussed by \citet{sommer2010_sok_outside}, who noticed that operators were reluctant to adopt ML-based anomaly detectors despite high research efforts within the area. This fact was attributed to fundamental differences between NID and other ML-application domains, driven by factors like very high cost of errors, enormous data variability, semantic gap (ML explainability issues), and difficulties with evaluation -- including the lack of benchmark data. Although published more than 10 years ago, most of these issues arose from the specific properties of cybersecurity and computer networking domains, affecting the research and development efforts to this day. This section elaborates on these problems and their impact on overall NID data availability and quality.

\subsection{When Hell Comes to Earth: Unpleasant Properties of the Network Intrusion Detection Domain}
\label{ssec:domain_specs_domain_char}

Network intrusion detection combines properties of both cyberthreat detection and computer networking domains. These properties significantly influence how the data are collected, handled, and interpreted. If machine learning is used for detection, an assumption on the data to be independent and identically distributed (\emph{i.\,i.\,d.}) is also introduced \citep{dundar2007_learning_classifiers_noniid}.

In essence, the \emph{i.\,i.\,d.} assumes the data to be uncorrelated, while the data for training a model to be similar, i.e., drawn from the same distribution, as the future observed data. If not met, the model's performance might be negatively affected. However, the intrinsic properties of cyberthreat detection and computer networking interfere with this assumption and thus impair ML deployment in real-world scenarios \citep{apruzzese2023_role_of_ml_cybersec}. In the following paragraphs, we dive deeper into these properties and outline their potential impacts, as summarized in Figure~\ref{fig:nid_properties}.

\iftrue
\begin{figure}[t]
    \includegraphics[width=.85\linewidth]{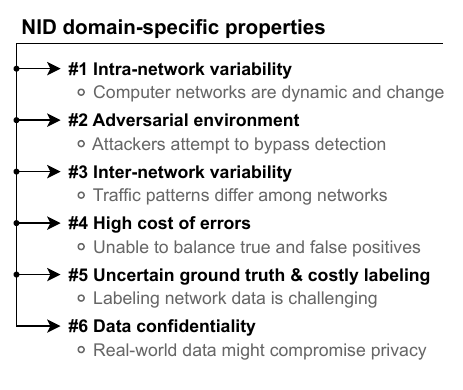}
    \vspace*{-1.25em}
    \caption{Summary of Network Intrusion Detection domain-specific properties affecting data collection, handling, and interpretation.}
    \label{fig:nid_properties}
\end{figure}
\fi

\begin{enumerate}[wide, label={\arabic*)}, topsep=0pt, itemindent=\parindent, itemsep=0pt, parsep=0pt, partopsep=0pt]
    \item \emph{Network traffic diversity and dynamicity (intra-network variability)}: Computer networks are highly dynamic by nature. Network environments constantly change as new devices are connected to the network, new applications and protocols are developed, or even new networking paradigms, such as the Internet of Things (IoT), are introduced. The temporal aspect further introduces periodicity patterns based on day-and-night cycles and office hours \citep{haffey2018_modeling_analysis_campus_edge}.

    \item \emph{Adversarial environment}: In addition to natural networks' dynamicity, the NID domain has to assume the implicit presence of adversaries. Due to the endless cybersecurity arms race, attacks continuously evolve, and new vulnerabilities are discovered. Moreover, the attackers may also attempt to bypass detection via adversarial samples crafted with obfuscation techniques or tiny perturbations of the input data \citep{he2023_adversarial_ml_nids_survey}.

    \item \emph{Inter-network variability}: Every network is unique \citep{sommer2010_sok_outside}. Each has its specific users, topology, and Internet service provider. For these reasons, traffic patterns may vary significantly among distinct networks. An attack with the same parameters executed in different networks will likely vary in its effects and observed traffic characteristics.
    
    \item \emph{High cost of errors}: Unlike other ML applications, the relative cost of any output error (i.e., misclassification) is very high. Regular applications can optimize for lower false alarm (false positive) or higher detection (true positive) rates. However, in NID, both need to be considered. Even a single missed anomaly might lead to a successful intrusion with dire consequences. On the other hand, a low relative false alarm rate can still create a tremendous amount of actual alarms due to a traffic imbalance, known as base-rate fallacy \citep{axelsson1999_base_rate_fallacy_ids}. For instance, with a 99\% true positive and 1\% false positive rate in a 1:100 class balance, we receive 100 false alarms for every 99 true attacks.

    \item \emph{Uncertain ground truth and costly labeling}: In many domains, e.g., computer vision, the ground truth is clear and stable -- ``A cat will always be a cat, whereas a dog will always be a dog'' \citep{apruzzese2022_sok_unlabeled_data}, and non-experts can often distinguish between classes very well \citep{law2011_human_computation}. Moreover, data augmentation \citep{shorten2019_surv_image_data_augmentation} can increase the effectiveness of existing labeled data. However, determining the ground truth in NID is complex. Due to its adversarial nature, a sample benign today might be malicious tomorrow (label shift). High inter-network variability can cause anomalies in one network to be considered benign elsewhere.
    
    In NID, even domain experts might struggle to verify ground truth, whereas attack detectors might return different and even contradicting results \citep{charlton2018_measuring_relative_accuracy}. These facts also disable crowdsourcing annotation, a common way to obtain large corpora of labeled data \citep{zhang2016_learning_from_crowdsourced}. Finally, even all the traffic of infected hosts cannot be considered malicious, as some might correspond to legitimate activities (e.g., ARP messages) \citep{apruzzese2022_sok_unlabeled_data}. For these reasons,
    accurately labeling network data is difficult and often infeasible.

    \item \emph{Data confidentiality}: Captures from real networks might contain sensitive and confidential information, such as IP~addresses or packet payloads, significantly limiting their sharing. Although anonymization techniques \citep{coull2009_anon_net_data_challenges}, like IP~address remapping or payload stripping, can be used, they decrease the data's value -- e.g., disabling payload-analysis NIDSs.
\end{enumerate}

Given the challenges of labeling and data sharing, the scarcity of high-quality real-world data becomes understandable. However, the problem persists even when such data are collected and labeled. Since data is captured at a specific point in time, future changes in the underlying distribution (data drift) or a relationship between input features and the output labels (concept drift) violate the \emph{i.\,i.\,d.} assumption. This phenomenon, known as model drift, leads to a gradual decline in a prediction model’s performance over time. Since NID is adversarial and highly dynamic, model drifting, amplified by the low tolerance of errors, is considered a major obstacle in ML utilization in real-world scenarios. While approaches like incremental \citep{masana2023_class_incremental_learning}, lifelong \citep{parisi2019_lifelong_learning_review}, or active learning \citep{settles2009_active_learning_survey} have been proposed to mitigate drift by leveraging newly observed data for model updates, applying ML for real-world NIDSs remains heavily constrained.

\subsection{Limitations of Existing Datasets}
\label{ssec:domain_specs_limitations}

\begin{figure}[t]
    \centering
    \includegraphics[width=.85\linewidth]{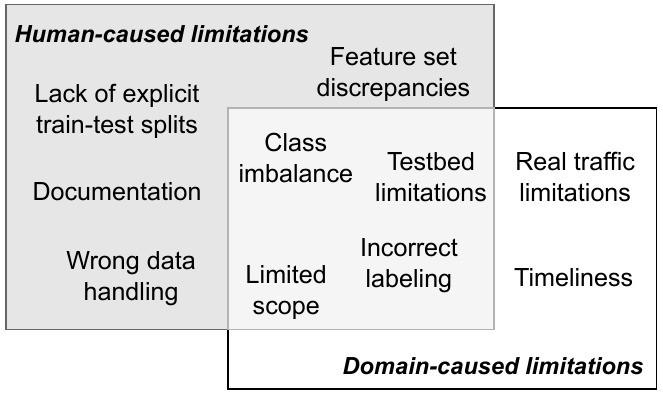}
    \caption{Limitations of datasets for network intrusion detection. As illustrated, some are a direct consequence of domain-specific properties and are challenging to address (e.g., timeliness), some are primarily caused by a human factor and can be mitigated completely (e.g., documentation), and those that lay in an intersection of both can be addressed only partially (e.g., class imbalance -- NID traffic is naturally imbalanced, but data authors might attempt to reduce it).}
    \label{fig:nid_data_limits}
\end{figure}

Despite the mentioned obstacles, various datasets for evaluating NIDSs have been proposed over the years. Nevertheless, all of them were directly or indirectly influenced by the domain properties, resulting in several limitations. Since these properties are unlikely to change, future datasets will likely exhibit some of these flaws as well, and no dataset will ever be ``perfect'' \citep{ring2019_nids_datasets_survey}. In the following paragraphs, we elaborate on these limitations and summarize them in Figure~\ref{fig:nid_data_limits}.

\begin{enumerate}[wide, label={\arabic*)}, topsep=0pt, itemindent=\parindent, itemsep=0pt, parsep=0pt, partopsep=0pt]

\item \emph{Timeliness}: The direct consequence of constantly evolving network and threat landscapes is data timeliness. Generally, it cannot be assumed that the behavior collected during a certain period will remain immutable over time \citep{viegas2017_trabid_dataset}. Therefore, static datasets become less representative of real-world scenarios as they age \citep{catillo2023_ml_public_ids_datasets}, affecting the soundness of the evaluation results. Although continual updates with the newest traffic patterns would mitigate this issue, immutable (static) datasets are required for fair benchmark comparisons between systems.

\item \emph{Real traffic limitations}: If the dataset authors want to utilize real-world traffic, data anonymization is required to ensure privacy. Nevertheless, anonymization decreases data realism and disables various use cases (e.g., payload analysis) \citep{abt2014_are_we_missing_labels}. Furthermore, capturing data from real-world enterprise networks might be unfeasible due to excessive traffic volume. For this reason, traffic sampling might need to be introduced, causing additional bias or incomplete network flows \citep{silva2017_inside_packet_sampling, meng2018_enhancing_trust_wids_sampling}.

\item \emph{Testbed limitations}: Due to the limitations of real-world data, most public datasets contain emulated traffic captured in a lab testbed environment. Despite some benefits of such an approach, its data are not guaranteed to resemble realistic network behavior and threats \citep{molinacoronado2020_survey_nids_kdd}. Such datasets tend to model over-simplified network conditions, failing to capture their real-world complexity and diversity \citep{catillo2023_ml_public_ids_datasets,flood2024_bad_design_smells_nids_datasets}. The lack of realism in benchmark data is considered a major obstacle in deploying anomaly-based NIDSs in real-world production environments \citep{sommer2010_sok_outside, viegas2017_trabid_dataset}. For instance \citet{catillo2021_demystifying_public_ids_data} have shown that many attacks from DoS public datasets do not resemble real-world characteristics due to being inefficient (not achieving their malicious intent) or assuming non-realistic environment configurations.

\item \emph{Class imbalance}: Modern machine learning classifiers are trained to maximize predictive accuracy by minimizing misclassification errors \citep{thabtah2020_data_imbalance}. Such a setting usually assumes roughly balanced classes, i.e., with similar prior probabilities. However, cyberthreat detection domains like NID typically deal with a huge imbalance in favor of the negative (benign) class, as only a few attack instances are observed along a vast amount of non-malicious communication \citep{catillo2023_ml_public_ids_datasets}. Although testing data should resemble real-world distribution to prevent spatial bias \citep{pendlebury2019_tesseract}, training a supervised classifier should be done on approximately balanced data. Nevertheless, most datasets are imbalanced, leaving to address the issue by their users \citep{bagui2021_resampling_imbalanced_nids_data}.

\item \emph{Documentation}: The dataset's accompanying metadata and documentation are crucial in determining the data value and its suitability for particular use cases. However, datasets often lack the documentation completely or omit necessary details. For this reason, the soundness of conclusions drawn by the data users is significantly hampered.

\item \emph{Lack of explicit train-test splits}: Many datasets are not explicitly split for training and testing. As a result, their users create arbitrary splits, causing their methods to be incomparable, as each may be trained and evaluated on a different data subset. Furthermore, applying methods like cross-validation or partitioning the data without respecting time dependencies may lead to temporal bias and inflated evaluation performance \citep{arp2022_dos_donts_ml_security, pendlebury2019_tesseract}. 

\item \emph{Feature set discrepancies}: As shown in our survey, many dataset authors propose their own feature sets, typically based on network flows. While some provide only 10\;--\;15 essential features, others offer over 100 advanced statistics. Several newer datasets adopt the CICFlowMeter tool \citep{lashkari2016_cicflowmeter} for feature extraction. An attempt to unify feature sets was made by \citet{sarhan2022_nids_feature_set,sarhan2021_netflow_datasets_ml_nids}, who proposed extracting 43 NetFlow~v9 features using nProbe \citep{ntop_nprobe} flow collector. Despite these efforts, the community has not yet agreed on a unified, widely accepted feature set, hampering the transferability and generalizability between datasets with different features.

\item \emph{Reproducibility:} In computer networking, it is infeasible to collect exactly the same data twice, even in identical environments. This property disallows the reproducibility of the raw packet data. Although it would not be a critical limitation by itself, reproducibility is further restricted for datasets publishing only extracted features (e.g., network flows) without packet data. Even if some datasets do provide raw data, they rarely publish source code and configuration for feature extraction and labeling. In all cases, reproducing the dataset (or even extracting the same features from packets) is infeasible, disallowing data validation and its possible extensions \citep{landauer2023_ait_ldsv2_dataset}.

\item \emph{Limited scope}: Due to the diversity of network technologies and cyberthreats, assembling a comprehensive dataset that covers the entire NID landscape is impractical. While specialized datasets (e.g., IoT) have been released recently, many scenarios and intrusions remain underrepresented. In particular, \citet{hindy2020_network_threats_taxonomy} note that public datasets cover only 33\% of known attacks. Given that new attacks and vulnerabilities continually emerge, a NIDS benchmarked on a static dataset will never be able to fully comprehend all potential threats in the wild.

\item \emph{Incorrect labeling}: Due to networking dynamicity and the common methods for traffic labeling (based on the attack time and the source and destination addresses of the attacker/victim \citep{guerra2022_datasets_labeling}), NID data are prone to mislabeling. The issue has been demonstrated by \citet{liu2022_error_prevalence_nids_data} and \citet{flood2024_bad_design_smells_nids_datasets} through analyses of several popular benchmark datasets. Such mislabeling can result in unrealistic results and benchmark destabilization \citep{northcutt2021_pervasive_label_errors}.

\item \emph{Methodically wrong data handling}: In addition to mislabeling, several other factors, primarily originating from human error in data collection and processing, affect the final dataset. A common issue we encountered was improper timestamping, as some datasets contain incorrect timestamps or time inconsistencies of events (e.g., attacks) with the documentation. We also observed disordered packets, undocumented capture gaps, and capture interruptions, resulting in improperly terminated files. Data processing issues, often due to bugs in feature extractor software \citep{catillo2023_ml_public_ids_datasets}, include missing values, duplicated features or entries, and zero-variance features (e.g., all zeros). These issues are frequently unreported, creating ambiguity for users and undermining the soundness of their conclusions.
\end{enumerate}

While many discussed limitations arising from the domain-specific properties are challenging to address, others (documentation, lack of explicit splits, feature set discrepancies, reproducibility, and data handling errors) are entirely within the control of dataset authors. Even when a limitation cannot be fully mitigated, proper documentation reduces ambiguity and significantly enhances data usability. For this reason, we consider documentation a crucial pillar of every dataset. However, most do not discuss any limitations at all. Supposing that authors do not release inherently flawed records on purpose, we theorize that they often neglect thorough analysis of their collected data.

From the perspective of data users, understanding these limitations aids in efficient data selection, proper handling, and avoiding evaluation errors. In Section~\ref{sec:recommendations}, we discuss recommendations to minimize or entirely prevent their impact.


\section{Network Intrusion Data and Its Properties}
\label{sec:data_props}

Depending on the dataset goal and capture attributes, we characterize network intrusion data by various properties that influence their use cases and the soundness of methods utilizing them. In this survey, we collected 13 essential properties to provide an unbiased, comprehensive view and comparison of available datasets for NID purposes. A portion of them was adopted from \citet{ring2019_nids_datasets_survey}'s survey. 

We split the extracted information into five categories: General information, Nature of data, Data volume, Network properties, and Evaluation. This section elaborates on these categories and individual properties to aid in data understanding, selection, as well as facilitate the evaluation of datasets not included in this survey in future research. Their summary is provided in Table~\ref{tab:collected_features}.

\begin{table*}[t]
    \centering
    \renewcommand{\arraystretch}{1.1}
    \caption{Summary of 13 collected properties of the surveyed NID datasets split into five categories. The table briefly describes each property, and the fourth column (``Possible values'') is a comma-separated list of values a particular property can attain. Curly brackets symbolize a set from which a single value can be drawn, whereas $\times$ is a Cartesian product of such sets in cases when multiple elements define a single property. Ampersand (``\&'') symbol is occasionally used to combine multiple values within a single entry.}
    \vspace*{0.4em}
    \begin{tabular}{p{1.9cm} p{1.85cm} p{5.9cm} p{7cm}}
    \textbf{Category} & \textbf{Property} & \textbf{Description} & \textbf{Possible values} \\ \toprule
    
    \multirow{4}{2cm}[-.5\baselineskip]{\centering General information}
     & Year & Year of the data collection & \{1998, 1999, $\dots$, 2023\}, Now\textsuperscript{1} \\
     & Focus & Focus (scenario) of the dataset & text \\
     & Normal & Normal (benign) traffic type & real, emul-\{p, v, pv\}, synth, no \\
     & Attack & Attacking traffic type, number of attacks, and attack categories & \makecell[tl]{\{real, emul-\{p, v, pv\}, synth\} $\times\ (\mathbb{N})\ \times$ \\Attack types (see Table~\ref{tab:att_classification})} \\ \midrule

    \multirow{3}{2cm}[-.5\baselineskip]{\centering Nature of data}
     & Format & Format of the data & packet (\{bin, txt\}), flows (\{uni, bi\}), logs (\{net, host, both\}), nethost, other \\
     & \# Features & Number of dataset features & $\{n\ |\ n \in \mathbb{N}\}$, ``-'' for raw data formats \\
     & Anonymized & Anonymized parts of the dataset & no, IPs, MACs, payload, time, names, n.\,s.\textsuperscript{2} \\ \midrule

    \multirow{4}{2cm}[-.5\baselineskip]{\centering Data volume}
     & Size & The amount of records and/or packets. Train and test sets split by the `$+$' sign. & $\mathbb{R}^+$ \{k, M, $\dots\}\textsuperscript{3}\ \times\ $  \{pkts, flows, recs.\}\\
     & Duration & Data capture time length and its continuousness. Train/test split by the `+' sign. Periodicity expressed with `x' (times). & $\{n\ |\ n \in \mathbb{R}^+\}\ \times\ $\{m, h, d, w, mo\}\textsuperscript{4} $\times$ \{cont., discont., periodic\}\\ \midrule

    \multirow{3}{2cm}[-\baselineskip]{\centering Network properties}
     & Network type & Type of the network environment & \{small, medium, large\} $\times$ \{academic, enterprise, industrial, military, wireless, cloud, honeypot, IoT, SDN, P2P, ISP, $\emptyset$\}, mixed, n.\,s. \\
     & Complete capture & Traffic from the entire network is provided & yes, no, n.\,s \\ \midrule

     \multirow{2}{2cm}[0pt]{\centering Evaluation}
      & Split & Train and test subsets are pre-split & yes, no \\
      & Labels & The ground-truth is provided & yes, no, indir.
     \\ \bottomrule
    \end{tabular}
    \label{tab:collected_features}
\begin{minipage}{0.49\textwidth}
{\small \vspace{0.4em}
\textsuperscript{1} signifies the capture was still ongoing at the time of writing

\textsuperscript{2} ``n.\,s.'' is an abbreviation for ``not specified''
}
\end{minipage}
\begin{minipage}{0.49\textwidth}
{\small \vspace{0.4em}
\textsuperscript{3} metric suffixes, i.e., k -- $10^6$, M -- $10^9$

\textsuperscript{4} time units (m\;--\;minute, h\;--\;hour, d\;--\;day w\;--\;week, mo\;--\;month)
}
\end{minipage}
\end{table*}

\subsection{General Information}
\label{ssec:dprops_general_info}

When selecting a dataset, our primary focus is on the scenarios it includes, along with other details like the data collection period, the types of present attacks, and the environment used to perform the capture:

\textbf{Year}: The specific year(s) or period of data collection. Since many datasets are published months or years after collection, reporting the actual collection date is crucial for understanding the captured threat landscape and traffic characteristics (e.g., popular applications). For continuously ongoing captures, such as MAWILab \citep{fontugne2010_mawilab}, we denote the value as \emph{Now}.

\textbf{Focus}: This property summarizes the dataset scenario and its potential use cases. Based on the focus, we distinguish between general- or special-purpose datasets. General-purpose datasets (\emph{General}) are captured in traditional local area or enterprise networks with various intrusive scenarios. Depending on the nature of the data, we include \emph{IDS} (both host and network information) or \emph{NIDS} (only network data are present) identifiers. In contrast, special-purpose datasets focus on a particular attack like DDoS or a specific environment such as IoT.

\textbf{Normal}: Indicates whether normal (benign) traffic is present and how it was generated. If absent, the value is recorded as \emph{no}. Otherwise, we distinguish between data obtained from \emph{real}, emulated (\emph{emul}), or synthetic (\emph{synth}) environments. Since various definitions of these terms are found in the literature (especially, emulated and synthetic terms are often used interchangeably), we establish the following definitions for this survey:

\begin{itemize}[leftmargin=1.3em]
    \item \emph{real} -- Traffic originating from real-world operational networks captured using tools like \texttt{tcpdump} or \texttt{nfdump}. It reflects genuine network interactions with purposes beyond traffic generation. Since it contains real information, it must be anonymized. Additionally, real data cannot be declared benign with absolute confidence, as undetected attacks might be present. Instead, it is referred to as \emph{background} traffic without evident anomalous or malicious behavior.

    \item \emph{emul} -- Emulated traffic is captured in a network environment the same way as \emph{real} traffic, however, with a different purpose. While \emph{real} traffic encompasses genuine network interactions (e.g., web browsing), the primary purpose of emulation is to simulate and capture a specific behavior via activity simulation scripts, automation frameworks like Selenium\footnote{Selenium is a framework for browser automation and web page functional testing. Webpage: \url{www.selenium.dev}.}, or traffic generators. We distinguish between three types of emulated traffic depending on the environment:

    \begin{itemize}[topsep=0pt, itemsep=3pt]
        \renewcommand{\labelitemii}{$\circ$}
        \item \emph{emul-p}: Traffic from physical networks with real network equipment like routers and end-point stations.
        
        \item \emph{emul-v}: Traffic from virtualized networks and virtual machines without actual physical representation, e.g., NS3 \citep{riley2010_ns3}, Mininet \citep{lantz2010_mininet}, or cloud-based networks.
        
        \item \emph{emul-pv}: Traffic from mixed environment combining physical equipment with virtualized elements.
    \end{itemize}

    \item \emph{synth} -- Similar to emulation, synthetic traffic is also designed for the sole purpose of simulating specific behaviors. Nevertheless, whereas emulation creates and captures actual packets traversing a network, synthetic traffic implies no network activity. Instead, the data is generated by models that mimic activities without their actual network representation.
\end{itemize}

We highlight these three traffic origins due to differences in data characteristics, especially realism. In general, real-world traffic is considered ideal as it offers genuine network interactions. However, it suffers from mentioned privacy concerns and may lack coverage of specific events, such as flash crowds \citep{ari2003_flash_crowds}, if they are not present during the capture.

In contrast, emulated and synthetic traffic offer greater flexibility, lower collection costs, and more reliable labeling. However, as outlined in Section~\ref{ssec:domain_specs_limitations}, their crucial limitation is realism. Due to the complexity of network traffic, replicating real-world scenarios in a laboratory environment is very challenging \citep{catillo2023_ml_public_ids_datasets}. Although several works have tried to tackle this issue, e.g., $\beta$-profile system \citep{shiravi2012_iscx2012}, realistic simulation of network behavior remains an important research direction \citep{hindy2020_network_threats_taxonomy}.

While traffic generation methods significantly impact realism, the environment type also plays a role. In general, physical deployments for traffic simulation are considered more realistic than virtual ones \citep{chou2020_datadriven_nids}. Therefore, purely virtual emulation is expected to produce less realistic results than physical, although the differences might be minor.

Naturally, synthetic traffic created without any network activity is considered the least realistic. However, its high flexibility and lack of privacy concerns make it a promising research direction to tackle highly dynamic and ever-changing domains such as intrusion detection. Synthetic traffic has already shown promising results in generating normal traffic with statistically similar properties to more realistic data \citep{schoen2024_tale_two_methods}. Section~\ref{sssec:ndatasurv_trends_traffic_generation} discusses trends in traffic generation in detail.

\textbf{Attack}: This property specifies three aspects regarding intrusions in the dataset:
\begin{enumerate*}[label={\arabic*)}]
    \item Origin environment,
    \item The number of distinguishable attacks, and
    \item Attack types.
\end{enumerate*}

The attack origin environment is defined the same as for normal traffic -- \emph{real}, \emph{emul-\{p,\,v,\,pv\}}, and \emph{synth}. Real attacks correspond to traffic with actual malicious intent captured in real-world production networks, i.e., by detecting and sampling an ongoing attack, as in the Booters dataset \citep{santanna2015_booters}. Emulated attacks are launched in controlled environments just to simulate attacks' behavior, usually using the same tools as real cyber-attackers (e.g., in Kali Linux\footnote{Kali Linux is a Linux distribution designed for digital forensic analysis and penetration testing. Homepage: \url{https://www.kali.org/}.}). Finally, synthetic attacks were not captured in a network but rather generated by specialized algorithms.

Although in between real and emulated categories, we classify traces of sandboxed malware connected to the Internet (e.g., CTU-13 \citep{garcia2014_ctu13}) as real traffic, assuming its source code was not altered. In this case, we consider its patterns (e.g., spreading, Command \& Control traffic) representative of real malware even when executed in a controlled environment.

The environment origin is followed by a value in brackets indicating the number of distinguishable attacks. This represents the number of malicious classes clearly separable in the data (either on a feature-based level based on labels or via raw packet filtering). For datasets like CUPID \citep{lawrence2022_cupid} that declare multiple attacks but provide only binary labels, i.e., distinguishable anomalies, the value is listed as one. Datasets with unspecified attack counts, e.g., MAWILab \citep{fontugne2010_mawilab}, have no associated value.

In the third part, we summarize the attack types to clarify the dataset's objectives and associated traffic.
Although existing cyberattack taxonomies \citep{santos2025_towards_robust_cyberattack_taxonomies,derbyshire2018_analysis_cybersec_attack_taxonomies} could be used, we developed a custom scheme (Table~\ref{tab:att_classification}) to provide a concise summary of common attack events in NID datasets. Therefore, rather than a comprehensive taxonomy, it serves as a quick reference for identifying attacks within datasets. In contrast to other taxonomies that aim for completeness and specificity \citep{santos2025_towards_robust_cyberattack_taxonomies}, our scheme intentionally includes the \emph{Unknown} category to account for potentially undiscovered attacks, as well as the \emph{Other} category for less common ones. Each attack type is then represented by an acronym (e.g., ``D'' for a (D)DoS attack), with multiple types forming a single acronym string.

For instance, the string \emph{emul-v (10) DRO} for the Bot-IoT dataset \citep{koroniotis2019_botiot_dataset} indicates 10 different attack classes generated by emulation in a virtualized environment. They belong to (D)DoS, Reconnaissance, and Other attack categories due to data theft and keylogging traces.

\begin{table*}[t]
    \centering
    \caption{The scheme for a quick reference of attacks present in the reviewed datasets. Each attack category is assigned an acronym, such as ``B" for Brute-force. A sequence of such acronyms (e.g., BCO) then represents the presence of given types of attacks within the dataset.}
    \vspace*{0.4em}
    \begin{tabular}{p{2.8cm} c p{12.2cm}}
    \textbf{Attack category} & \textbf{Acronym} & \textbf{Description} \\ \toprule
    Brute-force & B & Brute-force performs trial-and-error attempts to guess login credentials, passwords, or encryption keys. These trials are typically based on pre-compiled lists and dictionaries. \\ \midrule
    
    C\&C Botnet traffic & C & Command and Control (C\&C) traffic is used by infected machines (bots) to communicate and receive orders from the master server within the client-server botnet architectures. \\ \midrule

    Denial of service & D & Denial of Service (DoS) and Distributed DoS (DDoS) are deliberate attempts to disrupt a target machine or network, causing unavailability for regular users. \\ \midrule

    Reconnaissance & R & Reconnaissance (scanning, probing) is a process of gathering information about potential targets, vulnerabilities, and attack vectors. \\ \midrule

    Vulnerability \mbox{exploitation} & V & The act of exploiting various vulnerabilities, such as SQL injection or buffer overflow, typically to gain remote access or escalate privileges on the target machine. \\ \midrule

    Other & O & Environment-specific or uncommon attacks present in a minority of datasets -- e.g., SPAM, data exfiltration, ransomware, cryptomining, or ARP spoofing. \\ \midrule

    Unspecified & ? & Signifies that some of the previous attack categories may be present but are not explicitly acknowledged. This happens when the dataset contains data from a real network, so the presence of intrusion attempts in the data is uncertain. \\ \bottomrule
    \end{tabular}
    \label{tab:att_classification}
\end{table*}

\subsection{Nature of Data}
\label{ssec:dprops_data_nature}

Three properties in this category describe how the data are stored and whether they have been preprocessed.

\textbf{Format}: Network monitoring is primarily based on collecting three types of data: packets, network flows, and logs \citep{zhou2018_survey_network_data_collection}. Accordingly, NID datasets are typically distributed in these formats. Some also offer auxiliary data like end-host logs, routing tables, or SNMP databases. This survey outlines the four most common data formats and includes the category of \emph{other} for specific cases, as described below:

\begin{itemize}
    \item \emph{packet}: Data provided in a per-packet manner. Most such datasets contain raw (binary) packets distributed as Packet Capture (PCAP) files. However, some (e.g., AWID2 \citep{kolias2015_awid2_dataset}) provide already pre-extracted full packets or their specific fields in a textual format. We distinguish them by the \emph{packet (bin)} and \emph{packet (txt)} identifiers.

    \item \emph{flows}: A flow is a sequence of packets with some common properties that pass through a network device \citep{claise2004_rfc3954_netflow}. These properties are typically considered as a 5-tuple: source IP, destination IP, source port, destination port, and a transport protocol identifier, used for packet aggregation. Aggregation can either be uni-directional (\emph{flows (uni)}) or bi-directional (\emph{flows (bi)}), recording both sides of the connection in a single flow entry.
    
    Collected traffic features depend on flow exporters \citep{vormayr2020_why_are_my_flows_different}, including general-purpose tools like YAF \citep{inacio2010_yaf}, nProbe \citep{ntop_nprobe}, and NFStream \citep{aouini2022_nfstream}, as well as security-focused tools like Argus \citep{qosient_argus}, Zeek \citep{paxson1999_bro}, and CICFlowMeter \citep{lashkari2016_cicflowmeter}. These exporters rely on various flow protocols, primarily NetFlow v5/v9 \citep{claise2004_rfc3954_netflow}, IPFIX \citep{aitken2013_rfc7011}, sFlow \citep{phaal2004_sflow5}, and OpenFlow \citep{openflow_switch_specs151}. While essential flow features (Table~\ref{tab:flow_features}) are commonly found in most flow-based datasets, they are often insufficient for cybersecurity applications. As a result, extended versions with additional statistics, such as NetFlow v9 \citep{cisco_netflow_v9_format} or CICFlowMeter \citep{lashkari2021_cicflowmeter_features} feature sets, are commonly used. Most frequently, these datasets are distributed in a textual comma-separated-values (CSV) format.

    \item \emph{logs}: In addition to data directly captured from a network (packets or flows), many practical IDSs utilize logs and alerts from other systems to improve detection capabilities or minimize false alarms. Therefore, some datasets also provide event logs from end-host systems (\emph{logs (host)}), network (\emph{logs (net)}), such as firewall logs or signature-based NIDS alerts, or from both sources at once (\emph{logs (both)}). They can be used for indirect data labeling or evaluation of systems working with multiple data sources.

    \item \emph{nethost}: Instead of providing separate captures from a network and logs from end-host devices, some datasets provide pre-extracted feature sets combining the two. For instance, a dataset feature set can consist of flow data enriched with statistics from the host system of interest, such as a server that the attackers target. These datasets enable faster experimentation by eliminating the need for a custom feature extractor from multiple data sources, but at the cost of decreased flexibility.
    
    \item \emph{other}: Specific data formats not corresponding to any of the above. An example of this case is UNR-IDD \citep{das2023_unridd_dataset}, which provides feature vectors as per-port statistics.
\end{itemize}

\begin{table}[t]
    \small
    \centering
    \caption{Common features for uni-directional flow datasets. Note that bi-directional datasets include separate fields for features 8--11, tracking source $\rightarrow$ destination and destination $\rightarrow$ source traffic separately, resulting in 15 features. Additional features may be added for data labels. Identifiers and some features might vary across datasets and protocols -- e.g.,  NetFlow v9 \citep{claise2004_rfc3954_netflow} uses \texttt{FIRST\_SWITCHED} as the first flow timestamp and \texttt{LAST\_SWITCHED} for the last instead of defining duration explicitly. Nevertheless, despite naming differences, feature semantics remain identical across datasets.}
    \vspace*{0.4em}
    \begin{tabular}{l l l}
    \textbf{\#} & \textbf{Identifier} & \textbf{Description} \\ \toprule
    1 & \texttt{stime} & Timestamp of the first flow packet \\
    2 & \texttt{dur} & Flow duration \\
    3 & \texttt{srcip} & Source IP address \\
    4 & \texttt{dstip} & Destination IP address \\
    5 & \texttt{srcport} & Source transport protocol port \\
    6 & \texttt{dstport} & Destination transport protocol port \\
    7 & \texttt{proto} & Transport protocol \\
    8 & \texttt{tos} & Type of service \\
    9 & \texttt{pkts} & Destination to source packet count \\
    10 & \texttt{bytes} & Destination to source bytes sum \\
    11 & \texttt{flags} & Cumulative OR of TCP flags \\
    \bottomrule
    \end{tabular}

    \label{tab:flow_features}
\end{table}

In general, packet-based datasets provide more detailed information because they provide exact contents and timestamps of every packet in contrast to flow aggregates. If packet payloads are not stripped, they also enable deep packet inspection (DPI) \citep{finsterbusch2014_payload_based_classification_survey}, a technique for full packet analysis. However, with increasing traffic volumes and widespread encryption, flow-based NIDSs have gained popularity for being unaffected by encryption and offering higher traffic throughput \citep{umer2017_flowbased_ids_techniques,sperotto2010_flow_based_ids}.

\textbf{The number of features (\# features)}: Defines the length of the data feature vector, including labels. This information helps to determine whether the dataset uses a basic feature set as in Table~\ref{tab:flow_features} (i.e., 10--18 NetFlow features) or an extended one. Note the value is only relevant for datasets with pre-extracted features.

\textbf{Anonymized}: In some cases, such as captures from real-world networks, the data require anonymization to preserve privacy or hide network details. This property lists parts of the data its authors decided to anonymize. While anonymizing IP addresses via prefix-preserving algorithms like CryptoPan \citep{xu2002_cryptopan} is typically sufficient for flow-based data, full packet captures require additional actions. Oftentimes, MAC addresses are also obscured, and packet payloads trimmed to a certain length (e.g., to retain application headers) or removed entirely.

\subsection{Data Volume}
\label{ssec:dprops_data_volume}

This category evaluates the volume of the data from two perspectives: the number of data points (i.e., raw packets or feature vectors) and the data capture duration.

Although this information is typically declared in dataset documentation, we did not rely on it. Instead, we downloaded and analyzed the data ourselves due to occasional inconsistencies between the data and its documentation. For datasets with predefined training and testing splits, we report these volumes separately using a ``+'' sign. The properties are defined as follows:

\textbf{Size}: Specifies the number of packets (\emph{pkts}), flows (\emph{flows}), or records (\emph{recs}) in the dataset. For their representation, we use metric suffixes like kilo (k) = $10^6$ and mega (M) = $10^9$. A comma sign (``,'') represents a delimiter between thousands. In order to keep the table concise, we omit listing the size of associated log files, as the survey is focused on network data.

\textbf{Duration}: This property represents the dataset capture duration. While dataset documentation often defines duration as the time span between the first and last data timestamps, this approach may be misleading due to the presence of potential temporal gaps. Therefore, this survey aims to report the actual data duration by disregarding such gaps.

Supposing the dataset entries are sorted chronologically, with $t_i$ as a timestamp of record $x_i$ and a difference between two adjacent records as $\tau_i = t_i - t_{i-1}$, a temporal gap in the data occurs if $\tau_i > \epsilon$. If no temporal gap exists ($\forall i: \tau_i \leq \epsilon$), i.e., the network activity is present throughout the entire dataset time span, we refer to the dataset as continuous (\emph{cont.}). Otherwise, a dataset is discontinuous (\emph{discont.}) if at least one capture gap is present ($\exists i: \tau_i > \epsilon$). In this matter, an actual capture duration $D$ (Equation~\ref{eq:duration}) is calculated by ignoring gaps longer than $\epsilon$.
\begin{equation}
    D = \sum\limits_{i: \tau_i \leq \epsilon} \tau_i.
    \label{eq:duration}
\end{equation}

In this survey, data continuity is based on $\epsilon$ relative to the dataset span (the first and last timestamps difference), as, for instance, a 10-minute gap is more acceptable in a month-long dataset than in a one-hour dataset. Therefore, we set $\epsilon$ as 1\% of the analyzed dataset's absolute time span. Additionally, assessing continuity for datasets with a pre-defined train-test split is performed separately with different $\epsilon$ for each split, as merging them would be semantically incorrect.

A special case of continuity is periodicity (\emph{periodic}), which refers to continuous blocks recurring in a regular pattern, such as specific hours per day or days per week. Periodic captures are noted as \emph{periods x duration}. For instance, the DARPA 1998 dataset \citep{lippmann2000_darpa1998} was captured over nine weeks on weekdays. Since it also contains a train-test split, it is listed as \emph{7x5d + 2x5d periodic}, indicating seven five-day periods for training and two for testing, with each period being continuous.

In addition to determining how the data duration is computed, information about continuity also aids in assessing whether the data would be suitable for detection systems relying on time dependencies between samples. If the data was discontinuous, these systems could not utilize their dependency-analysis mechanism, likely resulting in biased evaluation results.

For the practical analysis, we utilized \texttt{capinfos} and \texttt{tshark} tools for raw PCAP data and \texttt{pandas} for CSV files. We relied on duration from data documentation if the dataset did not offer a raw PCAP and its extracted features did not contain timestamps. In cases where no duration could be inferred from data and its documentation, the value is listed as ``n.\,s.'' (not specified).

\subsection{Network Properties}
\label{ssec:dprops_net_props}

The General information category distinguished the character of environment as real-world, emulated, or synthetic for both types of traffic. In this category, we provide additional information about the environment -- the underlying network type and whether the capture covers the entire network.

\textbf{Network type}: This property defines the network's size and type to indicate expected traffic patterns. We categorize network sizes into three groups:
\begin{enumerate*}[label={\arabic*)}]
    \item \emph{small} (up to 99 hosts) -- a small office/home office (SOHO), small company, or a household network,
    \item \emph{medium} (100-499 hosts), and
    \item \emph{large} (500+ hosts) -- a large enterprise or an internet service provider (ISP).
\end{enumerate*}
In addition to size, we also report the network's type, with the most common being \emph{academic}, \emph{enterprise}, \emph{industrial}, \emph{military}, \emph{wireless}, \emph{cloud}, \emph{honeypot}, internet of things (\emph{IoT}), software-defined network (\emph{SDN}), peer-to-peer (\emph{P2P}), and internet service provider backbone (\emph{ISP}). This information is omitted if no specific type is provided in the documentation. Datasets with traces combined from multiple networks are assigned the value of ``\emph{mixed}''.

\textbf{Complete capture}: Specifies whether the dataset contains data from the entire underlying network. In general, it is desirable to capture all network communication to ensure traffic diversity and realism. However, some datasets capture only from a single host (e.g., honeypot) or a specific set of hosts due to privacy or practical reasons. Such a setup might skew the traffic patterns (e.g., honeypots would receive significantly more malicious traffic than regular hosts \citep{sethuraman2023_flowbased_honeypot}). Therefore, specialized evaluations or NIDS types should be employed when dealing with incomplete captures.

\subsection{Evaluation}

Finally, this category groups properties influencing NIDS evaluation, such as whether the data have already been pre-split for training and testing and whether the dataset is labeled.

\textbf{Split}: Defines whether the dataset offers pre-split train and test subsets. This is desirable, as it allows transparent comparisons of different NIDSs evaluated on the same data. On the other hand, datasets without a train-test split have to be split by their users, potentially leading to evaluation bias and selective reporting \citep{arp2022_dos_donts_ml_security, lipton2019_research_for_practice_troubling_trends}.

\textbf{Labels}: Specifies whether the dataset is labeled -- a crucial property in analyzing and comparing NIDSs. We assign the value \emph{yes} only if an unambiguous mapping between data instances and classes exists. This applies to direct labels in feature vectors but also when classes can be separated via filters (e.g., IP-based packet labeling in 2017-SUEE8 \citep{lukeseder2017_suee_dataset}) or directory structures (e.g., CAIDA DDoS 2007 \citep{caida_ddos_attack2007}).

In addition to being labeled or unlabeled (\emph{yes/no}), datasets can also be labeled indirectly (\emph{indir.}). In this case, no direct mapping between records and classes exists, but the traffic is differentiated alternatively, such as with IDS alerts or log files. While this provides some level of ground truth, it is hardly completely precise, requires additional effort from users to correlate logs with the data, and can lead to ambiguous interpretations. As a result, evaluating with indirectly labeled or unlabeled data is error-prone and should be approached cautiously.

This section has presented 13 properties of NID datasets across five categories. We used these properties to establish an extraction protocol and collect data from the reviewed datasets. The survey results are presented in the following section.


\section{Network Intrusion Datasets}
\label{sec:nid_data_survey}

This section, the core of this survey, lists and comparatively analyzes datasets for network intrusion detection. As outlined, the datasets were selected in accordance with the systematic literature review methodology discussed in Section~\ref{sec:methodology} and~\ref{assec:meth_search_criteria}. Afterward, 13 properties defined in Section~\ref{sec:data_props} were extracted through documentation review and data exploratory analysis. The analysis results -- Jupyter Notebooks and scripts, are available at our GitHub repository (linked in Section~\ref{sec:intro}).

The survey results are summarized in Table~\ref{tab:data_survey}, with each row representing a single dataset. In rare cases like NF-UQ-NIDS \citep{sarhan2022_nids_feature_set}, two related datasets are listed as one entry divided by a ``/'' symbol. The entries are sorted in ascending order by year and then alphabetically. Datasets spanning multiple years are ordered based on their last capture year but appear before datasets captured in a single year. Dataset download links (functional in early 2025) are in their associated references.

\begin{figure}[t]
    \centering
    \includegraphics[width=.9\linewidth]{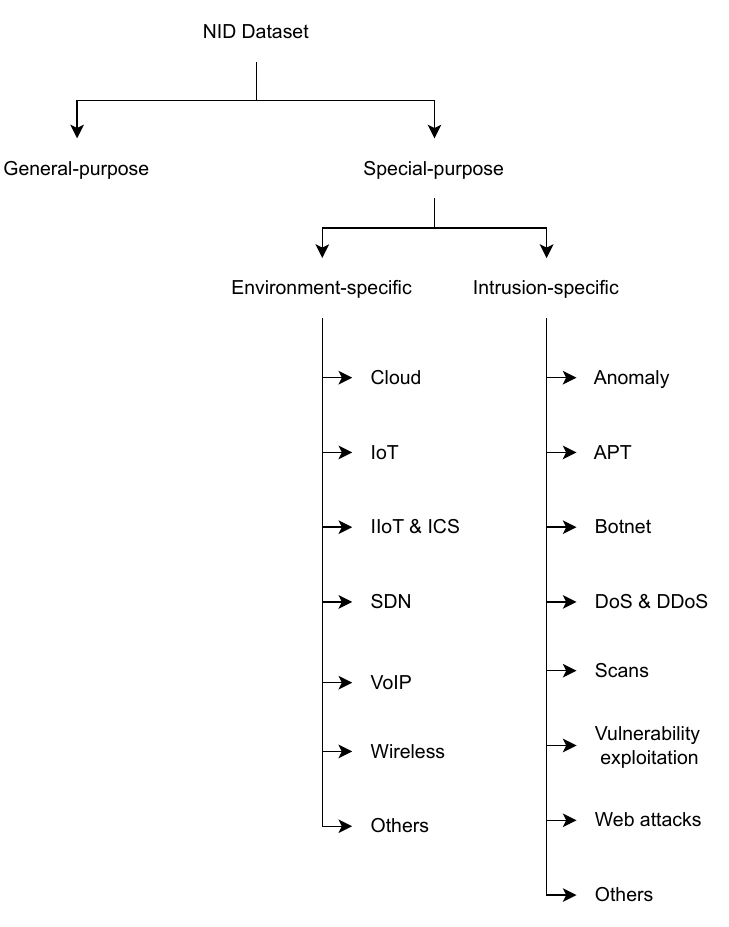}
    \vspace*{-0.5em}
    \caption{Taxonomy of datasets for Network Intrusion Detection based on their focus. Note that special-purpose categories are not exclusive, as a single dataset can both focus on a particular intrusion type captured in a specific environment.}
    \label{fig:datasets_taxonomy}
    \vspace*{-1em}
\end{figure}

The following subsections categorize and analyze the surveyed datasets in more detail. In order to group the data, we developed a taxonomy (Figure~\ref{fig:datasets_taxonomy}) classifying datasets as general-purpose (Section~\ref{ssec:ndatasurv_data_general}) or special-purpose (Section~\ref{ssec:ndatasurv_data_special}) based on their focus and intended use cases. Afterward, Section~\ref{ssec:ndatasurv_other_datasets} outlines sources for other datasets not included in this survey. In Section~\ref{ssec:ndatasurv_popularity}, we analyze the popularity of datasets in contemporary state-of-the-art NIDS research. Finally, Section~\ref{ssec:ndatasurv_trends} highlights observed trends in network intrusion data research.

\subsection{General-Purpose Datasets}
\label{ssec:ndatasurv_data_general}

General-purpose datasets, labeled as \emph{General}, provide data for benchmarking intrusion detection systems for traditional network environments such as local area networks (LANs) or enterprise networks. They offer a more diverse set of intrusions rather than focusing on specific attacks.

Although general-purpose datasets cannot be precisely divided into subcategories without ambiguity (Figure~\ref{fig:datasets_taxonomy}), we organize this subsection into five parts. Initially, we describe the first-generation datasets produced by DARPA. Next, we examine CIC datasets, which have largely replaced DARPA in recent research. Finally, we review other general-purpose datasets, including NetFlow, newly released, as well as already established but less commonly utilized.


\afterpage{%
\clearpage%
\newgeometry{vmargin=1.125in, hmargin=0.586in, nohead}%
\onecolumn
\begin{landscape}
\setlength\LTcapwidth{\linewidth}
\centering

\small


\renewcommand\cellalign{tl}

\tabcolsep=0.19cm

\begin{longtable}[!htbp]{>{\raggedright}p{2.15cm} >{\centering}p{0.6cm} >{\raggedright}p{2cm} >{\raggedright}p{1.15cm} >{\raggedright}p{1.78cm} >{\raggedright}p{1.75cm} p{1cm} >{\raggedright}p{1.15cm} >{\raggedright}p{1.78cm} >{\raggedright}p{1.54cm} >{\raggedright}p{1.4cm} p{0.85cm} p{0.7cm} p{0.8cm}}

\caption{Comparative table (spans multiple pages) of available public network intrusion datasets with 13 extracted properties. See Section~\ref{sec:data_props} for their explanation.}\\

\textbf{Dataset} & \textbf{Year} & \textbf{Focus} & \textbf{Normal} & \textbf{Attack} & \textbf{Format} & \textbf{\#\,Ftrs.} & \textbf{Anon.} & \textbf{Size} & \textbf{Duration} & \textbf{Net type} & \textbf{Compl.} & \textbf{Split} & \textbf{Labels} \\ \toprule \endfirsthead 

\textbf{Dataset} & \textbf{Year} & \textbf{Focus} & \textbf{Normal} & \textbf{Attack} & \textbf{Format} & \textbf{\#\,Ftrs.} & \textbf{Anon.} & \textbf{Size} & \textbf{Duration} & \textbf{Net type} & \textbf{Compl.} & \textbf{Split} & \textbf{Labels} \\ \toprule \endhead 

DARPA 1998 {\footnotesize\citep{lippmann2000_darpa1998}} & 1998 & General IDS (LANs) & emul-pv & \makecell{emul-pv\,(32)\\BDRVO} & \makecell{packet (bin)\\logs (host)} & - & no & \makecell{33.8M +\\2.2M pkts} & \makecell{7x5\,+\,2x5d\\periodic} & large military & yes & yes & yes \\ \midrule

GureKDDcup {\footnotesize\citep{perona2008_gurekddcup}} & 1998 & Payload analysis IDS & emul-pv & \makecell{emul-pv\,(27)\\BRVO} & nethost & 47 & no & 178.8k recs\footnote{Additional 2.76 millions records are provided, but unlabeled.} & \makecell{7x5\,+\,2x5d\\periodic} & large military & yes & no & yes \\ \midrule

KDD Cup 1999 {\footnotesize\citep{stolfo2000_costbased_modeling_ids_kdd99}} & 1998 & General IDS (LANs) & emul-pv & \makecell{emul-pv\,(38)\\BDRVO} & nethost & 42 & no & 4.9M + 311k recs & \makecell{7x5\,+\,2x5d\\periodic} & large military & yes & yes & yes \\ \midrule

NSL-KDD {\footnotesize\citep{tavallaee2009_nslkdd}} & 1998 & General IDS (LANs) & emul-pv & \makecell{emul-pv\,(32)\\BDRVO} & nethost & 43 & no & 126k + 22.5k recs & \makecell{7x5\,+\,2x5d\\periodic} & large military & yes & yes & yes \\ \midrule

DARPA 1999 {\footnotesize\citep{lipmann2000_darpa1999}} & 1999 & General IDS (LANs) & emul-pv & \makecell{emul-pv\,(58)\\BDRVO} & \makecell{packet (bin)\\logs (host)} & - & no & \makecell{57.5M +\\14.6M pkts} & \makecell{3x5\,+\,2x5d\\periodic} & large military & yes & yes & yes \\ \midrule

LLDOS {\footnotesize\citep{haines2000_darpa2000}} & 2000 & Infiltration \& DDoS attacks & emul-pv & \makecell{emul-pv\,(3)\\DRV} & \makecell{packet (bin)\\logs (host)} & - & no & 1.044M + 585k pkts & \makecell{3.3h\hspace{0.14em}+\hspace{0.12em}1.7h\\cont.} & large military & yes & yes & indir. \\ \midrule

MAWILab {\footnotesize\citep{fontugne2010_mawilab}} & 2001 -- Now & Network anomalies & real & \makecell{real\\DR?} & packet (bin) & - & IPs payload & Dozens of millions pkts/day & \makecell{15m/day\footnote{The project has been ongoing for over 20 years, capturing 15 minutes of traffic every day. Unfortunately, only a few days within a year were labeled since 2021.}\\periodic} & large internet backbone & yes & no & yes \\ \midrule

CAIDA DDoS Attack 2007 {\footnotesize\citep{caida_ddos_attack2007}} & 2007 & DDoS attacks & no & \makecell{real (1)\\D} & packet (bin) & - & IPs payload & 46M pkts & \makecell{1h 6m \\ cont.} & n.\,s & no & no & yes \\ \midrule

Twente {\footnotesize\citep{sperotto2009_twente_dataset}} & 2008 & NetFlow General NIDS & no & \makecell{real (1)\\BR?} & flows (uni) & 13 & IPs & 14.2M flows & \makecell{6d 10h\\cont.} & university honeypot & no & no & yes \\ \midrule

ASNM-CDX {\footnotesize\citep{homoliak2020_asnm_datasets}} & 2009 & Buffer overflow & emul-p & \makecell{emul-p (1)\\V} & flows (bi) & 194 & no & 5.7k flows & n.\,s. & small & yes & no & yes \\ \midrule

CDX {\footnotesize\citep{sangster2009_cdx_dataset}} & 2009 & General NIDS (LANs) & emul-p & \makecell{emul-p\\DRO?} & \makecell{packet (bin)\\logs (both)} & - & no & 24.4M pkts & \makecell{4 x 8h\\periodic} & small & yes & no & indir. \\ \midrule

ISOT Botnet {\footnotesize\citep{saad2011_isot_botnet}} & 2004 -- 2010 & Botnet traffic & emul-p & \makecell{emul-p (2)\\CO} & packet (bin) & - & no & 162M pkts & n.\,s. & small & yes & no & yes \\ \midrule

ISCX-IDS-2012 {\footnotesize\citep{shiravi2012_iscx2012}} & 2010 & General NIDS (LANs) & emul-p & \makecell{emul-p (5)\\BCDV} & \makecell{packet (bin)\\flows (bi)} & 21 & no & \makecell{120M pkts\\2.5M flows} & \makecell{7d \\ cont.} & small & yes & no & yes \\ \midrule

HogZilla {\footnotesize\citep{hogzilla2012_hogzilla_dataset}} & 2010 2011 & Botnet traffic & emul-p & \makecell{real (8)\\CDRO} & flows (bi) & 204 & no & 990k flows & n.\,s. & mixed & no & no & yes \\ \midrule

CTU-13 {\footnotesize\citep{garcia2014_ctu13}} & 2011 & Botnet traffic & real & \makecell{real (8)\\CDRO?} & \makecell{packet (bin)\\flows (uni)\\flows (bi)} & \makecell{12 (uni)\\15 (bi)} & payload & \makecell{856M pkts\\81.4M uni-f\\20.6M bi-f} & \makecell{5d 23h\\discont.} & large academic & yes & no & yes \\ \midrule

Booters~{\footnotesize\citep{santanna2015_booters}} & 2013 & \makecell{UDP ampli-\\fication DDoS} & no & \makecell{real (1)\\D} & packet (bin) & - & IPs & \makecell{119.6M pkts} & \makecell{2d \\ cont.} & mixed\footnote{Small target network attacked by a large botnet (30k unique IPs).} & no & no & yes \\ \midrule

Botnet {\footnotesize\citep{beigi2014_botnet_dataset}} & 2004 -- 2014 & Botnet traffic & emul-p & \makecell{real (8) \&\\ emul-p (8)\\CDRO} & packet (bin) & - & No & 9.4M + 5.1M pkts & n.\,s. & mixed & yes & yes & yes \\ \midrule

SSHCure {\footnotesize\citep{hofstede2014_sshcure}} & 2013 2014 & SSH brute-force & real & \makecell{real (1)\\B} & \makecell{flows (uni)\\logs (host)} & 10 & IPs & 360.87M flows & \makecell{2 x 1mo\\periodic} & honeypot \& large academic & yes & no & indir. \\ \midrule

ASNM-TUN {\footnotesize\citep{homoliak2020_asnm_datasets}} & 2015 & Buffer overflow & real \& emul-v & \makecell{emul-v (2)\\VO} & flows (bi) & 194 & IPs payload & 394 flows & n.\,s. & small \& academic & yes & no & yes \\ \midrule

AWID2 {\footnotesize\citep{kolias2015_awid2_dataset}} & 2015 & Wireless NIDS & emul-p & \makecell{emul-p (15)\\DRO} & packet (txt) & 156 & no & 37.8M + 4.57M recs & \makecell{96h + 12h\\cont.} & small wireless & yes & yes & yes \\ \midrule

Kyoto 2006 {\footnotesize\citep{song2011_kyoto_dataset}} & 2006 -- 2015 & General IDS & real & \makecell{real\\DRV?} & other & 24 & IPs & 806.1M flows & \makecell{9y 2mo\\time span\\discont.} & \makecell{honeypot \\\& legit\\server} & no & no & yes \\ \midrule

UNSW-NB15 {\footnotesize\citep{moustafa2015_unswnb15}} & 2015 & General IDS (LANs) & emul-pv & \makecell{emul-pv (9)\\DRVO} & \makecell{packet (bin)\\flows (bi)\\logs (net)} & 49 & no & \makecell{187M pkts\\ 2.54M flows} & \makecell{16h + 15h\\cont.} & small & yes & yes & yes \\ \midrule

DDoS 2016 {\footnotesize\citep{alkasassbeh2016_ddos2016}} & 2016 & DDoS attacks & emul-v & \makecell{emul-v (4)\\D} & packet (txt) & 28 & IPs & 2.16M pkts & n.\,s. & n.\,s. & n.\,s. & no & yes \\ \midrule

Kent 2016 {\footnotesize\citep{kent2016_cybersecdata_dynamicnet}} & 2016 & Anomaly-based IDS & real & \makecell{emul-p (-)\\?} & \makecell{flows (uni)\\logs (both)} & 9 & IPs, time, names & \makecell{130M flows\\40.8M net\\logs} & \makecell{29d (net)\\57d (all)\\cont.} & large enterprise & yes & no & indir. \\ \midrule

NDSec-1 {\footnotesize\citep{beer2017_ndsec1}} & 2016 & Attacks for salting & emul-v & \makecell{emul-v\,(12)\\BCDPVO} & \makecell{packet (bin)\\logs (host)} & - & no & 3.77M pkts & \makecell{4h 11m\\discont.} & small & yes & no & yes \\ \midrule

NGIDS-DS {\footnotesize\citep{haider2017_ngids_ds_dataset}} & 2016 & General IDS (LANs) & emul-pv & \makecell{emul-pv\,(7)\\DRVO} & \makecell{packet (bin)\\logs (host)} & - & no & 1.09M pkts & \makecell{5d 21h\\cont.} & small & yes & no & indir. \\ \midrule

UGR\textquotesingle16 {\footnotesize\citep{marciafernandez2018_ugr16}} & 2016 & \makecell{Cyclostatio-\\nary Net-\\Flow NIDS} & real & \makecell{real (4) \&\\emul-pv\,(3)\\CDRO?} & flows (uni) & 12 & IPs & 13,000M + 3,900M flows & \makecell{3mo 9d +\\1mo 3d\\cont.} & large ISP & yes & yes & yes \\ \midrule

Unified Host and Network~{\footnotesize\citep{turcotte2019_unified}} & 2016 & Anomaly-based IDS & real & \makecell{n.s.\\?} & \makecell{flows (bi)\\logs (host)} & 11 &  IPs, time, names & 17,883M flows & \makecell{3mo\\cont.} & large enterprise & yes & no & no \\ \midrule

2017-SUEE8 {\footnotesize\citep{lukeseder2017_suee_dataset}} & 2017 & HTTP(S) Slow DoS & real & \makecell{emul-pv\,(3)\\D?} & packet (bin) & - & IPs MACs & 19.3M pkts & \makecell{8d 7h\\cont.} & large academic & yes & no & yes \\ \midrule


CIC-IDS2017 {\footnotesize\citep{sharafin2018_cicids2017_csecic2018}} & 2017 & General NIDS (LANs) & emul-p & \makecell{emul-p (14)\\BCDRV} & \makecell{packet (bin)\\flows (bi)} & 79 & no & \makecell{56.4M pkts\\3.1M flows} & \makecell{5 x 8h\\periodic} & small & yes & no & yes \\ \midrule

CIC-IDS2017 Imp. {\footnotesize\citep{liu2022_error_prevalence_nids_data,engelen2021_cicids2017_troubleshooting}} & 2017 & General NIDS (LANs) & emul-p & \makecell{emul-p (15)\\BCDRV} & flows (bi) & 91 & no & \makecell{2.1M flows} & \makecell{5 x 8h\\periodic} & small & yes & no & yes \\ \midrule

CIDDS-001 {\footnotesize\citep{ring2017_cidds001}} & 2017 & NetFlow General NIDS & real \& emul-v & \makecell{emul-pv (4)\\BDR?} & flows (uni) & 16 & IPs & 32M flows & \makecell{28d 6h\\cont.} & small enterprise & yes & no & yes \\ \midrule

CIDDS-002 {\footnotesize\citep{ring2017_cidds002}} & 2017 & NetFlow Port scans & emul-v & \makecell{emul-v (1)\\R} & flows (uni) & 16 & IPs & 16.2M flows & \makecell{13d 19h\\discont.} & small enterprise & yes & no & yes \\ \midrule

ISOT HTTP Botnet {\footnotesize\citep{alenazi2017_isot_http_botnet}} & 2017 & Botnet traffic & emul-v & \makecell{emul-v (9)\\C} & packet (bin) & - & no & 10.6M pkts & \makecell{54d 17h\\discont.} & small & yes & no & yes \\ \midrule

IoT host-based ID dataset {\footnotesize\citep{bezerra2018_iot_hostbased_dataset}} & 2017 & IoT botnet activity & emul-p & \makecell{real (1)\\CDV} & \makecell{packet (bin)\\flows (bi)\\logs (host)} & 47 & yes\footnote{Anonymization details are unspecified in the documentation, but sanitization was detected in the associated PCAP file.} & \makecell{9.34M pkts\\1.72M flows} & \makecell{11h 55m\\discont.} & small & no & no & yes \\ \midrule

LYCOS-IDS2017 {\footnotesize\citep{rosay2022_lycos_ids2017}} & 2017 & General NIDS (LANs) & emul-p & \makecell{emul-p (13)\\BCDRV} & flows (bi) & 83 & no & 1.8M flows (661k+220k) & \makecell{5 x 8h\\periodic} & small & yes & yes & yes \\ \midrule

TrabID {\footnotesize\citep{viegas2017_trabid_dataset}} & 2017 & Network anomalies & emul-p & \makecell{emul-p (16)\\DR} & \makecell{packet (bin)\\packet (txt)} & 44 & no & 469.4M pkts & \makecell{8h\\discont.} & medium & yes & yes & yes \\ \midrule

ISOT-CID {\footnotesize\citep{aldribi2020_isotcid}} & 2016 2018 & Cloud IDS & real \& emul-pv & \makecell{real \&\\emul-pv\,(18)\\BDRVO?} & \makecell{packet (bin)\\logs (host)} & - & n.s.\footnote{No anonymization technique was specified, but IP anonymization \& payload trimming is suspected due to containing real traffic.} & 24.5M + 12.4M pkts & \makecell{5d + 6d\\discont.} & medium enterprise cloud & yes & yes & yes \\ \midrule

Kitsune {\footnotesize\citep{mirsky2018_kitsune}} & 2017 2018 & IoT NIDS & emul-p & \makecell{emul-p\,(9)\\BCDRO} & \makecell{packet (bin)\\other} & 115 & no & 2.77M pkts & \makecell{6h 50m\\discont.} & small IoT & yes & no & yes \\ \midrule

NetML {\footnotesize\citep{barut2020_netml_dataset}} & 2013 2017 2018 & General IDS & real & \makecell{real (20)\\CDO?} & flows (bi) & 63 & IPs & 484k flows & \makecell{n.\,s.\\discont.\footnote{Several merged captures from the Stratosphere Research Laboratory \citep{stratosphere2024_web}. The total duration of a few days or weeks is suspected, but the exact value is unknown.}} & large academic & yes & yes & yes \\ \midrule

NF-UQ-NIDS/ NF-UQ-NIDS-v2 {\footnotesize\citep{sarhan2021_netflow_datasets_ml_nids,sarhan2022_nids_feature_set}} & 2015 2017 2018 & NetFlow General NIDS & emul-pv & \makecell{emul-pv\,(20)\\BCDRVO} & flows (bi) & 15/45 & no & 11.99M/ 79.99M flows & \makecell{58d 12h\\discont.} & \makecell{mixed\\(3 small\\networks)} & yes & no & yes \\ \midrule

ASNM-NPBO {\footnotesize\citep{homoliak2020_asnm_datasets}} & 2018 & Buffer overflow & real \& emul-v & \makecell{emul-v (2)\\VO?} & flows (bi) & 194 & IPs payload & 11.4k flows & n.\,s. & small \& academic & yes & no & yes \\ \midrule

Bot-IoT {\footnotesize\citep{koroniotis2019_botiot_dataset}} & 2018 & IoT NIDS & emul-v & \makecell{emul-v (10)\\DRO} & \makecell{packet (bin)\\flows (bi)} & 47 & no & \makecell{549.8M pkts\\73.4M flows\\{\footnotesize(2.9M+0.7M)}} & \makecell{20d\\cont.} & small IoT & yes & yes & yes \\ \midrule

CSE-CIC-IDS2018 {\footnotesize\citep{sharafin2018_cicids2017_csecic2018}} & 2018 & General NIDS & emul-v & \makecell{emul-v (14)\\BCDRV} & \makecell{packet (bin)\\flows (bi)\\logs (host)} & 84 & no & \makecell{1,18M pkts\\16.2M flows} & \makecell{10d\\discont.} & large & yes & no & yes \\ \midrule

CSE-CIC-IDS 2018 Imp. {\footnotesize\citep{liu2022_error_prevalence_nids_data}} & 2018 & General NIDS & emul-v & \makecell{emul-v (14)\\BCDRV} & flows (bi) & 91 & no & 63.2M flows & \makecell{10d\\discont.} & large & yes & no & yes \\ \midrule

N-BaIoT {\footnotesize\citep{meidan2018_nbaiot_dataset}} & 2018 & IoT botnet activity & emul-p & \makecell{emul-p (10)\\DRO} & other & 115 & no & 7.06M pkts & n.\,s. & small IoT & yes & no & yes \\ \midrule

IoT-23 {\footnotesize\citep{garcia2020_iot23_dataset}} & 2018 2019 & IoT botnet activity & real & \makecell{real (9)\\CDRVO?} & \makecell{packet (bin)\\flows (bi)} & 23 & no & \makecell{814M pkts\\325M flows} & \makecell{21d\\discont.} & small IoT & yes & no & yes \\ \midrule

CIC-DDoS2019 {\footnotesize\citep{sharafaldin2019_cic_ddos2019}} & 2019 & Flooding DDoS & emul-p & \makecell{emul-p (12)\\D} & \makecell{packet (bin)\\flows (bi)} & 87 & no & \makecell{251M\,+\,61M\\pkts, 50M\,+\\20.3M flows} & \makecell{7h + 8h\\cont.} & small & yes & yes & yes \\ \midrule

CUPID~{\footnotesize\citep{lawrence2022_cupid}} & 2019 & General NIDS (LANs) & emul-pv & \makecell{emul-pv\,(1)\\BRVO} & \makecell{packet (bin)\\flows (bi)} & 84 & no & \makecell{50M pkts\\1.5M\,flows} & \makecell{7d 16h\\discont.} & small & yes & no & yes \\ \midrule

GTCS {\footnotesize\citep{mahfouz2020_gtcs}} & 2019 & General NIDS (LANs) & emul-v & \makecell{emul-v (4)\\BCDV} & flows (bi) & 84 & no & 517k flows & \makecell{9d\\cont.} & small & yes & no & yes \\ \midrule

IoT network intrusion {\footnotesize\citep{kang2019_iot_net_intrusion_dataset}} & 2019 & IoT NIDS & emul-p & \makecell{emul-p (9)\\BDRO} & packet (bin) & - & no & 2.99M pkts & \makecell{2h 1m\\discont.} & small IoT & yes & no & yes \\ \midrule

IoTID20 {\footnotesize\citep{ullah2020_iotid}} & 2019 & IoT NIDS & emul-p & \makecell{emul-p (8)\\BDRO} & flows (bi) & 86 & no & 625.7k flows & \makecell{2h 13m\\discont.} & small IoT & yes & no & yes \\ \midrule

TON\_IoT {\footnotesize\citep{alsaedi2020_toniot_dataset, moustafa2021_toniot_net}} & 2019 & IoT and IIoT IDS & emul-pv & \makecell{emul-pv\,(9)\\BDRVO} & \makecell{packet (bin)\\flows (bi)\\logs (host)} & 46 & no & \makecell{212M pkts\\22.34M flows} & \makecell{6d 12h\\discont.} & small IoT & yes & no & yes \\ \midrule

MedBIoT {\footnotesize\citep{guerramanzanes2020_medbiot}} & 2019 & IoT botnet activity & emul-pv & \makecell{emul-pv (3)\\CRV} & \makecell{packet (bin)\\packet (txt)} & 100 & no & 37.4M pkts & \makecell{10d 11h\\cont.} & small IoT & yes & no & yes \\ \midrule

WUSTL-IIoT-2021 {\footnotesize\citep{zolanvari2019_wustl_iiot2021}} & 2019 & IIoT NIDS & emul-pv & \makecell{emul-pv (4)\\DRV} & flows (bi) & 49 & no & 1.19M flows & \makecell{7h 2m\\cont.} & small IIoT & yes & no & yes \\ \midrule

LITNET-2020 {\footnotesize\citep{damasevicius2020_litnet2020}} & 2019 2020 & General NIDS & real & \makecell{emul-pv\,(12)\\DRVO?} & flows (bi) & 85 & IPs & 39.6M flows & \makecell{10mo\\cont.} & large academic & yes & no & yes \\ \midrule

InSDN {\footnotesize\citep{elsayed2020_insdn}} & 2019 2020 & SDN NIDS & emul-v & \makecell{emul-v (7)\\BCDRV} & \makecell{packet (bin)\\flows (bi)} & 84 & no & \makecell{15.3M pkts\\344k flows} & \makecell{5d 7h\\discont.} & small SDN & yes & no & yes \\ \midrule

X-IIoTID {\footnotesize\citep{alhawawreh2022_xiiotid}} & 2019 2020 & IoT and IIoT IDS & emul-p & \makecell{emul-p (18)\\BCDRVO} & nethost & 68 & no & 820.8k recs & \makecell{6d 7h 8m\\discont.} & small IoT & yes & no & yes \\ \midrule

BOUN DDoS {\footnotesize\citep{erhan2020_bogazici_ddos_dataset}} & 2020 & DDoS attacks & real & \makecell{emul-p (2)\\D?} & \makecell{packet (txt)} & 12 & \makecell{IPs\\time} & 17.38M pkts & \makecell{2 x 8m\\periodic} & large academic & yes & no & yes \\ \midrule

CCD-INID-V1 {\footnotesize\citep{liu2021_ccd_inid_v1}} & 2020 & Smart home IoT IDS & emul-p & \makecell{emul-p (5)\\BDO} & flows (bi) & 88 & no & 91.6k flows & \makecell{13h 45m\\discont.} & small IoT & n.\ s. & no & yes \\ \midrule

DAPT 2020 {\footnotesize\citep{myneni2020_dapt2020}} & 2020 & APT simulation & emul-pv & \makecell{emul-pv\,(12)\\BRVO} & \makecell{packet (bin)\\flows (bi)\\logs (host)} & 85 & no & \makecell{77.1M pkts\\86.7k flows} & \makecell{5 x 8h\\periodic} & small & yes & no & yes \\ \midrule

OPC UA {\footnotesize\citep{pinto2020_opc_ua_dataset}} & 2020 & CPS IDS for OPC UA & emul-p & \makecell{emul-p (3)\\DO} & flows (bi) & 32 & no & 108k flows & \makecell{2h 3m\\cont.} & small IIoT & yes & no & yes \\ \midrule

SDN Dataset {\footnotesize\citep{sarica2020_sdn_dataset}} & 2020 & SDN NIDS & emul-v & \makecell{emul-v (5)\\DRV} & flows (bi) & 33 & no & 27.92M + 30.2M flows & \makecell{44h + 42h\\discont.} & small SDN IoT & yes & yes & yes \\ \midrule

SR-BH 2020 {\footnotesize\citep{riera2022_sr_bh2020}} & 2020 & HTTP(S) web attacks & real & \makecell{real (13)\\BRVO?} & flows (bi) & 38 & no & 907.8k flows & \makecell{12d 10h\\discont.} & honeypot & no & yes & yes \\ \midrule

UKM-IDS20 {\footnotesize\citep{aldaweri2021_ukmids20}} & 2020 & General NIDS (LANs) & emul-v & \makecell{emul-v (7)\\DRVO} & flows (bi) & 48 & IPs & 10.3k + 2.6k flows & \makecell{1d + 1d\\cont.} & small & yes & yes & yes \\ \midrule

VOIP Enterprise -- Attack {\footnotesize\citep{alvares2021_voip_enterprise_attack_dataset}} & 2020 & VoIP NIDS & emul-p & \makecell{emul-p (6)\\BDRO} & packet (bin) & - & no & 4.48M pkts & \makecell{3h 38m\\discont.} & small VoIP & yes & no & yes \\ \midrule

AWID3~{\footnotesize\citep{chatzoglou2021_awid3_dataset}} & 2021 & Wireless NIDS & emul-p & \makecell{emul-p (13)\\BCDVO} & \makecell{packet (bin)\\packet (txt)} & 254 & no & 37M pkts & \makecell{2h 20m\\discont.} & small wireless & yes & no & yes \\ \midrule

CyberFORCE Scenario~{\footnotesize\citep{cybervan2021_cyberforce_scenario}} & 2021 & APT simulation & emul-v & \makecell{emul-v (5)\\CRVO} & \makecell{packet (bin)\\logs (both)} & - & no & 184k pkts & \makecell{2h 30m\\cont.} & small & yes & no & indir. \\ \midrule

Edge-IIoTset {\footnotesize\citep{ferrag2022_edge_iiotset}} & 2021 & IoT and IIoT NIDS & emul-p & \makecell{emul-p (14)\\BDRVO} & \makecell{packet (bin)\\packet (txt)} & 63 & no & \makecell{20.9M pkts} & \makecell{1d 23h\\discont.} & small IoT & yes & no & yes \\ \midrule

I-Sec {\footnotesize\citep{serinelli2023_isec_ids_dataset}} & 2021 & General NIDS (LANs) & emul-v & \makecell{emul-v (7)\\DR} & flows (bi) & 78 & \makecell{IPs\\time} & \makecell{532k flows} & n.\,s. & small P2P & no & no & yes \\ \midrule

PWNJUTSU {\footnotesize\citep{berady2022_pwnjutsu_dataset}} & 2021 & APT simulation & no & \makecell{emul-v\\BRVO} & \makecell{packet (bin)\\nethost\\logs (host)} & 160\footnote{Multiple files with different feature vectors. When merged, there are 160 different features, but the majority contains NULL values, making the data challenging to work with.} & no & \makecell{178.4M pkts\\45.1M recs} & \makecell{22\,x\,1-15d\\periodic\\116d total} & small & no & no & no \\ \midrule

Unraveled {\footnotesize\citep{myneni2023_unraveled}} & 2021 & APT simulation & emul-v & \makecell{emul-v (5)\\BCVO} & \makecell{packet (bin)\\flows (bi)\\logs (both)} & 89 & no & \makecell{252M pkts\\6.86M flows} & \makecell{40d 13h\\cont.} & small enterprise & yes & no & yes \\ \midrule

USB-IDS-1 {\footnotesize\citep{catillo2021_usbids1}} & 2021 & Layer 7 web DoS & emul-p & \makecell{emul-p (4)\\D} & flows (bi) & 84 & no & 4.5M flows & \makecell{16 x 10m\\\& 2d 19h\\discont.} & small & yes & yes & yes \\ \midrule

LUFlow {\footnotesize\citep{mills2022_luflow}} & 2020 -- 22 & Network anomalies & real & \makecell{real (2)\\?} & flows (bi) & 16 & IPs & 206.56M flows & \makecell{8mo\\cont.} & large academic & no & no & yes \\ \midrule

CIC IoT dataset 2022 {\footnotesize\citep{dadkhah2022_ciciot2022}} & 2021 2022 & Home IoT monitoring & emul-p & \makecell{emul-p (2)\\BD} & \makecell{packet (bin)\\other} & 48 & no & \makecell{120M pkts\\466k recs} & \makecell{22d 13h\\discont.} & small IoT & no\footnote{Some of the data is captured from the whole network, but a big portion was injected with individual captures from specific isolated devices.} & no & indir. \\ \midrule

HIKARI-2021 {\footnotesize\citep{ferriyan2021_hikari2021}} & 2021 2022 & Encrypted traffic NIDS & real \& emul-p & \makecell{emul-p (4)\\BRO?} & \makecell{packet (bin)\\flows (bi)} & 86 & IPs payload & \makecell{66.5M pkts\\783k flows} & \makecell{1d 4.5h\\discont.} & small & yes & no & yes \\ \midrule

UWF-Zeek Data22 {\footnotesize\citep{bagui2023_uwf_zeekdata22}} & 2021 2022 & General NIDS (LANs) & emul-v & \makecell{emul-v\,(2\footnote{Several other attack types and labels are present, but the amount of their samples is so negligible (under 20) that they cannot be practically utilized.})\\R} & flows (bi) & 23 & no & 18.56M flows & \makecell{1mo 22h\\discont.} & small & yes & no & yes \\ \midrule

VHS-22 {\footnotesize\citep{szumelda2022_vhs22_dataset}} & 2004 -- 2022 & General NIDS & \makecell{real \&\\emul-pv} & \makecell{real\;\&\;emul\\-pv (115)\\BCDRVO?} & flows (uni) & 47 & no & 27.7M flows & \makecell{5d 5h\\discont.\\} & mixed & yes & no & yes \\ \midrule

AIT Log Dataset 2.0 {\footnotesize\citep{landauer2023_ait_ldsv2_dataset}} & 2022 & General IDS (LANs) & emul-v & \makecell{emul-v\,(22)\\BRVO} & \makecell{flows (bi)\\logs (both)} & 145 & no & \makecell{224M pkts\\3.5M flows} & \makecell{8 x 4-6d\\periodic\\40d total} & small & yes & no & yes \\ \midrule

ICS-Flow {\footnotesize\citep{deghlaghi2023_ics_flow}} & 2022 & ICS NIDS & emul-v & \makecell{emul-v (5)\\DRO} & \makecell{packet (bin)\\flows (bi)} & 64 & no & \makecell{25.2M pkts\\45.7k flows} & \makecell{2h 20m\\cont.} & small industrial & yes & no & yes \\ \midrule

OD-IDS2022 {\footnotesize\citep{patel2023_od_ids2022}} & 2022 & General NIDS (LANs) & emul-pv & \makecell{emul-pv\,(28)\\BDVO} & flows (bi) & 82 & no & 103.2k flows & \makecell{1mo\\discont.} & small & yes & no & yes \\ \midrule

Simargl2022 {\footnotesize\citep{komisarek2022_simargl2022}} & 2022 & NetFlow General NIDS & real \& emul-pv & \makecell{emul-pv\,(3)\\CDR?} & flows (bi) & 31 & n.\,s.\footnote{No anonymization technique specified, but IP anonymization is suspected due to containing real traffic.\label{footnote:anon_suspected}} & 30.1M flows & \makecell{5d 1h\\discont.} & large academic & yes & no & yes \\ \midrule

UNR-IDD {\footnotesize\citep{das2023_unridd_dataset}} & 2022 & Port-based SDN NIDS & emul-v & \makecell{emul-v (5)\\DRO} & other & 34 & no & 37k recs & n.\,s. & small SDN & yes & no & yes \\ \midrule

CICIoT2023 {\footnotesize\citep{neto2023_ciciot2023}} & 2022 2023 & IoT NIDS & emul-p & \makecell{emul-p (33)\\BCDRVO} & \makecell{packet (bin)\\other} & 47 & no & \makecell{4,49M pkts\\46.7M recs} & \makecell{6mo 15d\\discont.} & medium IoT & yes & no & \makecell{yes \&\\indir.} \\ \midrule

TII-SSRC-23 {\footnotesize\citep{heryalla2023_tii_src23}} & 2022 2023 & General NIDS (LANs) & emul-p & \makecell{emul-p (26)\\BCDR} & \makecell{packet (bin)\\flows (bi)} & 86 & no & \makecell{43.6M pkts\\8.7M flows} & \makecell{1mo 1.5d\\discont.} & small & yes & no & yes \\ \midrule

Appraise H2020 {\footnotesize\citep{komisarek2023_appraise_h2020_dataset}} & 2023 & NetFlow anomalies & real & \makecell{emul-p (3)\\BDR?} & flows (bi) & 19 & n.\,s.\footref{footnote:anon_suspected} & 15.12M flows & \makecell{2mo 12.5d\\discont.} & large enterprise & yes & no & yes \\ \midrule

FLNET2023 {\footnotesize\citep{kumar2023_flnet2023}} & 2023 & Federated learning NIDS & emul-v & \makecell{emul-v (11)\\DVO} & \makecell{packet (bin)\\flows (bi)} & 83 & no & \makecell{370M\,+\,79M\\pkts\,||\,5.42M\\+\,747k flows} & \makecell{6d 2h +\\11d 4h\\discont.} & large ISP & yes & yes & yes \\ \midrule

ROSIDS23~{\footnotesize\citep{degirmenci2023_rosids2023}} & 2023 & Robotic arm ICS NIDS & emul-p & \makecell{emul-p (4)\\DO} & \makecell{packet (bin)\\flows (bi)} & 84 & no & \makecell{176.3M pkts\\136.6k flows} & \makecell{10h 6m\\discont.} & small industrial & yes & no & yes \\ \midrule

Westermo~{\footnotesize\citep{strandberg2023_westermo}} & 2023 & ICS NIDS & emul-pv & \makecell{emul-pv\,(6)\\BRO} & \makecell{packet (bin)\\flows (bi)} & 64 & no & \makecell{4.7M pkts\\68.7k flows} & \makecell{1h 30m\\cont.} & small industrial & no & no & yes \\ \bottomrule

\label{tab:data_survey}
\end{longtable}
\twocolumn
\restoregeometry
\end{landscape}
\clearpage
}


\subsubsection{DARPA \& KDD}
\label{sssec:ndatasurv_general_darpa}

Datasets published by the Defense Advanced Research Projects Agency (DARPA) are considered the first generation of public IDS benchmarks. The first dataset, \emph{DARPA 1998} \citep{lippmann2000_darpa1998}, emulating an Air Force base with thousands of hosts, includes a robust benign traffic profile along with 32 different attacks, categorized as Denial of Service (DoS), Probing, User-to-Root (U2R), and Remote-to-Local (R2L). Despite having different goals (escalating privileges vs. gaining an access to a local account from a remote machine), both U2R and R2L classes are mainly performed via brute-force and vulnerability exploitation. Its data collection lasted nine weeks during weekdays, with a train-test split after the seventh week.

Since DARPA 1998 provides only raw packet and host log data, additional feature extraction and labeling steps are required to utilize it. In order to simplify its usage, \emph{KDD Cup 1999 (KDD\textquotesingle99)} \citep{stolfo2000_costbased_modeling_ids_kdd99} introduced 41 pre-extracted features with a label by processing DARPA 1998 raw data. Its practicality, respected organization backing, and general lack of alternatives made it a de facto standard for IDS benchmarks for years to come. Its two derivatives, \emph{NSL-KDD} \citep{tavallaee2009_nslkdd} and \emph{GureKDDcup} \citep{perona2008_gurekddcup}, further improve data quality by removing duplicate records and reducing the dataset size (NSL-KDD) or enriching records with packet payload data (GureKDDcup).

Other DARPA-family datasets -- \emph{DARPA 1999} \citep{lipmann2000_darpa1999}, extending the DARPA 1998, and \emph{LLDOS} (DARPA 2000) \citep{haines2000_darpa2000}, focusing on the infiltration and DoS attacks using the DARPA 1999 environment, were also published soon after. Nevertheless, because these datasets are provided only in raw format,  they have never gained significant attention from the community.

Due to their popularity, DARPA 1998/1999, as well as their derived versions KDD\textquotesingle99 and NSL-KDD, have been extensively studied and criticized by the community. The criticism points toward questionable data validity \citep{mchugh2000_darpa_data_critique}, the existence of artifacts \citep{mahoney2003_darpa1999_critique}, lack of real-world attack stealthiness \citep{silva2020_attackers_statistical_analysis_kdd}, or inconsistencies between DARPA 1998 data and KDD\textquotesingle99 features indicating labeling issues or record duplication \citep{altobi2018_kdd99_genfaults}. Given these issues and the datasets' age (over 25 years), DARPA datasets are no longer recommended for IDS  benchmarking, though many recent studies still rely on them.

\subsubsection{Canadian Institute for Cybersecurity}
\label{sssec:ndatasurv_general_cic}

Although several datasets have been published since the DARPA-family release, its first major competition for multi-purpose IDS benchmarking was \emph{ISCX-IDS-2012} \citep{shiravi2012_iscx2012}, by the Canadian Institute for Cybersecurity (CIC). Its authors formalized and systematized dataset generation using $\alpha$- and $\beta$-profiles to describe intrusions and abstract distribution models for benign network traffic. The dataset was created in a small testbed network with 21 Windows workstations and several vulnerabilities, capturing seven days of traffic as raw packets and bi-directional flows with a basic feature set.

Building on the concept of profiles, \citet{sharafaldin2018_towards_reliable_ids_dataset} introduced the B-profile and M-profile systems and published the dataset \emph{CIC-IDS2017} \citep{sharafin2018_cicids2017_csecic2018}, spanning 40 hours over five days. The dataset introduced encrypted benign traffic and new widespread intrusions like Heartbleed, SQL injection, or ARES Botnet. A similar scenario was executed a year later on a virtualized AWS environment with hundreds of hosts, resulting in the \emph{CSE-CIC-IDS2018} \citep{sharafin2018_cicids2017_csecic2018} dataset. A characteristic property of both datasets was the employment of CICFlowMeter \citep{lashkari2016_cicflowmeter} for feature extraction, introducing a novel feature set for network intrusion detection of approximately 80 (the exact number depends on the tool version) features suitable for machine learning algorithms. The datasets and the feature set itself have gained much attention in the community.

However, the quality of these datasets was not appropriate, as their users reported many errors, including duplicated features, duplicated packets, incorrect flow construction, incoherent timestamps, labeling issues, or the usage of deprecated tools for attack execution \citep{engelen2021_cicids2017_troubleshooting, liu2022_error_prevalence_nids_data, rosay2022_lycos_ids2017, lanvin2023_errors_cicids2017}. In efforts to reduce the evaluation bias caused by faulty data, the researchers have repaired \citep{liu2022_error_prevalence_nids_data} or reimplemented \citep{rosay2022_lycos_ids2017} the CICFlowMeter tool and rerun it on the available raw packet data, resulting in \emph{Improved CIC-IDS2017} and \emph{Improved CSE-CIC-IDS2018} datasets \citep{liu2022_error_prevalence_nids_data}, as well as the dataset \emph{LYCOS-IDS2017} \citep{rosay2022_lycos_ids2017}.

\subsubsection{NetFlow Datasets}
\label{sssec:ndatasurv_general_netflow}

The following paragraphs cover general-purpose datasets utilizing features exclusively from the NetFlow specification. While they could also be categorized under the next subsections (\ref{sssec:ndatasurv_general_established},~\ref{sssec:ndatasurv_general_new}), we discuss them separately due to their real-world relevance. In contrast to most NID datasets with custom features (unavailable in production networks without additional tools), NetFlow datasets rely on standardized features of the NetFlow architecture already deployed in many networks for monitoring purposes. Therefore, NetFlow-based NIDSs are considered more practical due to lower implementation costs and more realistic benchmark results.

The first NetFlow NID dataset, \emph{Twente} \citep{sperotto2009_twente_dataset} (2009), was collected over six days from a university honeypot server. Its authors identified real brute-force and scanning attempts and labeled the flows via alert correlation techniques. However, since data is collected from a honeypot, it lacks benign traffic, thus limiting its usage to signature-based systems.

A decoy server connected to the Internet was also utilized in \emph{CIDDS-001} \citep{ring2017_cidds001}. In addition, the authors created a virtual network to communicate with the server to emulate benign traffic and attacks. The dataset thus contains both emulated benign and attack traffic, as well as real-world traffic labeled either as suspicious or unknown. The same environment without an external server was also used in \emph{CIDDS-002} \citep{ring2017_cidds002}, purely focused on scanning attacks.

Instead of obtaining real traffic from honeypots, \emph{UGR-16} \citep{marciafernandez2018_ugr16} captures NetFlow traces from a Tier-3 ISP network over three months. The dataset aims to provide periodic network activity patterns such as day-and-night cycles or weekdays and weekend cycles. It includes only flow-detectable attacks such as DoS, port scans, or botnet traffic, either emulated or injected from other sources. Since the capture contains real traffic, its authors used signature- and anomaly-based methods to detect and label malicious traffic, identifying UDP scanning, SSH scanning, and spam attacks.

The works by \citet{sarhan2021_netflow_datasets_ml_nids, sarhan2022_nids_feature_set} have identified the issue of lacking data with real-world features and provided practical datasets based on the NetFlow protocol to address it. For this purpose, popular datasets -- UNSW-NB15, Bot-IoT, ToN-IoT, and CSE-CIC-IDS2018 -- were regenerated using nProbe \citep{ntop_nprobe}, providing basic (15) and extended (45) feature sets. While available individually, the authors also merged them, resulting in the \emph{NF-UQ-NIDS} and \emph{NF-UQ-NIDS-v2} \citep{sarhan2022_nids_feature_set} datasets.

The most recent NetFlow dataset, \emph{Appraise H2020} \citep{komisarek2023_appraise_h2020_dataset}, also intends to tackle the lack of real-world, real-time features suitable for network intrusion detection. It captures traffic from a shopping mall network and simulates DoS, reconnaissance, and brute-force attacks. However, it uses a minimalist feature set (19) while lacking details on the network environment and attack generation. The same research group released a similar NetFlow dataset, \emph{Simargl2022} \citep{komisarek2022_simargl2022}, simulating reconnaissance, DoS, and botnet intrusions in an academic network. Both datasets were collected using nProbe with NetFlow~9.

\subsubsection{Established Datasets}
\label{sssec:ndatasurv_general_established}

One of the most prominent datasets for NIDS research is \emph{UNSW-NB15} \citep{moustafa2015_unswnb15} (2015). Emulated using the PerfectStorm platform\footnote{PerfectStorm: \url{https://www.keysight.com/br/pt/products/network-test/network-test-hardware/perfectstorm.html}}, it includes nine intrusion classes, such as DoS, backdoors, reconnaissance, and worms. The data is provided as raw PCAP, host logs, and 49 flow-based features. However, its documentation lacks details on data generation, leading to ambiguities in its interpretation. For instance, there is no description of the emulated vulnerabilities, types of attacks, or their parameters. Recently, issues with mislabeling, simulation artifacts, and lack of diversity have also been discovered \citep{flood2024_bad_design_smells_nids_datasets}. For this reason, conclusions drawn by evaluations using this dataset might be less sound. PerfectStorm was also used to generate NGIDS-DS \citep{haider2017_ngids_ds_dataset}, which also shares these documentation issues. In addition, it only provides raw data labeled indirectly via host logs, further limiting its usability.

Other notable datasets include \emph{GTCS} \citep{mahfouz2020_gtcs} and \emph{LITNET-2020} \citep{damasevicius2020_litnet2020}. GTCS, generated in a simple virtualized topology for over a week, includes four attack scenarios: botnet, brute-force, DDoS, and infiltration, along with legitimate traffic from Ostinato\footnote{Ostinato is a traffic generation tool primarily used for network testing and throughput measurements. Homepage: \url{https://ostinato.org/}}. Its~84 features were extracted using CICFlowMeter. In contrast to small-scale datasets, LITNET-2020 was captured from a large academic network over 10 months. Part of its flow-based features is based on the NetFlow protocol, but a few additional were calculated via custom scripts. The authors emulated 12 types of intrusions, but similar to previous datasets, they provide no information about used tools and their parameters. Datasets \emph{CDX} \citep{sangster2009_cdx_dataset}, \emph{Kyoto 2006+} \citep{song2011_kyoto_dataset}, and \emph{NetML} \citep{barut2020_netml_dataset} have also been around for a while, but their labeling or feature sets are somewhat limited, thus not receiving much attention from the community.

\subsubsection{New Datasets}
\label{sssec:ndatasurv_general_new}

In recent years, several new general-purpose datasets have been released. Due to their novelty, they are expected to offer the newest benign traffic patterns and current intrusions. Therefore, the results obtained using them as benchmarks should be closer to reality. Since they were released only recently, the community has not yet thoroughly analyzed them and ``decided'' on their popularity, as they are not yet commonly referenced in recent scientific publications. For this reason, a separate paragraph is provided with an introduction and a brief description of each newly-released dataset.

The \emph{CUPID} dataset \citep{lawrence2022_cupid} (2019) emulates a small physical network with several virtualized systems. Its objective is to provide both scripted and human-generated traffic produced by professional penetration testers, allowing researchers to investigate their differences. The dataset contains raw packets and pre-extracted traffic features via CICFlowMeter. Unfortunately, it provides only binary labels (benign/attack), thus lacking those of specific intrusions.

\emph{UKM-IDS20} \citep{aldaweri2021_ukmids20} (2020) authors emulated two small office networks within the Hyper-V virtualization platform over two weeks. The first week is attack-free, capturing two individuals accessing the network's services. During the second week, DoS, scanning, exploits, and ARP poisoning attacks were introduced. The dataset contains 12.9 thousand flows of 48 features extracted by a custom algorithm.

\emph{I-Sec} \citep{serinelli2023_isec_ids_dataset} (2021) was captured within a Peer-to-Peer (P2P) topology emulated in VirtualBox, featuring DoS and scanning attacks performed via \texttt{hping3} and \texttt{nmap} tools. The traffic features were extracted using CICFlowMeter. However, due to a simplistic topology and lack of a strategy for benign traffic generation, its value for reliable benchmarks remains questionable.

\emph{AIT Log Dataset 2.0 (AIT LDS2.0)} \citep{landauer2023_ait_ldsv2_dataset} (2022) focuses on collecting log data from various systems and services while simulating multi-step attacks. Such attacks span multiple stages of a cyber kill chain \citep{hutchins2011_cyber_kill_chain}, involving diverse activities like active scans, vulnerability exploitation, and data exfiltration. Normal behavior is emulated via automation tools, implementing formally described state machines. The data were captured in a configurable virtual testbed across eight scenarios. Despite focusing on logs, the later released AIT NetFlow dataset supports benchmarking NIDSs as well. A related work, the \emph{Kyoushi Log dataset} \citep{landauer2021_kyoushi_dataset}, is a subset of AIT LDS2.0.

\emph{OD-IDS2022} \citep{patel2023_od_ids2022} (2022) offers a diverse emulation of 28 modern attacks captured in a virtualized network. The environment contains external and internal attacker networks targeting an Ubuntu Apache HTTP server. The features were extracted using CICFlowMeter. However, details of the environment and normal traffic profiles are lacking.

\emph{UWF-ZeekData22} \citep{bagui2023_uwf_zeekdata22} (2022) contains traces from a cyber wargaming course at the University of West Florida. During the course, red (offensive) and blue (defensive) teams were tasked to attack/defend the parts of the university network. The traffic features were extracted by Zeek and labeled based on the MITRE ATT\&CK \citep{mitre2024_mitre_attck} phases using participants' logs. However, 99.9999\% of malicious traffic involves reconnaissance and discovery, leaving too few samples for other phases. Therefore, the dataset consists mostly of scanning traffic, making it unsuitable for systems focusing on other intrusion types.

\emph{VHS-22} \citep{szumelda2022_vhs22_dataset} (``Very Heterogenous Set of Network Traffic Data'', 2004--2022) combines five datasets -- ISOT Botnet, CTU-13, Booters, CICIDS-2017, and Malware Traffic Analysis (MTA) project captures \citep{duncan2024_malware_traffic_analysis}. Their traffic was replayed, and timestamps reset on 2022-01-01, thus resembling a single capture (including gaps) with 115 classes. The data is provided as unidirectional flows extracted via a custom network probe. Its high diversity will likely prove challenging for some ML methods, as anomaly detectors might struggle to establish a sense of a traffic normality profile.

\emph{FLNET2023} \citep{kumar2023_flnet2023} (2023) emulates a complex network with ten routers while collecting traffic on each to facilitate federated learning (FL). In contrast to traditional ML algorithms, which suppose that all data is available at one centralized location during training, federated learning aims to train the model in a decentralized, privacy-preserving manner \citep{zhang2021_survey_federated_learning}. If datasets are not created with FL in mind, different data locations need to be simulated, often leading to an experimental bias. In FLNET2023, the authors mimicked a real-world network and emulated slow DoS, volumetric DDoS, web, and infiltration attacks, extracting features via CICFlowMeter and adding an extra location feature for FL.

Finally, \emph{TII-SSRC-23} \citep{heryalla2023_tii_src23} (2023) promotes data diversity by providing eight distinct traffic types. They are divided into six benign (network use cases like audio or video streaming, text, and background) and 26 malicious subtypes (brute-force, DoS, information gathering, and Mirai botnet). The data are provided as the raw PCAP and bidirectional flows produced by CICFlowMeter. However, the dataset is heavily imbalanced, with benign traffic comprising only 0.015\% (1300 out of 8.7 million) flows, posing challenges for its utilization.

\subsection{Special-Purpose Datasets}
\label{ssec:ndatasurv_data_special}

In contrast to general-purpose datasets, special-purpose datasets focus on specific aspects of intrusion detection -- a single intrusion type (intrusion-specific) with multiple variants and parameters or intrusions within a particular network environment (environment-specific), such as IoT or industrial control networks. This subsection introduces existing special-purpose datasets within the NID domain.

\subsubsection{Intrusion-Specific}

\textbf{DoS/DDoS:} Denial of Service (DoS) and Distributed DoS (DDoS) are the among the most prevalent intrusion-specific datasets. These attacks have been around since the beginning of the Internet and remain a major threat to this day \citep{mittal2023_dl_detecting_ddos}. The first DoS dataset, \emph{LLDOS} \citep{haines2000_darpa2000} (2000), was published by DARPA. Valuable real-world samples were published in \emph{CAIDA DDoS Attack 2007} \citep{caida_ddos_attack2007} and \emph{Booters} \citep{santanna2015_booters} datasets.

Seeking to address challenges in obtaining real-world traffic, \emph{DDoS 2016} \citep{alkasassbeh2016_ddos2016} was emulated within a virtualized environment, providing selected per-packet features. Similar per-packet features are also present in \emph{Bo\u{g}azi\c{c}i University DDoS Dataset} \citep{erhan2020_bogazici_ddos_dataset}, mixing emulated DDoS attacks with real campus traffic. \emph{CIC-DDoS2019} \citep{sharafaldin2019_cic_ddos2019} emulates diverse flooding attacks in raw packets and CICFlowMeter features. In contrast to flooding attacks, the \emph{2017-SUEE-data-set} \citep{lukeseder2017_suee_dataset} and \emph{USB-IDS-1} \citep{catillo2021_usbids1} focus on layer 7 slow DoS, with the latter evaluating its effectiveness against in-built webserver defensive modules.

\textbf{Botnet:}
Botnet detection is another crucial branch of NID research. The first dataset, \emph{ISOT Botnet} \citep{saad2011_isot_botnet} (2004-2010), mixes Storm and Walowdac botnet captures with non-malicious traffic from the Ericsson Research Lab in Hungary. A comprehensive set of 13 real-world botnet scenarios is provided in \emph{CTU-13} \citep{garcia2014_ctu13} by running unrestricted malware in a sandbox environment within an academic network. Selected CTU-13 traces and legitimate ISCX-IDS-2012 traffic were combined into the \emph{HogZilla} \citep{hogzilla2012_hogzilla_dataset} dataset with custom flow-based features. ISCX-IDS-2012 and CTU-13 data were also used with the ISOT Botnet and merged into the \emph{Botnet} \citep{beigi2014_botnet_dataset} dataset. In contrast to analyzing entire network traffic, \emph{ISOT HTTP Botnet} \citep{alenazi2017_isot_http_botnet} focuses on detecting botnets only via DNS data. Botnet detection is also closely linked to IoT, as IoT devices were historically poorly secured, making them easy targets for compromise and subsequent participation in botnets. We cover them in the environment-specific datasets subsection.

\textbf{Anomaly detection:}
Instead of identifying specific intrusions, anomaly detectors learn patterns of regular communication and then detect deviations from them. While any dataset can be used to benchmark anomaly-based systems, some were specifically designed for this purpose. \emph{Comprehensive, Multi-Source Cyber-Security Events} (known as \citet{kent2016_cybersecdata_dynamicnet}) and \emph{Unified Host and Network} \citep{turcotte2019_unified} datasets, collected from the Los Alamos National Laboratory enterprise network, provide flow and end-host log data with a base minimalistic feature set. In addition, heavy anonymization and the absence of direct labeling significantly limit their usability.

\emph{MaWILab} \citep{fontugne2010_mawilab}, the longest-running intrusion detection capture (since 2001), utilizes the MAWI Working Group Traffic Archive \citep{mawi2024_mawi_wg_traffic_archive} for daily 15-minute backbone traces. These traces are automatically labeled via combined anomaly detectors. However, the project's inactivity since 2015 has likely rendered these detectors outdated, hurting the ground truth accuracy. Furthermore, although the MAWI Archive continues to provide daily traces, MAWILab includes limited recent data.

Automatic labeling, this time based on external third-party Cyber Threat Intelligence, was also utilized by \emph{LuFlow} \citep{mills2022_luflow}. It provides flow-based traffic from the Lancaster University network, labeling its data as benign, malicious, or outlier. The \emph{TrabID} dataset \citep{viegas2017_trabid_dataset} aims to realistically evaluate NIDSs' adaptability by dividing the data into three types likely encountered in operational networks: known, similar, and unknown. Unfortunately, its environment details were never released.

\textbf{APT:}
Conventional IDS datasets typically focus on a certain set of attacks executed in a non-systematic way. In contrast, Advanced Persistent Threat (APT) datasets simulate real-world cyber incidents, which typically follow a well-defined sequence of steps, as outlined in the MITRE ATT\&CK Matrix \citep{mitre2024_mitre_attck}. APT datasets aim to replicate realistic APT behavior, typically starting with reconnaissance and progressing toward attacker's objectives like data exfiltration, while often trying to keep such activities stealthy.

\emph{DAPT 2020} \citep{myneni2020_dapt2020} captures all APT stages performed by a red team (experienced cyber-attackers) in a virtualized environment. Realism was further increased by heavily unbalancing the traffic and including stealthy attacks, causing baseline semi-supervised detectors to perform poorly. The \emph{Unraveled} \citep{myneni2023_unraveled} dataset expands on this work with a more complex network, longer duration, defender's responses, attack tracks covering, and different attacker types. Professional red teamers were also used in the \emph{PWNJUTSU} \citep{berady2022_pwnjutsu_dataset} dataset, which is, unfortunately, unlabeled. Another APT activity, spanning 2.5 hours, is also simulated by the \emph{CyberFORCE Scenario} \citep{cybervan2021_cyberforce_scenario}. Unfortunately, its data are labeled indirectly, and the corresponding documentation lacks many details.

\textbf{Others:}
In addition to the mentioned datasets, several others focused on specific intrusions are also available. \citet{homoliak2020_asnm_datasets} published \emph{ANSM-CDX}, \emph{ANSM-TUN}, and \emph{ANSM-NPBO} for detecting buffer overflow attacks using custom flow-based features. The \emph{SSHCure} \citep{hofstede2014_sshcure} dataset captures real-world SSH brute-force attempts via a university honeypot. \emph{SR-BH 2020} \citep{riera2022_sr_bh2020} offers flow-based data for evaluating HTTP(S) multi-label intrusion detection models. In contrast to standalone datasets, \emph{NDSec-1} \citep{beer2017_ndsec1} is designed for salting by injecting its traffic into other datasets. Finally, \emph{HIKARI-2021} \citep{ferriyan2021_hikari2021} promotes the usage of encrypted traffic for NID purposes by employing application-layer attacks via HTTPS, extracting its features using CICFlowMeter and Zeek.

\subsubsection{Environment-Specific}

Unlike datasets collected from traditional computer networks, the following paragraphs focus on those originating from specialized environments. These include network infrastructures such as Software Defined Networking (SDNs) or the Internet of Things (IoT). As a result, they produce unique benign traffic patterns and are susceptible to environment-specific attacks, requiring tailored intrusion detection systems.

\textbf{IoT (botnet):}
The Internet of Things (IoT) is the most prevalent specialized environment among the public NID datasets, featuring IoT devices such as sensors, cameras, or smart home appliances. Often connected to traditional networks via Wi-Fi (IEEE 802.11) \citep{crow1997_ieee802_11} or IoT technologies like ZigBee (IEEE 802.15.4) \citep{ieee_ieee802_15_4}, these devices were manufactured without cybersecurity considerations. Some are thus vulnerable to IoT malware infections and consequent participation in botnets.

Reflecting this trend, \emph{IoT host-based ID dataset} \citep{bezerra2018_iot_hostbased_dataset}, \emph{N-BaIoT} \citep{meidan2018_nbaiot_dataset}, \emph{IoT-23} \citep{garcia2020_iot23_dataset}, and \emph{MedBIoT} \citep{guerramanzanes2020_medbiot} focus on botnet activity. Captured in controlled environments, these datasets simulate a botnet by infecting testbed hosts with popular malware families like Mirai and monitoring their behavior. In addition to botnet activity, the \emph{IoT network intrusion dataset} \citep{kang2019_iot_net_intrusion_dataset} includes other attacks like scanning and ARP spoofing. Since this dataset provides only raw traffic, Ullah et al. processed its data using CICFlowMeter, producing the \emph{IoTID} \citep{ullah2020_iotid} dataset.

\textbf{IoT (general):}
In addition to botnets, several datasets also explore other intrusions within IoT. \emph{Bot-IoT} \citep{koroniotis2019_botiot_dataset} emulates various virtual IoT devices and performs scanning, denial of service, and information theft attacks like keylogging via Kali Linux. \emph{TON\_IoT} \citep{alsaedi2020_toniot_dataset, moustafa2021_toniot_net} offers various scenarios involving physical and virtual IoT devices with a large corpus of modern attacks like ransomware. The \emph{Kitsune Network Attack Dataset} \citep{mirsky2018_kitsune} emulates reconnaissance, man-in-the-middle, and DoS attacks within a small IoT environment with an IP surveillance camera, including a Mirai botnet scenario.

\emph{CCD-INID-V1} \citep{liu2021_ccd_inid_v1} captures DoS, man-the-middle, and brute-force attacks within a network of four Raspberry Pi devices, using NFStream for flow-based feature extraction. Traffic from 60 physical IoT devices is gathered within the \emph{CIC IoT dataset 2022} \citep{dadkhah2022_ciciot2022}. Although primarily designed for monitoring purposes, it also includes flood and brute-force intrusion scenarios. Its data are provided as raw PCAP and custom per-packet statistical features. \emph{CICIoT2023} \citep{neto2023_ciciot2023} extends its scope to 105 devices and 33 attacks across seven categories, including the Mirai botnet. Along with the raw data, pre-extracted fixed-size packet window features are also available.

\textbf{IIoT/ICS:}
Industrial IoT (IIoT) is an ecosystem of devices, sensors, and applications that collect, monitor, and analyze data from industrial operations \citep{boyes2018_industrial_iot}, oftentimes integrating Industrial Control Systems (ICS) technologies, e.g., Supervisory Control and Data Acquisition (SCADA). IIoT datasets aim to capture the behavior of industrial networks but some also contain regular, non-industrial-specific attacks.

\emph{WUSTL-IIoT-2018} \citep{teixeira_wustl_iiot2018} and \emph{WUSTL-IIOT-2021} \citep{zolanvari2019_wustl_iiot2021} present reconnaissance and exploit attacks (the latter also DoS and command injection) within a small SCADA setup with a water storage tank. Their features were extracted by Argus in a bi-flow format. \emph{X-IIoTID} \citep{alhawawreh2022_xiiotid} designed connectivity- and device-agnostic emulation in a physical testbed for various attacks categorized by the ATT\&CK framework. \emph{Edge-IIoTset} \citep{ferrag2022_edge_iiotset} proposed a small but diverse multi-layer testbed with numerous device types. It provides a rich set of attacks, including man-in-the-middle, malware, injection, information gathering, and DoS, offering both raw and pre-extracted data.

\textbf{ICS:}
While previous datasets mixed both IIoT and ICS traffic, \emph{ICS-Flow} \citep{deghlaghi2023_ics_flow} focuses solely on the ICS environment, simulating a bottle-filling factory within a virtual testbed. Four attack types: reconnaissance, replay, DDoS, and man-in-the-middle were executed, and their activity was captured via ICSFlowGenerator as custom flow-based features. Using the same setup, the \emph{Westermo} \citep{strandberg2023_westermo} dataset added anomaly detection scenarios like IP address misconfiguration. Since three different capture points were used, the data is also suitable for federated learning. \emph{ROSIDS23} \citep{degirmenci2023_rosids2023} provides attacks against the Robot Operating System (ROS), utilizing CICFlowMeter for feature extraction. Finally, the \emph{OPC UA} \citep{pinto2020_opc_ua_dataset} dataset captures data from an OPC UA cyber-physical system with DoS, man-in-the-middle, and spoofing attacks.

\textbf{Other:}
Various other datasets cover additional environments. \emph{InSDN} \citep{elsayed2020_insdn}, \emph{SDN Dataset} \citep{sarica2020_sdn_dataset}, and \emph{UNR-IDD} \citep{das2023_unridd_dataset} present regular, as well as environment-specific attacks for Software Defined Networking (SDN). Wireless-specific attacks like disassociation and deauthentification are captured by \emph{AWID2} \citep{kolias2015_awid2_dataset} and \emph{AWID3} \citep{chatzoglou2021_awid3_dataset}, emulating small-office-home-office (SOHO) networks. \emph{ISOT-CID} \citep{aldribi2020_isotcid} offers cloud-specific traces with network and end-host logs collected from three hypervisors in a production environment. Finally, the \emph{VOIP Enterprise – Attack Dataset} \citep{alvares2021_voip_enterprise_attack_dataset} focuses on Voice over IP (VoIP)-specific attacks like VoIP-DoS.

\subsection{Datasets Beyond This Survey}
\label{ssec:ndatasurv_other_datasets}

For this survey, we established explicit, non-ambiguous criteria to filter relevant works, leading to the exclusion of several well-known datasets (e.g., LBNL \citep{pang2005_lbnl_dataset} or ADFA \citep{creech2013_adfa_dataset}). Additionally, we excluded dozens of relevant datasets due to being publicly unavailable. As these datasets might be suitable for specific use cases, we provide a complete list of exclusions and their reasons on our GitHub repository (see Section~\ref{sec:intro}).

Although network intrusion datasets backed by scientific publications (the focus of this survey) are a preferred way for benchmarking NIDSs, other sources might also provide relevant data. This subsection outlines these alternatives.

\subsubsection{Data Repositories}
\label{sssec:ndatasurv_other_repositories}

Data repositories are centralized locations to store, share, and categorize the data. For cybersecurity and intrusion detection purposes, the most relevant include \emph{IEEE DataPort}\footnote{IEEE DataPort: \url{https://ieee-dataport.org/}}, \emph{Kaggle}\footnote{Kaggle: \url{https://www.kaggle.com/}}, and \emph{Zenodo}\footnote{Zenodo: \url{https://zenodo.org/}}, hosting many datasets from scientific publications. Kaggle and Zenodo offer open access, while data on DataPort might require a paid subscription. Additional general-purpose repositories are listed on Wikipedia\footnote{Wikipedia datasets list: \url{https://en.wikipedia.org/wiki/List_of_datasets_for_machine-learning_research}}.

In addition to general-purpose repositories with unrestricted upload, several repositories specifically focused on network and malware traces also exist. These are typically maintained by individuals or groups, and therefore, their uploads are often restricted. We list some of the most relevant below:

\begin{itemize}
    \item \emph{ANT Datasets}\footnote{ANT datasets: \url{https://ant.isi.edu/datasets/index.html}} -- maintained by the ANT lab at the University of Southern California. It contains various benign and malicious captures since 2006.

    \item \emph{AZSecure-data}\footnote{AZSecure-data: \url{https://www.azsecure-data.org/home.html}} -- a collection of various cybersecurity datasets, including several NID datasets as well.

    \item \emph{CAIDA Data}\footnote{CAIDA: \url{https://www.caida.org/catalog/datasets/overview/}} -- Center for Applied Internet Data Analysis offers over 100 datasets, typically raw captures from diverse locations with malicious traffic, such as DDoS.

    \item \emph{Contagio}\footnote{Contagio: \url{https://contagiodump.blogspot.com/}} -- a blog with historical and contemporary malware samples, threats, observations, and traffic captures from emulated environments.

    \item \emph{IMPACT}\footnote{IMPACT: \url{https://www.impactcybertrust.org/}} -- hosts over 450 datasets in 14 classes, with a few relevant to NID. However, the data is only available to researchers in the U.S. and a few approved locations.

    \item \emph{Malware Traffic Analysis (MTA)}\footnote{MTA: \url{https://www.malware-traffic-analysis.net/}} -- a blog sharing malware samples and their raw network behavior since 2013.

    \item \emph{Security Datasets}\footnote{Security Datasets: \url{https://securitydatasets.com/}} -- an open-source project with dozens of malicious and benign datasets from various platforms.

    \item \emph{Stratosphere Laboratory Datasets}\footnote{Stratosphere Datasets: \url{https://www.stratosphereips.org/}} -- captures malware and benign samples from controlled conditions. Hosted by a research group from Czech Technical University.

    \item \emph{UMASS Trace Repository}\footnote{UMASS Traces: \url{https://traces.cs.umass.edu/}} -- contains various network datasets primarily for traffic classification.
\end{itemize}

\subsubsection{Data Lists}

In contrast to repositories, data lists aggregate datasets from various locations without storing them. This part covers lists published in a non-scientific manner. For scientifically-backed ones, i.e., surveys, see Related Work (Section~\ref{sec:related_work}).

A list of publicly available PCAP files, primarily sourced from malware activity, is maintained by \emph{NETRESEC}\footnote{NETRESEC list: \url{https://www.netresec.com/?page=PcapFiles}}. In contrast, researcher \emph{Laurens D'Hooge} focuses on pre-processed NID datasets in a textual format\footnote{L. D'Hooge data: \url{https://www.kaggle.com/dhoogla/datasets}}. Several network and malware captures are also aggregated by the \emph{SecRepo project}\footnote{SecRepo: \url{https://www.secrepo.com/}}. The \emph{Outlier Detection Datasets (ODDS)} project\footnote{ODDS: \url{https://odds.cs.stonybrook.edu/}} collects datasets for anomaly detection, including some for cybersecurity. Finally, the GitHub repositories \emph{Awesome Cybersecurity Datasets}\footnote{Awesome Cybersecurity Datasts: \url{https://github.com/shramos/Awesome-Cybersecurity-Datasets}} and \emph{Real-CyberSecurity-Datasets}\footnote{Real-CyberSecurity-Datasets: \url{https://github.com/gfek/Real-CyberSecurity-Datasets}} compile numerous cybersecurity datasets, including those for NID.

\subsection{Dataset Popularity}
\label{ssec:ndatasurv_popularity}

As outlined in this paper, the Network Intrusion Detection domain has long been plagued by the lack of quality datasets. For this reason, many researchers have relied on DARPA-based datasets -- KDD\textquotesingle99 and NSL-KDD, for over two decades. While many recent studies \citep{abdulganiyu2023_slr_nids, yang2022_slr_anids, ahmed2022_nids_dataset_survey, hindy2020_network_threats_taxonomy} still consider them the most popular, such studies tend to analyze extended time periods, sometimes dating back to 2008 \citep{hindy2020_network_threats_taxonomy}. While  KDD\textquotesingle99 and NSL-KDD remain extensively used, we argue that analyzing across such long time spans fails to reflect contemporary trends in the field properly.

\begin{figure}[t]
    \centering
    \includegraphics[width=\linewidth]{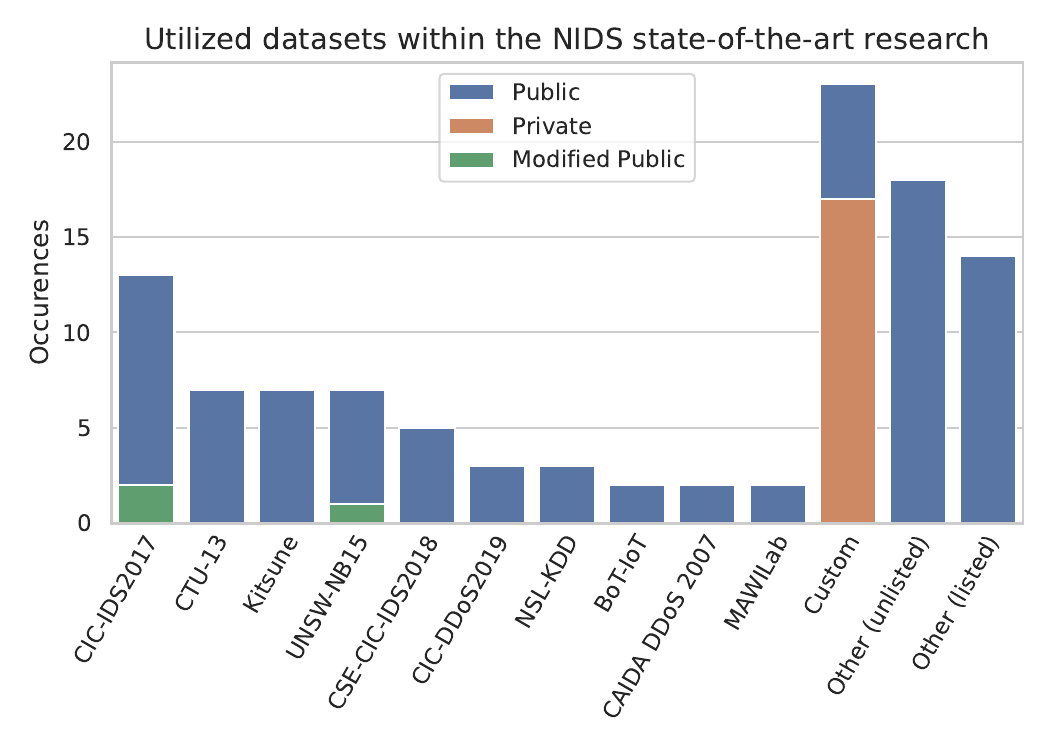}
    \vspace*{-2em}
    \caption{Analysis of the utilized datasets from the contemporary NIDS research. Most papers used custom datasets collected specifically for the given research (category \emph{Custom}), whereas the majority of them were not publicly shared. The category \emph{Other (listed)} represents datasets included in the survey (Table~\ref{tab:data_survey}) but used only once, so they were grouped. The category \emph{Other (unlisted)} groups datasets not included in the survey.}
    \label{fig:data_popularity_graph}
\end{figure}

For this reason, we conducted our own data popularity analysis of NID papers published between 2020 and 2023 at top-tier security conferences (details in~\ref{assec:meth_popularity_determination}). In total, we identified 45 relevant papers and summarized their utilized datasets in Figure~\ref{fig:data_popularity_graph}. Most papers evaluate using more than one dataset (often three or four), causing the total sum to be higher than 45. We discuss key findings below, with a complete list of papers and their analysis on our GitHub.

\textbf{Custom data to get the job done:} This survey shows that while many datasets exist, most current research prefers to collect its own data (category \emph{Custom}). However, the lack of data sharing often renders the associated research non-replicable.

Due to challenges in generating normal traffic, many works collected only malicious traffic themselves and then utilized normal (i.e., assumed benign) traces from other sources, such as MAWI \citep{mawi2024_mawi_wg_traffic_archive} or CAIDA \citep{caida_anonymized_traces}, to create a mixed custom dataset.

\textbf{Not a clear winner:} Even if researchers decide to utilize public datasets, our findings indicate that their usage is relatively evenly distributed. While CIC-IDS2017 emerged as the most popular, its utilization only in 29\% of papers brings it far from being a ``standardized'' benchmark in the domain. In addition, most of its users rely on the original version with known errors, potentially undermining their research outcomes and indicating their unawareness of the dataset's flaws and fixes.

Other datasets exceeding five uses include CTU-13, Kitsune, and UNSW-NB15. Datasets listed in this survey (Table~\ref{tab:data_survey}) but utilized only once were aggregated under the \emph{Other (listed)} category, whereas \emph{Other (unlisted)} groups datasets not included in the table. This category mainly refers to reused datasets from other papers that produced data as a research side-effect. However, it also includes datasets like \emph{MACCDC} \citep{netresec_maccdc}, published as a part of the NETRESEC data list, and \emph{CIC DoS 2017} \citep{jazi2017_cic_dos2017_dataset}, unavailable at the time of writing. For more details, refer to our GitHub.

Concerning the above results, we reason that creating custom datasets (\emph{Custom}) or relying on custom datasets from similar research (\emph{Other (unlisted)}) stems from the diversity of NID tasks. Although many datasets exist, novel research often targets very specific scenarios (e.g., attacks or environment setups) not covered by public datasets. This results in a significant dataset usage variety and the need for custom data.

\textbf{Are DARPA-based datasets finally dead?} Unlike other studies highlighting the dominance of DARPA-based datasets, our analysis did not confirm such claims -- NSL-KDD was used only three times, whereas KDD\textquotesingle99 only once, contradicting other surveys. We explain this discrepancy in two ways:

\begin{enumerate}[label=\alph*), topsep=0pt, itemsep=2pt]
    \item \emph{Surveyed years}: In contrast to other surveys, our analysis focused only on recent papers (2020--2023). As many datasets have been released recently and the DARPA flaws are now better understood, newer datasets are preferred.

    \item \emph{Surveyed venues:} While data popularity analyses in other surveys considered diverse literature sources regardless of their ranking, our analysis selected only papers from Tier~1 and~2 security conferences (CORE A*/A ranks). As their review process is known to be more strict and emphasizes novelty, presented works are expected to be evaluated more rigorously. Given the well-known flaws of DARPA data, we assume that researchers aiming for rigorous evaluation would indeed prioritize newer datasets.
\end{enumerate}

\subsection{Trends in NID Dataset Research}
\label{ssec:ndatasurv_trends}

This subsection analyzes trends in datasets and their properties based on Table~\ref{tab:data_survey}. We primarily focus on property changes over time, as well as differences between new datasets (published since 2020) and older ones. For this analysis, we consider the data publication date rather than its collection year.

\subsubsection{Increase in Number While Narrowing the Scope}
\label{sssec:ndatasurv_trends_more_and_narrow}

\begin{figure}[t]
    \centering
    \includegraphics[width=\linewidth]{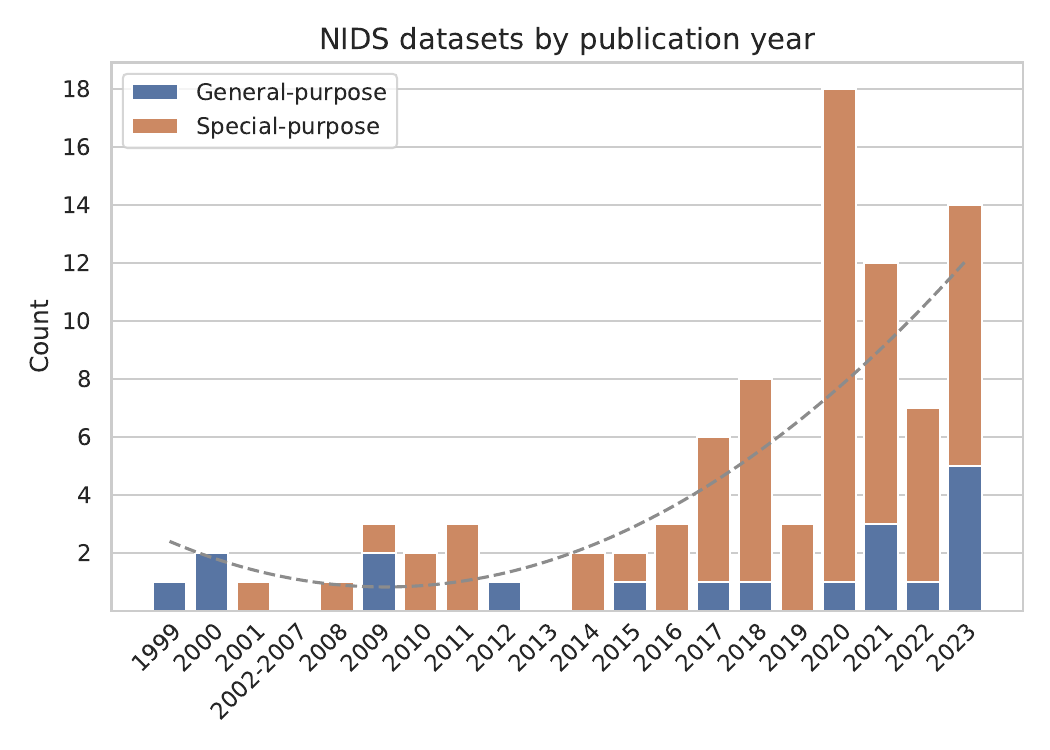}
    \vspace*{-2em}
    \caption{The number of NID datasets by the year of publication. As shown, the number of released datasets has grown polynomially in the past years.}
    \label{fig:data_publications_yearly}
\end{figure}

At the dawn of the NIDS domain (1990s, 2000s), public datasets were scarce, forcing researchers to collect their own data to evaluate the experiments. While this practice is still widely utilized (Figure~\ref{fig:data_popularity_graph}), recent years have witnessed a surge in dataset publications. Notably, 51 out of 89 datasets included in this survey were collected since 2020, with a steady rise in publications since 2016, as shown in Figure~\ref{fig:data_publications_yearly}.

Along with publishing more data, the datasets are also becoming more specialized. Since 2016, special-purpose datasets have dominated the field (83\% of published datasets), a significant shift from the previous balance with general-purpose ones. Among environment-specific datasets, IoT is the most popular environment, present in 15\% of all datasets (20\% with IIoT and industrial networks). From an attack perspective, most prevalent intrusion-specific datasets focus on botnet traffic (10\%) and DoS/DDoS (9\%). Additionally, since 2020, four Advanced Persistent Threat (APT) datasets have been released, emphasizing the growing significance of these threats. 

\subsubsection{Traffic Generation Strategies}
\label{sssec:ndatasurv_trends_traffic_generation}

Providing realistic normal and attack traffic should be a primary objective of every dataset. As discussed throughout this paper, the network traffic can be acquired through real capture, emulation, and synthetic generation -- each with distinct advantages and drawbacks. This part elaborates on these techniques and explores their applications in existing datasets.

\textbf{Normal traffic:} While capturing normal traffic from a production network offers the most realism, the associated privacy concerns and anonymization needs can be highly discouraging. For this reason, only 17\% of datasets provide purely \emph{real} normal traffic, while an additional 8\% combine it with emulation.

Aiming to overcome the issues with real traffic, many datasets utilize emulation: 71\% rely solely on it, while additional 8\% include it alongside other traffic types. This trend is even stronger for new datasets -- 84\% include emulation since 2020. While physical emulation is the most common (41\% of emulated datasets), virtual emulation (32\%) and their combination (27\%), are also used. Notably, virtual emulation is gaining more prominence (60\% of emulated datasets since 2020) due to improved computing and virtualization technologies.

In addition to the emulation environment (physical or virtual), normal traffic is also characterized by its generation method. In this regard, we observed four common patterns:

\begin{enumerate}[topsep=0pt]
    \item \emph{Emulation based on profiles}: Pioneered by DARPA datasets and formalized by \citet{shiravi2012_iscx2012}, this mechanism is based on creating profiles simulating user behaviors. The profiles, implemented via automata and scripts, perform various tasks such as interacting with applications, browsing websites, or checking e-mails. This method could create realistic traffic patterns depending on profile robustness. Datasets using this method include CIC-based data \citep{shiravi2012_iscx2012, sharafin2018_cicids2017_csecic2018}, CIDDS-001/-002 \citep{ring2017_cidds001, ring2017_cidds002}, Unraveled \citep{myneni2023_unraveled}, AWID3 \citep{chatzoglou2021_awid3_dataset}, or HIKARI \citep{ferriyan2021_hikari2021}.
    
    \item \emph{Emulation via traffic generators}: Traffic generators and performance testers are a straightforward way to simulate background activity. However, they typically produce overly simplistic (e.g., monotonous) traffic that fails to reflect real network complexity. This fact makes it easier for detectors to separate such traffic from anomalies and malicious patterns, leading to an overestimation of their real capabilities. Examples employing traffic generators include FLNET2023 \citep{kumar2023_flnet2023}, UNR-IDD \citep{das2023_unridd_dataset}, and GTCS \citep{mahfouz2020_gtcs}, using hping3, IPerf, and Ostinato tools respectively. Further information and a list of 92 traffic generators are provided by \citet{adeleke2022_network_traffic_generation_survey}.
    
    \item \emph{Incorporating humans}: The most realistic -- yet most expensive -- way of traffic emulation involves humans performing regular tasks on computers or interacting with the network (e.g., using IoT devices). Consequently, this approach also requires to specify and follow pre-defined user profiles to ensure traffic diversity. Datasets with this emulated behavior include CUPID \citep{lawrence2022_cupid} or CICIoT2023 \citep{neto2023_ciciot2023}.
    
    \item \emph{Disregarding normal traffic}: Another way of dealing with normal traffic is to exclude it entirely. These datasets, i.e., Booters \citep{santanna2015_booters}, Twente \citep{sperotto2009_twente_dataset}, and PWNJUTSU \citep{berady2022_pwnjutsu_dataset}, are usually suitable only for specific purposes, such as testing signature-based detection systems or need to be mixed with normal traffic from another source.
\end{enumerate}

Although not commonly present in public datasets, \emph{synthetic} traffic generation is another option to produce network traffic without privacy concerns. In the past, synthetic traffic, often narrowed to a network traffic matrix modeling, was generated using simple statistical methods \citep{nucci2005_synthetically_generating_ip_matrices}. However, recent advancements in generative artificial intelligence (GenAI) have improved its sophistication significantly.

\citet{ring2019_flowbased_generation_gans} were the first to use Generative Adversarial Networks (GANs) to generate network flows synthetically. Their method used a novel technique, IP2Vec \citep{ring2017_ip2vec}, to learn meaningful vector representations of IP addresses by transforming them into a continuous feature space. \citet{manocchio2021_flowgan} introduced FlowGAN, a model based on Manifold Guided GAN architecture to generate NetFlow v9 data with more realistic data distributions. In addition to GANs, the authors of STAN \citep{xu2021_stan} proposed an autoregressive neural network with convolutional layers to capture realistic spatio-temporal network traffic characteristics.

Generated traffic, whether emulated or synthetic, should aim for a high degree of realism. Synthetic generation methods are often evaluated by statistical tests that compare the distribution of generated traffic with other data, typically from the distribution the synthetic generator was trained on. For instance, \citet{schoen2024_tale_two_methods} demonstrated a robust validation approach using eight metrics, statistical tests, and domain knowledge to assess the realism, diversity, novelty, and compliance of the synthetically generated network flow data. As presented by \citet{cha2007_distance_measures_survey}, various other distance metrics for comparing probability density functions can also be applied.

\textbf{Malicious traffic}: In contrast to normal traffic, malicious traffic is generally easier to obtain. While capturing and labeling \emph{real} attacks (e.g., Booters \citep{santanna2015_booters}) might be challenging due to ground truth uncertainty, some types of real-world malicious behavior like botnets can be contained and captured in controlled environments (e.g., CTU-13 \citep{garcia2014_ctu13} or IoT host-based ID dataset \citep{bezerra2018_iot_hostbased_dataset}). Based on our survey, real malicious traffic is present in 19\% of datasets. However, this trend is decreasing, as only 13.7\% of datasets contain real attacks since 2020.

Similar to normal traffic, \emph{emulation} was the most popular way for generating malicious traffic, used in 85\% of all datasets (80\% for sole emulation without other traffic types). This trend has strengthened in recent datasets, reaching 90\% (86\% for sole emulation). Physical emulation remains the most popular, with virtual emulation gaining in popularity. The most frequent attacks were DoS (74\% of all datasets), Reconnaissance (67\%), and Other (57\%), rising to 65\% for datasets published since 2020 due to the increase in special-purpose datasets.

In general, emulating malicious traffic is less complex than normal traffic, so rigorous statistical tests are often omitted. Instead, both physical and virtual attacks are typically simulated using real-world tools, such as in the Kali Linux suite. However, since attack realism remains crucial, efforts should focus on designing diverse attack scenarios and varying attack parameters, as in the AIT Log Dataset \citep{landauer2023_ait_ldsv2_dataset}.

Although varying attack parameters improves realism, some argue it is still insufficient for modeling real-world cyber-attacks. For this reason, researchers seek to model realistic attack processes by employing professional penetration testers to perform the attacks (often aiming to be stealthy) and capture the resulting traffic. This strategy is especially popular within APT datasets like DAPT 2020 \citep{myneni2020_dapt2020} or PWNJUTSU \citep{berady2022_pwnjutsu_dataset}.

\emph{Synthetic} attack traffic is uncommon in public datasets, yet its generation and utilization are well-studied. In general, it can be used in two ways: a) oversampling or b) injection. While all synthetic traffic must comply with network functionality (align its timing characteristics with other traffic and generate valid data -- e.g., a network flow cannot contain zero packets), attacks should also account for interactions with other network activity. For instance, a successful DDoS attack would affect the timing characteristics of all simultaneous communication or introduce packet drops. For this reason, ensuring data realism with synthetic network attacks requires a delicate approach.

The first use case -- oversampling, addresses the problem of large class imbalance in NID data by generating samples for underrepresented classes. Training attack detectors on balanced data is expected to enhance their performance \citep{liu2021_id_imbalanced_network_traffic_ml_dl}. While early NID oversampling included SMOTE or ADASYN \citep{cieslak2006_combating_nid_imbalance,hu2020_wireless_nids_adasyn_cnn,bagui2021_resampling_imbalanced_nids}, modern approaches utilize GANs \citep{kumar2023_synthetic_attack_model_gans} or even transformer-based models \citep{wolf2024_benchmarking_synthetic_network_data}.

Unlike oversampling, which enlarges the sample size of existing classes, injecting synthetic attacks introduces new classes not previously present in the data. This enables the simulation of attack scenarios that were never executed on the network, offering a cost-efficient way of benchmarking NIDSs, as crafting synthetic samples is typically cheaper than emulation. The Intrusion Detection Dataset Toolkit (ID2T) \citep{cordero2021_id2t,vasilomanolakis2016_id2t} allows the injection of synthetic attacks into raw PCAP files consisting of background traffic while accounting for potential attack effects like packet drops or delays. Similar tools, including FLAME (2008) \citep{brauckhoff2008_flame} and MACE (2004) \citep{sommers2004_mace}, were also developed but are no longer maintained, thus not supporting newer attacks.

\subsubsection{Popular Formats and Feature Sets}
\label{sssec:ndatasurv_trends_formats_features}

As outlined in Section~\ref{sec:data_props}, the most common data distribution formats include network flows and packets. Our results show that flows are more popular, present in 60\% of surveyed datasets, 87\% of which are bi-directional. The trend of flow-based NID has been on the rise lately, with 76.5\% of datasets published since 2020 using this format. However, some argue that packet-based approach is more suitable for real-time detection and subsequent attack mitigation due to lower latency and comparable detection capabilities \citep{goldschmidt2024_windower}. Notably, 23\% of datasets include both formats.

Most packet-based datasets are distributed as raw packets (94\%), while a few extract headers (and payloads) in a textual format, with features depending on the captured packet types and contents. In contrast, flow-based datasets require careful feature selection, making feature engineering an important research focus \citep{sarhan2022_evaluting_ftr_sets_nids,ngo2024_mlbased_features_sel_extr}.

As no standardized feature set for flow-based data in the domain exists, dataset authors typically engineer their own features. This causes incompatibilities and degrades generalizability. Nevertheless, CICFlowMeter \citep{lashkari2016_cicflowmeter}, introduced in CIC-IDS2017, challenges this trend, as 31\% of surveyed flow-based datasets used it, making it the most popular tool for feature extraction. It is also adopted by many non-CIC datasets such as GTCS \citep{mahfouz2020_gtcs}, TII-SSRC-23 \citep{heryalla2023_tii_src23}, and FLNET2023 \citep{kumar2023_flnet2023}, and has also been used to transform other existing datasets like ToN\_IoT and Bot-IoT \citep{sarhan2022_nids_feature_set}.

Although a step towards feature standardization is positive in terms of generalizability and transferability, relying a single feature set can also bring downsides. \citet{komisarek2022_simargl2022} criticized CICFlowMeter features as unsuitable for real-time intrusion detection. \citet{sarhan2022_nids_feature_set} investigated the standardization of NetFlow v9 features for NID and found that they outperform CIC-based features in attack detection.

\subsubsection{Other Observations}
\label{sssec:ndatasurv_trends_others}

As final observations, we analyzed how the data was captured. Our results show that the majority of datasets (60\%) model small networks (under 100 hosts), rising to 72.5\% for datasets released since 2020. This trend reflects a preference for simpler, cost-efficient testbeds over larger environments.

From a temporal perspective, 29\% of datasets are captured in a single continuous run, 17\% are periodic, and 43\% are discontinuous ( gaps within the data exist), with the latter increasing to 55\% since 2020. Similar to simulations in small networks, we argue that discontinuous captures are easier to perform, allowing flexible adjustments and less strict scheduling. Therefore, they pose a convenient choice for data creators.

Looking at the trends in evaluation, 84\% of datasets offer a complete capture of the whole network, while the rest focus on honeypots or specific network parts. 89\% of surveyed datasets are directly labeled, 8\% labeled indirectly, and 3\% unlabeled. Only 23\% (16\% since 2020) provide predefined train-test data splits. As mentioned, this design offers greater flexibility but is prone to biased evaluations and selective reporting \citep{arp2022_dos_donts_ml_security,lipton2019_research_for_practice_troubling_trends}.


\section{Recommendations: Data Selection, Creation, \& Usage}
\label{sec:recommendations}

After reviewing existing datasets, this section outlines best practices on data-related NID topics: selecting the most appropriate dataset for one’s needs (Section~\ref{ssec:recommendations_selection}), creating a custom dataset (Section~\ref{ssec:recommendations_creation}), and proper data handling to mitigate biases during experimentation (Section~\ref{ssec:recommendations_usage}).

\subsection{Choosing a Dataset}
\label{ssec:recommendations_selection}

Using a public dataset is the most straightforward and time-efficient approach for NID experimentation. With dozens of such datasets available, the following paragraphs guide the selection of the most suitable ones for specific use cases. However, we refrain from suggesting concrete datasets, as this survey does not aim to assess data quality but rather help others to make informed decisions.

\textbf{The goal guides your way:} The key criterion for dataset selection is its alignment with research goals. Therefore, defining these goals first (e.g., detecting DDoS attacks on large networks) helps in analyzing properties like the dataset's focus, attacks, format, and network type to select a suitable dataset.

\textbf{Have it while it is fresh:} As discussed in Sections~\ref{sec:domain_specs} and~\ref{sec:data_props}, the traffic patterns evolve over time, making datasets age rather quickly. For this reason, selecting up-to-date datasets with recent traffic patterns is desirable to accurately assess system performance in real-world scenarios.

\textbf{Quality estimation:} While we do not assess dataset quality in this paper, we strongly recommend performing it as a part of the data selection process. Despite the previous research on the topic \citep{gharib2016_evaluation_framework_ids_data,haider2017_ngids_ds_dataset,soukup2021_towards_nids_data_quality_eval}, no standardized formal metric for NID data quality has been established within the domain \citep{kenyon2020_public_ids_datasets_fit}. As ``quality'' itself is vague, and user needs may vary, formalizing such a general metric is challenging. Therefore, we discuss data quality in a more practical sense and attempt to gauge it by analyzing
\begin{enumerate*}[label=\alph*), topsep=0pt]
    \item the realism of the data based on the rigorousness of its creation process and
    \item the contents of the data itself.
\end{enumerate*}

A quality dataset should be realistic and error-free. Since documentation (metadata) is one of the few ways to learn about the data creation, its analysis remains a primary resource for assessing realism. Its key details include data purpose, creation process, and related limitations (more in Section~\ref{sssec:recommendations_creation_publication}). Given this information, datasets with robust traffic generation and realistic environments are preferred over more simplistic ones.

However, our experience shows that analyzing documentation alone is insufficient. Even with decent documentation, numerous datasets were found to contain discrepancies like incorrect entries or timestamps. For this reason, thorough data analysis (e.g., \emph{N/A} values, incorrect timestamps, packet errors, labeling issues) is essential in addition to documentation analysis to gauge data correctness and realism.

While many studies merely point out limitations of NID data,~\citet{flood2024_bad_design_smells_nids_datasets} present a practical framework for auditing network datasets. Given network flows, their approach proposes various metrics to detect mislabeling, artifacts, and poor data diversity. Alongside automated tests, they also conduct several manual checks. Replicating these steps can help identify a dataset's drawbacks and assess its suitability.

\textbf{Other considerations:} Additional criteria, such as data popularity, might be used during dataset selection. For instance, when comparing a new proposal to existing solutions, it is important to use the same data for evaluation. However, since popular datasets are typically older, it is beneficial to include other newer datasets in addition to the popular ones.

When no suitable dataset is available, another possibility is to find the data in repositories (Section~\ref{sssec:ndatasurv_other_repositories}) or to collect a custom dataset, as outlined in the following subsection.

\subsection{Creating a Dataset}
\label{ssec:recommendations_creation}

\begin{figure*}[t]
    \centering
    \includegraphics[width=0.95\linewidth]{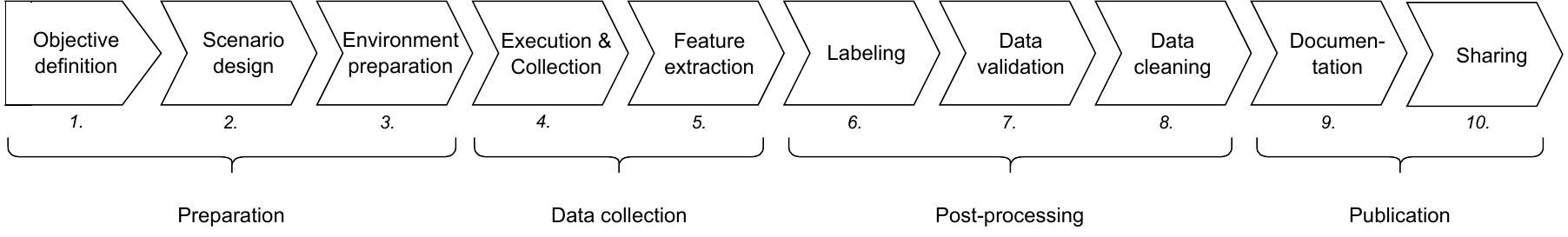}
    \caption{The process of creating a NID dataset. The ten steps are further grouped into four phases -- preparation, data collection, post-processing, and publication. Ideally, the process is linear, and the data is collected and processed in one run. However, in practice, the process is often iterative, such as re-designing and adding new scenarios or collecting and post-processing the data in multiple turns. Continually-captured datasets have to reiterate steps 2--10 periodically by design.}
    \label{fig:dataset_creation}
\end{figure*}

In essence, creating a dataset involves ten steps, from defining objectives and designing scenarios to documenting and sharing the data, as shown in Figure~\ref{fig:dataset_creation}. In the following paragraphs, we elaborate on the datasets' desirable properties (Table~\ref{tab:dataset_desirable_props}) and recommendations for individual data creation steps to mitigate common dataset limitations outlined in Section~\ref{ssec:domain_specs_limitations}.

\begin{table*}[ht!]
    \small
    \centering
    \renewcommand\cellalign{tl}
    \caption{Summary of desirable dataset properties. Dataset's compliance with more properties typically increases data quality. The creation step column outlines which steps during the dataset creation influence the given property. If a dataset needs to stay timely, steps 1--10 need to be periodically reiterated.}
    \vspace*{1em}
    \begin{tabular}{p{2.5cm} p{4cm} p{1.8cm} p{8.2cm}}
    \textbf{Property} & \textbf{Mentioned in} & \textbf{Creation step} & \textbf{Description} \\
    \toprule
    
    Correctness & \makecell{\citet{viegas2017_trabid_dataset} \\ \citet{malowidzki2015_network}} & 1--10 & Error-free implementation of the dataset creation steps -- e.g., no packet losses, correct feature extraction and labeling, documentation corresponds to the data. \\ \midrule
    
    Diversity & \makecell{\citet{catillo2023_ml_public_ids_datasets}\\ \citet{ring2019_nids_datasets_survey}\\ \citet{marciafernandez2018_ugr16}} & 2, 3 & Diversity and variability of both malicious traffic (multiple launches of the same attack with differing configurations) and benign traffic (rich usage profiles). \\ \midrule
    
    Documentation & \makecell{\citet{landauer2023_ait_ldsv2_dataset}\\ \citet{ring2019_nids_datasets_survey}\\ \citet{marciafernandez2018_ugr16}} & 9 & Documentation (metadata) provides detailed and unambiguous information about the dataset. \\ \midrule
    
    Extensibility & \citet{kenyon2020_public_ids_datasets_fit} & 2, 3, 9, 10 & New scenarios, attacks, and traffic capture events can be added on demand. \\ \midrule
    
    Label completeness & \makecell{\citet{apruzzese2022_sok_unlabeled_data}\\ \citet{ring2019_nids_datasets_survey}\\ \citet{marciafernandez2018_ugr16}\\ \citet{malowidzki2015_network}} & 6 & All records are labeled directly, correctly, and unambiguously. \\ \midrule
    
    Pre-split & \citet{ring2019_nids_datasets_survey} & 2--4 & Provides a predefined train-test split to support fair and objective benchmarking. \\ \midrule
    
    \makecell{Privacy\\preservation} & \makecell{\citet{ferriyan2021_hikari2021}\\ \citet{ring2019_nids_datasets_survey}\\ \citet{viegas2017_trabid_dataset}} & 2--5, 8 & The data does not compromise the privacy (reveal sensitive information) of users. Captures from real networks are properly anonymized. \\ \midrule
    
    Realism & \makecell{\citet{apruzzese2023_role_of_ml_cybersec}\\ \citet{myneni2023_unraveled}\\ \citet{viegas2017_trabid_dataset}\\ \citet{shiravi2012_iscx2012}} & 1--5, 7 & The dataset contents and their characteristics reflect real-world networks. The environment is properly and completely configured. The data does not exhibit any unintended properties (e.g., artifacts) as a result of simulation. \\ \midrule
    
    Relevance & \makecell{\citet{landauer2023_ait_ldsv2_dataset}\\ \citet{ring2019_nids_datasets_survey}\\ \citet{ring2017_cidds002}} & 1--6 & Dataset scenarios (e.g., network topology, attacks, normal behavior) and the data itself (format, features, labels) are appropriate and relevant for the specified dataset objectives. \\ \midrule
    
    Reproducibility & \makecell{\citet{catillo2021_demystifying_public_ids_data}\\ \citet{landauer2023_ait_ldsv2_dataset}\\ \citet{lewandowski2023_guidelines_nids_datasets}} & 2, 3, 9, 10 & The dataset contents can be regenerated on demand. \\ \midrule
    
    Timeliness & \makecell{\citet{apruzzese2023_role_of_ml_cybersec}\\ \citet{ring2019_nids_datasets_survey}\\ \citet{malowidzki2015_network}} & 1--3, 10 & Contains up-to-date modern attacks and traffic profiles resembling the current threat landscape and the behavior of contemporary computer networks. \\ \bottomrule
    \end{tabular}
    \label{tab:dataset_desirable_props}
\end{table*}

\subsubsection{Preparation}

The preparation phase encompasses all activities prior to the actual data collection. Its steps involve
\begin{enumerate*}[label=\alph*)]
    \item defining the dataset's objective,
    \item designing relevant scenarios, and
    \item preparing the capture environment.
\end{enumerate*}

\begin{enumerate}[nosep, label=\textbf{\arabic*}., topsep=0pt, wide, labelindent=\parindent, itemsep=0pt, parsep=0pt]
\setcounter{enumi}{0}

\item \textbf{Objective definition:} The first step is to define the purpose of the dataset, which then guides its entire creation process -- selecting scenarios, features, and post-processing. Since many different attacks and network environments exist, creating a universal ``perfect'' dataset is infeasible. Instead, dataset authors should narrow down to the specific problem (e.g., attack class or adversary tactics). Its objective should cover at least three aspects:
\begin{enumerate*}[label={\arabic*)}]
    \item malicious traffic class(es),
    \item environment type and its size, and
    \item time span.
\end{enumerate*}
For instance, ``\emph{to benchmark NIDSs for Slow DoS detection on medium-sized enterprise networks over several weeks}. A clear objective helps to determine the dataset's scope and delimitations.

\item \textbf{Scenario design:} Following the specified objective, the next step is to design the scenarios simulated during the execution and collection. Key considerations include planning on obtaining normal background traffic, malicious traffic, and their interactions. This phase also schedules network events -- e.g., specific attacks, exact commands to generate them, and their relative start and end timestamps. The dataset should reflect current traffic patterns and attacks in real-world operational networks, making \emph{timeliness} and \emph{realism} top priorities. Designs are also advised to be \emph{extensible}, allowing new scenarios and traffic patterns to be incorporated on demand \citep{viegas2017_trabid_dataset,apruzzese2022_sok_unlabeled_data}.

As discussed in Section~\ref{ssec:recommendations_selection}, documentation is crucial for assessing data realism. Nevertheless, realism is primarily shaped during scenario design and environment setup, further influenced by nearly every other action during the dataset creation. Regarding design, data authors should create scenarios aligned with the patterns observable in real-world networks. To prevent spatial bias \citep{pendlebury2019_tesseract}, datasets can also reflect realistic traffic ratios, typically by a disbalance favoring background traffic.

Further considerations should be given to \emph{diversity}, ensuring the dataset includes diverse and variable normal and malicious traffic profiles. Insufficient variability can make the distinction between normal and malicious traffic trivial, leading to unrealistically inflated NIDS performance. \emph{Extending capture duration} by covering day-night or workweek-weekend cycles, as in UGR\textquotesingle16 \citep{marciafernandez2018_ugr16}, is one way to enhance traffic variability. Although emulation and synthetic traffic generation methods are becoming more sophisticated, some argue that \emph{real background traffic} must be used for realism \citep{ring2019_nids_datasets_survey, marciafernandez2018_ugr16}. However, it presents challenges like \emph{privacy preservation} and establishing reliable ground truth \citep{landauer2023_ait_ldsv2_dataset}.

\item \textbf{Environment setup:} The last preparation step includes setting up a capture environment that aligns with the defined objectives and scenarios. In emulated setups, this includes configuring network equipment, end-hosts, as well as capture infrastructure, such as port mirroring or NetFlow probes, with realistic configuration and ensuring the capture will not bias the data (e.g., port mirroring introducing significant delays). Real environments utilize existing infrastructure but might add specific monitoring devices, vulnerable servers, or malicious hosts just for simulation purposes. For synthetic data, this setup refers to training or configuring a model to generate the traffic.

With the aim of maximizing data validity, some researchers suggest the datasets should be \emph{reproducible} \citep{landauer2023_ait_ldsv2_dataset}, \emph{extensible}, and \emph{dynamic} \citep{ferriyan2021_hikari2021, ring2017_cidds001, shiravi2012_iscx2012}, allowing their regeneration and updates with new traffic patterns and topologies. While feasible for virtual and synthetic environments, this is often impractical in physical setups. A step towards emulation reproducibility and systematization of the generation process is provided by standardized emulation platforms like MITRE Caldera \citep{mitre2024_caldera}.
\end{enumerate}

\subsubsection{Data Collection}

Collecting the data comprises
\begin{enumerate*}[label=\alph*)]
    \item executing scenarios and collecting the traffic and
    \item extracting relevant features.
\end{enumerate*}

\begin{enumerate}[nosep, label=\textbf{\arabic*}., topsep=0pt, wide, labelindent=\parindent, itemsep=0pt, parsep=0pt]
\setcounter{enumi}{3}

\item \textbf{Execution \& collection:} After setting up the environment, the scenarios are run according to the execution plan, and relevant traffic is captured. The data format depends on the dataset goals, but \emph{multiple data formats (and sources)}, such as packets, flows, and logs, are preferred to support benchmarking of various NIDS types \citep{myneni2023_unraveled}. However, collecting data from multiple sources presents challenges, such as clock synchronization and the need for event correlation.

\item \textbf{Feature extraction:} Feature extraction now processes the collected data into feature vectors suitable for intrusion detection. When creating a raw packet dataset or extracting features directly from NetFlow, this step is implicit during execution and collection. Otherwise, a \emph{relevant feature set} needs to be selected in accordance with the dataset's objective, i.e., effectively describing attacks while distinguishing them from normal background traffic. Despite efforts to standardize NID feature sets \citep{sarhan2022_nids_feature_set}, the optimal feature set remains an open issue \citep{sarhan2022_evaluting_ftr_sets_nids}. We thus encourage researchers to experiment with feature sets while ensuring they are:
\begin{enumerate*}[label=\alph*)]
    \item \emph{practical} for real-time detection (computable online), and
    \item accompanied by the extraction source code and raw data to facilitate \emph{reproducibility}.
\end{enumerate*}
\end{enumerate}

\subsubsection{Post-Processing}

The post-processing phase modifies the collected data to accommodate its usage as benchmarks. It involves three steps, namely data:
\begin{enumerate*}[label=\alph*)]
    \item labeling,
    \item validation, and
    \item cleaning.
\end{enumerate*}

\begin{enumerate}[nosep, label=\textbf{\arabic*}., topsep=0pt, wide, labelindent=\parindent, itemsep=0pt, parsep=0pt]
\setcounter{enumi}{5}

\item \textbf{Labeling:} \emph{Complete and direct labeling} is a crucial property of NID benchmark datasets, ensuring that traffic classes are distinguishable via a simple look-up or filter upon the data. Although log files benefit certain NIDS types, they should not be used for indirect labeling due to introduced uncertainties and added effort in determining the labels. Instead, logs can supplement existing direct labels. Depending on the dataset setup, labeling can be automated or human-guided. \citet{guerra2022_datasets_labeling} provide an overview of NID labeling techniques.

\item \textbf{Data validation}: Validating the data ensures \emph{correctness} and \emph{realism} -- key factors in quality assurance. \emph{Data correctness} refers to error-free capture and alignment with the designed scenarios. This involves investigating for packet losses or malformations, verifying proper attack execution and its desired effects (e.g., a DoS attack causing a service disruption), correct feature extraction (e.g., no unexpected N/A values, empty columns), the post-processing not introducing any artifacts or biases, and correct labeling. These checks are typically performed manually based on the data's nature and semantics, although automated tests, as proposed by \citet{flood2024_bad_design_smells_nids_datasets}, are also viable. In addition to verifying correctness, data \emph{realism} can be validated using statistical tests along with domain-knowledge checks, as presented by \citet{schoen2024_tale_two_methods}.

\item \textbf{Data cleaning:} In NID data creation, cleaning primarily refers to \emph{privacy preservation} -- i.e., \emph{anonymizing} IP addresses and packet payloads to address ethical, privacy, and legal concerns for data collected from real-world networks. Additional information, like DNS queries, timestamps, or transport layer ports, may also need to be stripped. In some cases, removing entire erroneous entries might even be necessary. However, excessive anonymization can reduce realism, limiting the dataset's value \citep{kenyon2020_public_ids_datasets_fit}. For this reason, we advocate for minimal changes to avoid the degradation of realism.
\end{enumerate}

\subsubsection{Publication}
\label{sssec:recommendations_creation_publication}

After collecting and processing the dataset, the final phase involves its publication. Public datasets need to be
\begin{enumerate*}[label=\alph*)]
    \item documented and
    \item shared,
\end{enumerate*}
as further discussed in the following paragraphs. We also strongly advocate \emph{specifying a static train-test evaluation split} to enable objective method comparisons and prevent introducing biases.

\begin{enumerate}[nosep, label=\textbf{\arabic*}., topsep=0pt, wide, labelindent=\parindent, itemsep=0pt, parsep=0pt]
\setcounter{enumi}{8}
\item \textbf{Documentation:} \emph{Documentation (metadata)} is one of the most important yet often neglected aspects of public datasets. Although typically finalized at the end of the process, it should record all the steps and decision-making in dataset creation. At a minimum, proper documentation should cover:

\begin{itemize}[topsep=0.1em,itemsep=0.4em,parsep=0cm]
    \item \emph{Goal of the dataset:} Types of NIDS is the dataset designed for. Attacks and scenarios captured within it.
    
    \item \emph{Data generation:} A description of the procedures to generate background traffic and attacks, including specific tools, algorithms, and their parameters.

    \item \emph{Environment:} Network topology, IP addressing, end-hosts, network devices and their configuration, placement of capture devices within the network, exact capture software/hardware, identification of attackers and victims.

    \item \emph{Data properties:} Format of the data, size of the capture, list and description of extracted features.
    
    \item \emph{Capture timestamps:} Exact timestamps when the data capture process started and ended. Exact timestamps of events (e.g., attacks) happening on the network.

    \item \emph{Labeling:} The process of how the data was labeled.

    \item \emph{Data handling:} Data transformations, anonymization, and cleaning after the data collection.

    \item \emph{Data limitations:} Out-of-scope considerations. Known faults or deficiencies discovered during data validation.

    \item \emph{Analysis of the captured traffic:} A brief captured data analysis: label and protocol distributions and traffic trends (e.g., peak hours). A benchmark detection performance with simplistic models can also be provided.
\end{itemize}

Comprehensive documentation elaborating on the above points would clarify the dataset's properties, use cases, and limitations. Despite strict space constraints in scientific papers, we encourage providing this information as additional supplementary material. Although not perfect, AWID3 \citep{chatzoglou2021_awid3_dataset} serves as an example of decent documentation.

\item \textbf{Sharing:} The final step in dataset creation is sharing its data and metadata. We recommend uploading them to multiple data storage locations, including at least one public repository (Section~\ref{sssec:ndatasurv_other_repositories}) for redundancy. In addition, publishing source code and configurations of tools for data creation (i.e., environment setup, traffic generation and collection, feature extraction, and labeling) is also highly desirable to facilitate data \emph{reproducibility} and validation \citep{lewandowski2023_guidelines_nids_datasets}. If possible, both raw packets and extracted features should be shared.
\end{enumerate}

The above 10-step process produces a static dataset reflecting network behavior within a specific period of time. However, as discussed in this paper, static datasets can quickly become outdated. For this reason, it is desirable to continually update datasets with new scenarios and traffic patterns by re-iterating these steps or designing specialized methodologies for collecting continual datasets, as elaborated in Section~\ref{sec:future_directions}.

\subsection{Using a Dataset}
\label{ssec:recommendations_usage}

After obtaining relevant data, the paper now briefly covers their proper handling. Similar to other domains, the key is to understand their semantics via exploratory data analysis and documentation, followed by standard ML procedures for data preparation like encoding and scaling \citep{garcia2014_data_preprocessing}. However, additional NID-specific considerations should be made. While this subsection summarizes them only briefly, works by \citet{arp2022_dos_donts_ml_security} and \citet{apruzzese2023_sok_pragmatic} provide more detailed analysis.

\textbf{You cannot see the future:} Given the dynamic nature of networks (Section~\ref{ssec:domain_specs_domain_char}), benign and malicious traffic patterns can change over time. Therefore, it is crucial to respect the temporal properties of the data during evaluation -- training data timestamps strictly precede evaluation data to avoid \emph{temporal data snooping}. Evaluation methods ignoring this aspect (including cross-validation) might introduce bias, leading to over-optimistic results \citep{arp2022_dos_donts_ml_security,pendlebury2019_tesseract}.

\textbf{Some shortcuts lead to hell:} Network data contain several features that might contaminate the process of data-driven learning (e.g., ML-based algorithms). These include IP addresses, port numbers, timestamps, flow IDs, or even Time to live (TTL) values. If used for training, the learner might \emph{spuriously correlate} these artifacts with a specific activity, causing a \emph{shortcut learning} phenomenon \citep{geirhos2020_shortcut_learning} and overestimating the model's performance \citep{arp2022_dos_donts_ml_security,catillo2023_ml_public_ids_datasets}. For instance, \citet{dhooge2022_contaminating_metadata_ml_nids} demonstrated 70\%--100\% accuracy on popular NID datasets by considering the destination port as the sole feature.

For this reason, possible artifacts should be removed or encoded into general features (e.g., classifying IP addresses as local, global, or other) to preserve some of their informational value while preventing spurious correlations. However, encoding must be done cautiously, and the presence of spurious correlations should be verified. Identification of potential artifacts is task-specific, requiring domain knowledge and data understanding. Nevertheless, the features mentioned above ought to be problematic in most datasets and should be addressed. Heuristic identification of highly dependent features (potential artifacts) can be performed by training a model to distinguish between the attack and the background traffic using a single feature \citep{flood2024_bad_design_smells_nids_datasets}. \citet{flood2024_bad_design_smells_nids_datasets} suggest justifying the model's performance via ML explainability techniques \citep{nadeem2023_sok_explainable_ml_security,jacobs2022_aiml_network_sec} and connecting important features to attacks' properties.

\textbf{The heaviest hammer is not always the best:} As shown by numerous studies \citep{apruzzese2022_sok_unlabeled_data,zhang2022_comparative_nids_ml,vinayakumar2019_dnn_ids}, even simplistic tree-based models like Random Forest or Gradient Boosting achieve near state-of-the-art performance with significantly faster execution. \citet{dhooge2023_castles} demonstrated that even single-rule supervised ML models perform well on most cybersecurity datasets due to the over-correlation of features with labels \citep{silva2022_netsec_datasets_bias}. Therefore, a general recommendation is to start with simple models before employing deep learning \citep{catillo2023_ml_public_ids_datasets}. For practical real-time intrusion detection, it is desirable to keep models simple so their inference (i.e., attack detection) is quick, enabling efficient processing of large traffic volumes. Therefore, we suggest adopting a data-centric perspective \citep{zha2025_datacentric_survey}, focusing on understanding and enhancing the data with simple, explainable models rather than maximizing a specific metric with complex black-box models without understanding the data.

\textbf{All that glitters is not gold:} As discussed in the previous paragraph, even simplistic detection models can achieve decent performance, such as accuracy above $0.99$. However, this does not imply their actual usefulness. Consider an example of imbalanced traffic with a malicious-to-benign ratio of 1:100. Even if we classify all samples as benign, we receive $0.99$ accuracy while not detecting any attack. On the other hand, if a detection rate was 100\% with only 1\% of false alarms, every second alarm produced by the model would be incorrect. Due to this phenomenon, known as the base rate fallacy \citep{axelsson1999_base_rate_fallacy_ids,alahmadi2022_99fp_study_soc_alarms}, and the typical NID traffic imbalance, selecting the right metrics and their correct interpretation is crucial to gauge the overall benefit of the proposed system.

Regarding specific metrics, \citet{arp2022_dos_donts_ml_security} suggest selecting them based on practical domain needs -- primarily false alarm and detection rates for NID. To tackle the base rate fallacy, it is beneficial to incorporate metrics that account for class imbalance, namely precision, recall, precision-recall curve, as well as the Receiver Operating Characteristics (ROC) curve and the area under it (AUC-ROC). Therefore, it is crucial to utilize multiple metrics rather than relying on a single one.

\textbf{Once? Not enough!} Drawing conclusions based on a single evaluation run on a single dataset is generally insufficient yet relatively common in the literature \citep{apruzzese2023_sok_pragmatic}. Ideally, models should be evaluated on multiple datasets with various splits. Conducting multiple runs (respecting temporal dependencies) allows for statistically significant comparisons using methods such as the t-test \citep{student1908_probable_error_of_a_mean} or the Mann-Whitney U-test \citep{mann1947_u_test}.

\textbf{Mix them with caution:} Same as with alcohol, you should also think twice before mixing different datasets. Since their data come from different networks and time periods, their traffic patterns will likely differ. Therefore, combining datasets is incorrect by design since bias would likely be introduced. This can cause automated learners to focus on dataset-specific properties (e.g., inter-packet arrival times) rather than malicious patterns as intended. Therefore, we strongly advise against mixing data from different sources unless necessary.

Despite its limitations, merging datasets in NID research is not uncommon (e.g., \citet{fu2021_realtime_malicious_traffic_detection}). This typically includes merging isolated attack traces with background traffic from other sources like MAWI \citep{mawi2024_mawi_wg_traffic_archive} or CAIDA \citep{caida_anonymized_traces}. To minimize potential bias from the dataset merge, their data should be
\begin{enumerate*}[label=\alph*), topsep=0pt]
    \item collected in similar time periods,
    \item from similar networks, and
    \item mixed by replaying them in a local environment \citep{adeleke2022_network_traffic_generation_survey}.
\end{enumerate*}

Tools for network traffic replay like \emph{tcpreplay} \citep{tcpreplay} preserve packet contents while unifying time dependencies within a single environment, thus reducing time-related biases. However, merging distinct datasets can still introduce artifacts. For instance, data from different systems (e.g., Linux vs. Windows) might present differences in packet contents, such as in encryption protocols. Additionally, replaying complex attacks requiring both sides of the communication might be challenging, as most replay tools are stateless \citep{adeleke2022_network_traffic_generation_survey}. These factors can introduce additional hidden biases during data merging. For this reason, NIDS methods should also be benchmarked on other datasets originating from a single source to prevent potential biases from skewing the results.


\section{Future Domain Directions}
\label{sec:future_directions}

Throughout this survey, we outlined the limitations of NID datasets as inherited properties from the unique intersection of cyber threat detection and computer networking domains. While Section~\ref{sec:recommendations} discussed best practices to tackle these limitations for immediate, personal benefits (e.g., better quality data, more sound results), this section explores long-term research and development directions to benefit the broader community.

\begin{figure*}[t]
    \centering
    \includegraphics[width=0.925\linewidth]{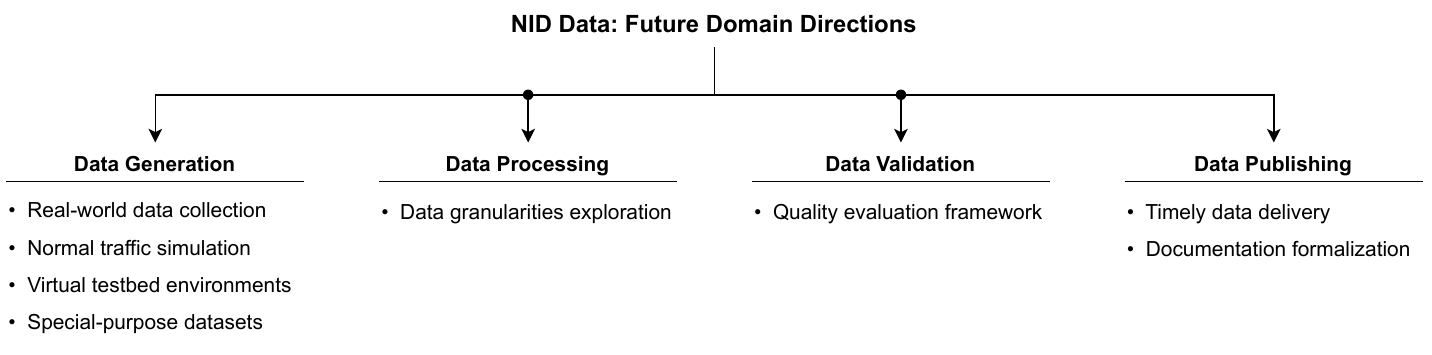}
    \vspace*{-1em}
    \caption{Future NID data research and development directions. We divide them into areas focused on generating, processing, validating, and publishing the data.}
    \label{fig:future_research}
\end{figure*}

In broad terms, future NID data research will primarily focus on the following central research question: \emph{How to facilitate \underline{access to} \underline{correct}, \underline{relevant}, and \underline{realistic data} reflecting the \underline{current} threat landscape and computer network characteristics of \underline{interest}?} This question can be divided into four areas: data generation, processing, validation, and publishing, as summarized in Figure~\ref{fig:future_research} and discussed in the following subsections.

\subsection{Data Generation}

The research branch of data generation focuses on generating timely and realistic data -- either by collecting real-world traces or improving traffic emulation and synthesis techniques.

\textbf{Collecting timely real-world data:} One of the key NID domain limitations is the lack of realistic real-world samples. As our survey revealed, only one in five datasets contains real traffic, with this ratio decreasing in recent years. This limitation impacts NIDS generalization, leading to overfitting to unrealistic traffic patterns and reduced performance in operational environments. For this reason, \emph{more data with contemporary real-world network patterns are required}. While datasets like CTU-13 \citep{garcia2014_ctu13}, Appraise H2020 \citep{komisarek2023_appraise_h2020_dataset}, and LITNET \citep{damasevicius2020_litnet2020} incorporate real traffic, they remain static and thus quickly become outdated as network traffic patterns evolve.

In this matter, a key open problem is how to enable \emph{continual real-world network data capture and annotation} while preserving privacy and ensuring label accuracy. Previous works such as Kyoto 2006+ \citep{song2011_kyoto_dataset}, MAWILab \citep{fontugne2010_mawilab}, and LuFlow \citep{mills2022_luflow} have explored this approach but are no longer updated. Currently, we are not aware of any similar project for continuous data capture providing labeled traces for NID purposes. For this reason, future research should explore \emph{automated, privacy-preserving data collection methods from real-world environments with accurate labeling}.

Arguably, the most critical factor for reliable continual datasets is the accuracy of labels. While labeling real-world traffic cannot be fully precise, more research on \emph{systematic continual labeling of real-world traffic} -- i.e., labeling traffic with confidence in a drifting environment, is needed. While traditional labeling approaches require manual interventions, continual labeling must be automated to handle a never-ending stream of traffic. Although \emph{alert correlation} techniques \citep{kotenko2022_slr_security_event_correlation} have improved labeling accuracy, adapting to concept drift remains an open problem \citep{agrahari2022_concept_drift_detection_review}. Since fully automated accurate labeling is difficult, integrating MLOps and human-in-the-loop methods could be a practical intermediate step \citep{guerra2022_datasets_labeling}.

\textbf{Improving simulated normal traffic:} Although obtaining real-world traffic is generally considered ideal, it is often difficult or impossible due to privacy concerns, the absence of desired attack events, or operational constraints. For this reason, realistic traffic emulation remains crucial for constructing diverse and representative NID datasets. While attack traffic can be emulated via existing tools used by real-world attackers, creating realistic benign traffic remains the primary challenge.

Existing datasets typically emulate benign traffic using profile-driven simulations (see Section~\ref{sssec:ndatasurv_trends_traffic_generation}). However, multiple studies \citep{layeghy2024_benchmarking_benchmark,engelen2022_pillars_of_sand,silva2022_netsec_datasets_bias} have demonstrated that this traffic lacks real-world complexity, making it easily distinguishable from intrusions even with simplistic methods \citep{dhooge2023_castles}. This is likely due to limitations in user behavior modeling, low traffic diversity, and the lack of attack stealthiness.

To enhance traffic realism and robustness, future research should focus on \emph{bridging the gap between simulated (emulated and synthetic) benign traffic and its real-world counterpart}. Although models like Generative Adversarial Networks (GANs) \citep{ring2019_flowbased_generation_gans} and Bayesian Networks (BNs) \citep{schoen2024_tale_two_methods} have already been explored, \emph{investigating and comparing novel generative methods} like diffusion models \citep{yang2023_diffusion_models_survey} \emph{for benign network traffic synthesis} offer promising research directions for diverse, realistic, and privacy-preserving traffic creation. Furthermore, the synthetic traffic must respect real-world constraints, making it important to \emph{formalize domain-specific conditions for synthetic traffic validity} (e.g., network flows cannot have zero bytes).

Regarding emulation, the community would greatly benefit from a \emph{public extensible framework for profile-based traffic generation} that could be tweaked with custom profiles to streamline the process of benign traffic creation. Methodologies \emph{combining emulated, synthetic, and real traffic} (e.g., augmenting simulated data with real traffic) could further enhance data realism while reducing privacy risks.

\textbf{Enhancing virtual testbed environments:} Advances in computing, cloud, and virtualization technologies have made virtual emulation for NID data creation more feasible and popular. In addition to its cost-efficiency, virtual emulation offers many desired properties like extensibility and reproducibility, as its configuration can be fully documented and exported. Virtual environments thus help address the problem of data timeliness by allowing datasets to be updated or completely regenerated with newer traffic patterns.

Despite their benefits, reproducibility and extensibility have received little attention from the community. Therefore, most virtualized testbed datasets lack environment setup files, configuration details, or information on customizing the environment. Although AIT-LDSv2.0 \citep{landauer2023_ait_ldsv2_dataset} emphasizes reproducibility, its environment customizability is limited. Future research should thus focus on proposing standardized \emph{methods, frameworks, and templates for fully replicable and extensible NID data creation in virtualized and cloud environments} to enable dataset validation, modification, customization, and extension.

\textbf{Creating special-purpose and scenario-based datasets:}  Despite the increasing number of NID datasets, many subdomains and scenarios remain underrepresented, hindering comparability and research progress. For instance, all surveyed \emph{cryptojacking detection} publications compiled their own datasets, highlighting the lack of a standardized one. Similarly, we are not aware of any public dataset for \emph{NID in an environment with concept drift}. Although ENIDrift NIDS \citep{wang2022_enidrift} includes such data, it was published as a research side-effect -- thus lacking comprehensive documentation and statistical analysis, limiting its usability for benchmarking. \emph{Long-term NID datasets incorporating cyclostationary patterns} are also needed, as UGR\textquotesingle16 modeling this scenario is now older.

In addition to previous scenarios, \citet{hindy2020_network_threats_taxonomy} and \citet{keersmaeker2023_survey_public_iot_datasets} pointed out that many attacks and IoT environments are not covered within public NID datasets at all. Therefore, future research should prioritize \emph{covering specific scenarios and environments, along with ensuring robust coverage of emerging threats} to streamline NIDS research in these areas. However, without continually captured and extensible datasets is made, new datasets will remain susceptible to getting outdated due to the ever-changing nature of NID.

\subsection{Data Processing}

Data processing focuses on handling the data after collection. While often discussed in NIDS methods research, we consider one aspect also deeply tied to the data itself.

\textbf{Exploring data granularities for intrusion detection:} As shown in our survey, most datasets are flow-based, while nearly half lack packet data. Flow-based NID has become the de facto standard for NID research. Similarly, many studies have explored the selection of optimal flow-based features \citep{dimauro2021_supervised_fs_nids_review,yin2023_igrf_rfe,turukmane2024_m_multisvm_ftr_select_nids}, but little research has been done on their effectiveness across different intrusion detection scenarios.

The focus on flow-based data has led to a relative neglect of research on alternative data granularities, such as packet-level features, time/packet window statistics, and cross-packet contents. These could help detect attacks missed by flow-based methods \citep{umer2017_flowbased_ids_techniques}. Future research should thus explore these alternatives and their fusion to improve detection performance, robustness, and adaptability to evolving threats.

\subsection{Data Validation}

Data validation evaluates collected (and processed) data to ensure its quality and correctness. It involves a single direction:

\textbf{Formalizing frameworks for quality evaluation:} While some metrics for NID data quality evaluation exist \citep{gharib2016_evaluation_framework_ids_data,haider2017_ngids_ds_dataset}, they primarily serve to compare datasets rather than assess their intrinsic quality (i.e., realism, correctness, and usability). Furthermore, their interpretation is often ambiguous due to the lack of usage instructions. Additionally, \citet{soukup2021_towards_nids_data_quality_eval} employed completeness and reliability measures with an evolutionary algorithm to assess dataset quality. However, their study presents only preliminary results and lacks detailed information on the proposed framework, limiting its practical applicability. These deficiencies make evaluating and selecting suitable NID datasets difficult, leading to frequent utilization of flawed datasets that undermine research reliability.

A recent paper by \citet{flood2024_bad_design_smells_nids_datasets} proposed manual and automated tests to audit common limitations of NID data, such as mislabeling, artifacts, and lack of diversity. Although this represents a significant step toward data quality evaluation and the selection of suitable datasets, their heuristics focus only on a subset of typical problems, while the proposed manual analysis requires access to both flow-based and raw data.

In order to improve NID data evaluation, future research should \emph{formalize methods and develop frameworks for evaluating NID data validity and quality}, i.e., ensure datasets are both realistic and error-free. Current data validation heavily relies on manual analysis, as seen in studies on CIC-IDS2017 and CSE-CIC-IDS2018 issues \citep{engelen2021_cicids2017_troubleshooting, liu2022_error_prevalence_nids_data, lanvin2023_errors_cicids2017}, with \citet{engelen2022_pillars_of_sand} recommending manual analysis by domain experts before public data release. Such validations are time-consuming and dependent on the availability of experts, posing additional challenges for data creators. Therefore, research on NID data quality should also integrate expert knowledge into (semi-)automated validation methods, potentially building on the \citet{flood2024_bad_design_smells_nids_datasets}'s work. Standardized approaches to verify other aspects of data quality -- e.g., the validity of features via domain knowledge, documentation via Large Language Models (LLMs) \citep{ginermiguelez2024_llms_enrich_docs_datasets}, or the detection of common issues -- would benefit data creators to validate their datasets and assist users in selecting suitable data, enhancing NID research relevance, trustworthiness, and explainability.

\subsection{Data Publishing}

Future directions in data publishing should focus on timely data delivery and standardization of data documentation.

\textbf{Timely data delivery:} Most NID datasets are uploaded to different data storage sites and later published as research papers, causing significant delivery delays for their users. However, in rapidly evolving fields like NIDS, even a few-month delay can render the data outdated. For this reason, beyond continually captured datasets discussed previously, \emph{reducing the lag between data collection and publication} is essential.

One of the potential solutions is the development of a centralized indexing system for NID datasets, similar to the digital object identifier (DOI), as proposed by \citet{abt2014_are_we_missing_labels} and later emphasized by \citet{ring2019_nids_datasets_survey}. Such a system would simplify data discovery, provide persistent availability, and reduce delivery delays. Despite these benefits and multiple calls, we are not aware of any indexing system, data storage, or publicly maintained repository of (network) intrusion datasets.

\textbf{Formalizing data documentations:} While reviewing the metadata of surveyed datasets, we observed that each was different. Although documentation of all scientifically-backed datasets follows a standard paper structure, some provide detailed data descriptions while others mention the data only briefly. This leads to under-documented datasets, reducing their credibility and usability. In addition, some datasets lack scientific backing or any documentation entirely.

For this reason, future research should \emph{formalize data documentation through guidelines, checklists, and templates to standardize NID dataset metadata}, thus enhancing data credibility and usability. Meanwhile, data creators can refer to existing informal guidelines, such as those in Section~\ref{sssec:recommendations_creation_publication}.


\section{Related Work}
\label{sec:related_work}

This section compares the presented survey with other studies discussing network intrusion detection datasets. While data-specific NID surveys (this paper) are rare, general NID surveys often mention datasets as well. Naturally, they typically lack detailed discussion on data-specific topics yet still provide a relevant reference point for many readers. We thus compare our work to those focused explicitly on NID data and five general surveys with data-related sections, summarized in Table~\ref{tab:related_work_comparison}. Based on the related work analysis, we conclude that our paper:

\begin{table*}[t]
\tabcolsep=0.09cm
    \centering
    \small
    \caption{Comparison of related work with this study. As depicted, our study covers most NID datasets (with the bonus property that all of them are publicly available), while it is the only study to perform a Systematic Literature Review (SLR) while specifically focusing on the data.}
    \vspace*{0.8em}
    \begin{tabular}{>{\raggedright\arraybackslash}m{4cm}  >{\centering\arraybackslash}m{1.33cm}  >{\centering\arraybackslash}m{1.35cm}  >{\centering\arraybackslash}m{0.9cm}  >{\centering\arraybackslash}m{1.2cm}  >{\centering\arraybackslash}m{1.45cm}  >{\centering\arraybackslash}m{0.95cm}  >{\centering\arraybackslash}m{1.7cm}  >{\centering\arraybackslash}m{1cm}  >{\centering\arraybackslash}m{1.9cm}}
    
    \textbf{Paper} & \textbf{Covered until} & \textbf{\# NID datasets} & \textbf{SLR-based} & \textbf{Data-focused} & \textbf{Compar-ative} & \textbf{Data limits} & \textbf{Popularity analysis} & \textbf{Data trends} & \textbf{Recommen-dations} \\ \toprule
    
    \citet{ring2019_nids_datasets_survey} & 2018 & 34 & \ding{55} & \ding{51} & \ding{51} & \ding{55} & \ding{55} & \ding{55} & \ding{51} \\ \midrule
    
    \citet{molinacoronado2020_survey_nids_kdd} & 2018 & 11 & \ding{55} & \ding{55} & \ding{51} & \ding{51} & \ding{51} & \ding{55} & $\circledbullet$ \\ \midrule

    \citet{ferrah2020_deep_learning_ids} & 2018 & 23 & \ding{55} & \ding{55} & \ding{55} & \ding{55} & \ding{51} & \ding{55} & \ding{55} \\ \midrule

    \citet{hindy2020_network_threats_taxonomy} & 2018 & 21 & \ding{55} & \ding{51} & \ding{51} & \ding{51} & \ding{51} & $\circledbullet$ & \ding{51} \\ \midrule

    \citet{kenyon2020_public_ids_datasets_fit} & 2019 & 34 & \ding{55} & \ding{51} & \ding{51} & \ding{51} & \ding{55} & $\circledbullet$ & \ding{51} \\ \midrule

    \citet{gumusbas2021_survey_db_dl_cybersec_ids} & 2020 & 25 & \ding{55} & $\circledbullet$ & $\circledbullet$ & $\circledbullet$ & \ding{55} & \ding{55} & \ding{55} \\ \midrule

    \citet{thakkar2022_ids_survey} & 2020 & 13 & \ding{51} & \ding{55} & \ding{55} & \ding{55} & \ding{55} & \ding{55} & \ding{55} \\ \midrule

    \citet{yang2022_slr_anids} & 2021 & 52 & \ding{51} & \ding{55} & \ding{51} & \ding{55} & \ding{51} & \ding{55} & \ding{55} \\ \midrule

    This paper & 2023 & 89 & \ding{51} & \ding{51} & \ding{51} & \ding{51} & \ding{51} & \ding{51} & \ding{51} \\
    \bottomrule
    \end{tabular}
    \label{tab:related_work_comparison}
    \vspace*{0.75em}
    {\raggedleft \small Symbols description: \ding{51}: fulfilled, $\circledbullet$: partially fulfilled, \ding{55}: not fulfilled \par}
\end{table*}

\begin{enumerate}[topsep=0.1em,itemsep=0.4em,parsep=0cm]
    \item Is the only Systematic Literature Review (SLR)-based study specifically focused on NID data,
    
    \item Lists the highest number of public datasets, all of which are publicly available with provided download links,
    
    \item Collects the most detailed properties (e.g., type of normal/attack traffic) for the listed datasets,
    
    \item Performs data popularity analysis in a fully transparent way based exclusively on Tier 1 security conferences.
\end{enumerate}

The closest study, and our inspiration in many aspects, is \citet{ring2019_nids_datasets_survey}'s work. This foundational work on NID data formalized data properties, compared existing datasets, and provided data-related recommendations. However, it does not include newer datasets. Our study builds on it by refining some  of its properties (e.g., removing the data balance property, as NID data are naturally imbalanced) and incorporating newly released datasets. We also introduce sections on data limitations and trends to address recent findings and emphasize data quality, aiming to provide a holistic view of NID data.

The work of \citet{kenyon2020_public_ids_datasets_fit} also overlaps with our paper, as it looks at NID data from different perspectives. The paper discusses various data properties and limitations, compares 34 datasets, and closes with recommendations for NID datasets. In contrast, our study covers more datasets with detailed properties in an SLR-based approach, as well as discusses dataset popularity to better aid in selecting suitable data.

Although \citet{hindy2020_network_threats_taxonomy} cover aspects similar to this paper, they mainly focus on the network threat landscape and its relation to 21 NID datasets (until 2018). The study further discusses NIDS algorithms' and datasets' popularity, threat taxonomy, IDS limitations, and recommendations for future research. In contrast, our study includes more datasets, compares them by their properties rather than by attacks, and provides a more detailed discussion of NID data-related aspects.

\citet{yang2022_slr_anids} conducted an SLR on the NID domain as a whole, highlighting popular ML-based methods and datasets for benchmarking. Although not explicitly data-focused, the study presents the largest published dataset list at the time (52), compares them, and analyzes their popularity, with KDD\textquotesingle99 and NSL-KDD coming as the most popular. In contrast, our SLR specifically focuses on the data, listing more datasets and elaborating on data-specific topics absent in Yang et al.'s work.

As mentioned in Section~\ref{ssec:meth_scope}, this survey focuses on network intrusion datasets regardless of the environment, with the condition that the data must be captured from the network. Therefore, we exclude IDS datasets where most of the data originates from non-network sources like end-host devices or sensors, as typical for Industrial Control Systems (ICS) or Internet of Things (IoT) datasets. Although relevant for environment-specific intrusion detection, we consider such datasets out-of-scope. For this reason, we recommend exploring other, more specialized surveys for data in specific areas. For instance, \citet{conti2021_survey_ics_testbeds_datasets} and \citet{koay2023_ml_ics_landscape} review ICS testbeds and datasets, whereas IoT-specific datasets are covered by \citet{keersmaeker2023_survey_public_iot_datasets} and \citet{kaur2023_iot_security_dataset_evolution}.


\section{Conclusions}
\label{sec:conclusions}

The cybersecurity environment evolves rapidly, with new attacks and vulnerabilities emerging daily. In an effort to keep up with these trends, many datasets and data-related findings for intrusion detection were published recently. Since they are scattered across multiple places and scientific resources, this study unifies the latest findings on network intrusion detection (NID) data, provides links for data access, outlines its common issues and limitations, and offers recommendations to mitigate them. We expect the research presented in this paper to be relevant for both researchers and practitioners in the NIDS domain seeking to benchmark methods or collect their own data.

In total, this paper has \emph{surveyed 89 NID datasets}, extracting 13 properties via manual analysis of their documentation and data. The resulting \emph{comparative table} is a valuable resource to aid in data selection for benchmarking new NIDS proposals. Analyzing these properties, we revealed several data creation trends, such as the growing number of published datasets each year, a preference for emulation for data generation, and the rising popularity of CICFlowMeter features.

Our analysis of contemporary state-of-the-art NIDS research revealed a shift in benchmark \emph{dataset popularity:} DARPA-based datasets are no longer the most prevalent. While CIC-IDS2017 is now frequently used, researchers still prefer to collect their own data, most of which are not publicly shared. This trend, while understandable due to the limitations of existing datasets, hinders reproducibility and validation. We emphasize the importance of following the \emph{best practices} discussed in this paper, as well as sharing the data to improve the quality and reliability of future NIDS research.

Given the NID \emph{domain-specific properties} and resulting \emph{data limitations} outlined in this paper, we state that creating a perfect dataset is infeasible. We consider data timeliness as a key issue, as research relying on static datasets will inevitably lag behind ever-evolving attacks and network trends. A step towards mitigating the issue could be achieved by adopting continuously updated datasets, as discussed in Section~\ref{sec:future_directions}.


\section*{Acknowledgments}

\noindent This work was supported by the OZ BrAIn association.

\section*{Declaration of Generative AI and AI-Assisted Technologies in the Writing Process}

\noindent \textbf{Statement:} During the preparation of this work, the authors used generative AI tools, GPT-4o and Gemini 1.5 Flash, to enhance writing in the paper -- most notably, shorten its contents and improve the overall flow of the text. After using these tools, the authors reviewed and edited the content as needed and take full responsibility for the content of the publication.

\bibliographystyle{elsarticle-harv} 
\biboptions{authoryear,colon,sort}
\bibliography{references}

\appendix

\section{Survey Methodology in Detail}
\label{asec:survey_methodology}

As outlined at the paper's beginning, the survey was conducted using the Systematic Literature Review (SLR) process to achieve comprehensive, unbiased, and replicable results.

The whole SLR process, most notably the specification of a review protocol, resource search, filtering, and results synthesis, was performed by two researchers. The first author (Ph.D. student) conducted the process under the supervision of the second author (Ph.D. supervisor), who helped in fine-tuning the review protocol and validated a random sample (10\%) of the studies. This appendix elaborates on the SLR details.

\subsection{Data Search and Filtering Process}
\label{assec:meth_search_criteria}

In order to ensure the replicability of our research, we outline the process used to identify and filter the datasets included in the survey (Figure~\ref{fig:slr_process}). The steps were as follows:

\begin{figure}[t]
    \centering
    \includegraphics[width=0.675\linewidth]{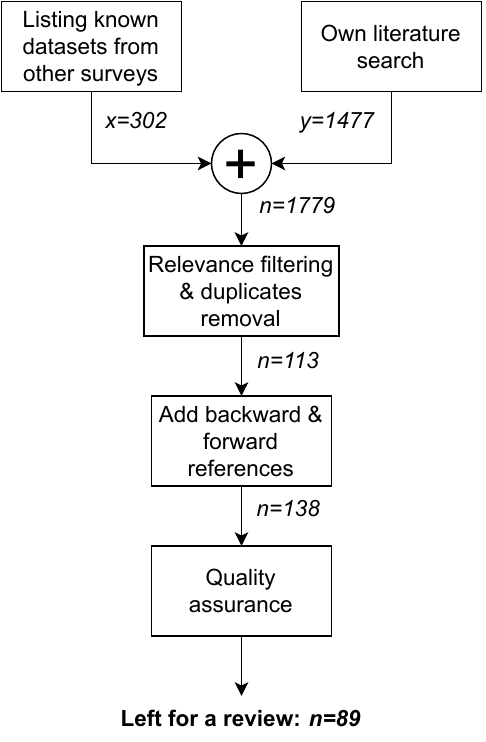}
    \vspace*{-0.5em}
    \caption{The diagram illustrates the steps of searching, filtering, and selecting relevant studies during the systematic literature review process. Specifically, we identified the objects of interest from three sources: existing survey papers, our own search via paper titles and keywords, and an analysis of backward and forward references. To ensure the feasibility of the reference analysis, it was conducted after the initial filtering. Finally, all identified sources underwent a quality assurance process to include only relevant datasets.}
    \label{fig:slr_process}
\end{figure}

\begin{enumerate}[topsep=2pt]
    \item \emph{Listing known datasets:} We extracted known datasets from existing network intrusion surveys listed in Table~\ref{tab:related_work_comparison}, resulting in 302 datasets (with duplicates).

    \item \emph{Own literature search:} We conducted a search with the following query: \textit{network (intrusion $\cup$ anomaly $\cup$ outlier $\cup$ attack $\cup$ threat) (dataset $\cup$ data\,set)}. The publication year was limited to 2023, so papers published since 2024 are not included. Matching on titles or author keywords, we used the following search engines and databases:
    
    \begin{itemize}[topsep=1pt, itemsep=1pt]
        \item SCOPUS: 656 results
        \item ACM Digital Library: 42 results
        \item IEEEXplore: 179 results
        \item Google Scholar: 600 results\footnote{Since Google Scholar returns thousands of results, we reviewed only the first 20 search pages (200 results) for all years, along with the first ten pages for 2020, 2021, 2022, and 2023 (400 results) to identify new datasets not covered in previous surveys. We acknowledge that this process is not entirely replicable. However, omitting it would leave over a dozen datasets undiscovered.}
    \end{itemize}
    
    \noindent In total, this step produced 1477 papers (with duplicates).

    \item \emph{Relevance filtering:} After collecting 302 + 1477 items from the previous steps, we excluded duplicates. Next, we selected only dataset-specific resources, as most articles from the previous step only used existing datasets. This filtering was performed based on title, abstract, and full-text analysis using the following criteria:

\noindent \textbf{Inclusion criteria:}
\begin{itemize}[topsep=1pt, itemsep=3pt, parsep=0pt]
    \item Published until 2023 (inclusive),
    \item Presents novel network data traces specifically for network intrusion or anomaly detection in the cyberthreat detection context.
\end{itemize}

\noindent \textbf{Exclusion criteria:}
\begin{itemize}[topsep=1pt, itemsep=3pt, parsep=0pt]
    \item Presents an IDS method using existing data,
    \item Focuses on host-based intrusions or an intrusion detection subdomain containing environment-specific data without network traces (packets, flows),
    \item Purposed for network traffic classification or non-cyberthreat anomaly detection.

\end{itemize}

    After the filtering, 113 objects were left.

    \item \emph{Backward and forward reference search:} Aiming to compile the most comprehensive contemporary dataset list, we also reviewed papers that cited (forward search) and were cited by (backward search) those obtained in the third step using Google Scholar. This process was limited to papers from the past five years (2018-2023). Applying the same filtering criteria as in the previous step yielded 25 additional papers, bringing the total to 138.

    \item \emph{Quality assurance:} Lastly, we aimed to ensure the quality of the selected papers. While datasets backed by scientific publications are preferable, a non-negligible portion of the data is only available via informal sources (e.g., websites), so including them is also desirable.
    
    Hence, we work with a mixture of scientific papers and other resources that do not follow a standard scientific paper format or provide experimental results, making common research quality assessments \citep{yang2021_qa_slr} impractical. Since we do not aim to propose new NIDS data quality metrics, our quality check is based on a simple premise: \emph{Is the dataset useful for potential users?} We thus specify:

    \noindent \textbf{Quality-related exclusion criteria:}
    \begin{itemize}[topsep=1pt, itemsep=3pt, parsep=0pt]
        \item Was collected as auxiliary material to other research not primarily focused on data,
        \item Is heavily pre-processed, thus preventing meaningful analysis and sound conclusions,
        \item Is not publicly available.
    \end{itemize}
\end{enumerate}

\noindent Failing to reject all criteria results in the exclusion of a dataset. After their application, \emph{89 datasets remained for a final review}.


\subsection{Data Popularity Determination}
\label{assec:meth_popularity_determination}

In order to aid in the dataset selection, we also analyze dataset popularity in contemporary NIDS research. For this purpose, we focus on network intrusion/anomaly detection from Tier 1 and Tier 2 cybersecurity conferences, according to the MLSec group list\footnote{MLSec is a research group at Technical University (TU) Berlin focused on the intersection of cybersecurity and machine learning. They list top-tier cybersecurity conferences at \url{https://mlsec.org/topnotch/}. Similar rankings are also maintained by others, e.g., Guofei Gu: \url{https://people.engr.tamu.edu/guofei/sec_conf_stat.htm}.}. In particular, we surveyed the following 12 conferences: S\&P (Oakland), CCS, Security, NDSS, ACSAC, AsiaCCS, CSF, ESORICS, EuroS\&P, PETS, RAID, and DIMVA, in addition to the MLSec list.

Aiming to maximize the relevance of our findings, we analyzed only papers from the past four years (2020-2023). This was done by manually reviewing conference proceedings and conducting full-text analyses to identify used datasets.

\vspace*{3cm}

\begin{wrapfigure}{l}{0.33\linewidth}
    \includegraphics[width=\linewidth]{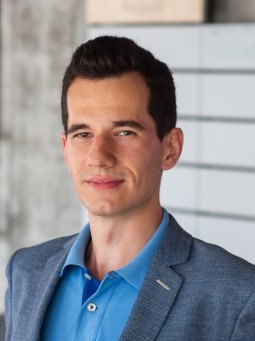}
    \vspace{-15pt}
\end{wrapfigure}
\noindent
\textbf{Patrik Goldschmidt} is a Ph.D. candidate at the Faculty of Information Technology (FIT) at Brno University of Technology (BUT), Czechia and a research assistant at Kempelen Institute of Intelligent Technologies (KInIT), Slovakia. With an interest in cybersecurity and computer networking since high school, he graduated with Bachelor's and Master's degrees from FIT BUT, focusing on DDoS detection and mitigation. In his Ph.D., Patrik studies the application of artificial intelligence and machine learning for cybersecurity, notably Network Intrusion Detection Systems (NIDSs), to enhance their performance, robustness, and trustworthiness while aiming to bridge the gap between academic research and practice.

\vspace{1em}

\begin{wrapfigure}{l}{0.33\linewidth}
    \vspace{-12pt}
    \includegraphics[width=\linewidth]{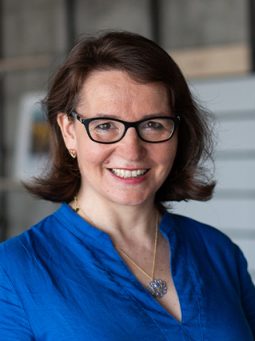}
    \vspace{-15pt}
\end{wrapfigure}
\noindent
\textbf{Daniela Chud\'a} is an associate professor currently working at Faculty of Electrical Engineering and Information Technology (FEI), Slovak University of Technology (STU) and as an expert researcher and a PhD supervisor at the Kempelen Institute of Intelligent Technologies (KInIT). She has authored or co-authored more than 60 publications in scientific journals and conferences, and regularly reviews submissions for international conferences and scientific journals. Her research is focused on  information security and privacy, in particular behavioral biometrics in the context of user authentication, malware analysis and detection, network anomaly detection, phishing and malicious behavior.

\end{document}